\documentclass[12pt,a4paper]{article}
\usepackage[utf8]{inputenc}        
\usepackage[T1]{fontenc}
\usepackage[english]{babel}
\usepackage{lmodern}               
\usepackage{geometry}              
\usepackage{xcolor}
\usepackage{booktabs}
\usepackage{longtable}
\usepackage{array}
\usepackage{amsmath,amssymb,amsfonts,amsthm}
\usepackage{bm}                    
\usepackage{mathtools}
\usepackage{graphicx}
\usepackage{float}                
\usepackage{xcolor}                
\usepackage{booktabs}             
\usepackage{caption}
\usepackage{subcaption}         
\usepackage{siunitx}               
\usepackage[numbers]{natbib}
\usepackage{url}
\usepackage[colorlinks=true,
linkcolor=blue!60!black,
citecolor=blue!60!black,
urlcolor=blue!60!black]{hyperref}
\usepackage{titlesec}
\titleformat{\section}{\large\bfseries}{\thesection.}{1em}{}
\titleformat{\subsection}{\normalsize\bfseries}{\thesubsection.}{0.5em}{}
\usepackage{setspace}
\usepackage{enumitem}     
\usepackage{csquotes}        
\usepackage{microtype}  
\usepackage{authblk}
\usepackage[font=footnotesize]{caption}

\begin{document}
\title{Inflation as an emergent phenomenon}
\author[*]{Alessio Emanuele Biondo}
\author[$\ddagger$]{Mauro Gallegati}
\affil[*]{Department of Economics and Business, University of Catania, Italy - \texttt{ae.biondo@unict.it} }
\affil[$\ddagger$]{Faculty of Economics, Marche Polytechnic University, Italy - \texttt{mauro.gallegati@univpm.it} }
\date{}
\maketitle

\begin{abstract}
We develop an agent-based model in which inflation emerges from decentralized price-setting and credit-financed production in an endogenous-money economy. Firms operate under working-capital constraints, form market-based price expectations through heterogeneous adaptive learning, and set prices via cost-plus rules with endogenous mark-ups. Bank lending simultaneously creates deposits, while heterogeneous lending rates and credit rationing shape firms’ financing costs and, through unit costs, their pricing decisions. The economy features interacting production and credit networks: intermediate-input linkages propagate cost shocks across supply chains, while bank--firm relationships transmit financial conditions across firms. The interaction of network-based pass-through, state-dependent pricing incentives, and evolving credit conditions generates inflationary regimes, including episodes driven by pricing cascades and feedback loops.
\end{abstract}

\section{Introduction}

Inflation is often discussed as if it were the direct and relatively transparent outcome of a small set of aggregate forces, such as monetary policy, aggregate demand, unit labour costs, exchange rates, or ``supply shocks''. Yet, the empirical object that policy targets and statistical agencies measure is an \emph{emergent} macroeconomic regularity: a persistent change in a price index that aggregates heterogeneous price-setting decisions, executed by firms operating in different markets, facing different cost structures, different demand conditions, and different informational constraints. The last few years have further highlighted that inflation dynamics can be abrupt, persistent, and sectorally uneven, and that propagation from upstream costs to downstream consumer prices may be both delayed and state-dependent. These features are difficult to reconcile with representative-agent narratives or with purely aggregate Phillips-curve reasoning, but they are natural in economies that are best understood as complex, networked systems in which micro-level interactions generate macro-level patterns. In this sense, an ``inflation process'' is not a primitive object: it is a statistical summary of decentralised decisions taken in environments in which feedbacks, externalities, and dispersed interactions are core structural features and not residuals \citep{Caiani2016}.

A large body of work in macroeconomics has long emphasized that inflation is intrinsically tied to micro price adjustment frictions and to the distribution of price changes across firms and goods. Time-dependent price setting \citep{Calvo1983,Taylor1979} and state-dependent price setting (as in menu-cost models and their modern empirical reassessments) imply different propagation of shocks and different degrees of inflation persistence, even when the same aggregate disturbance hits the economy. The micro evidence that prices are neither perfectly flexible nor uniformly sticky \citep{KlenowKryvtsov2008,NakamuraSteinsson2008} motivates a view of inflation as the outcome of heterogeneous adjustments interacting with heterogeneous shocks. In this perspective, since different demand and supply factors affect inflation dynamics \citep{Ciambesi2025}, the key question becomes not only \emph{whether} prices adjust, but \emph{which} prices adjust, \emph{when}, and \emph{through what structural channels} cost and demand disturbances spread across producers and markets. When micro adjustment is heterogeneous, macro inflation inherits a composition problem: the same aggregate impulse can yield different inflation outcomes depending on which nodes, sectors, and cost components move first, and on how quickly the rest of the economy reacts.

Once one takes heterogeneity seriously, the architecture of production becomes central. Intermediate inputs account for a large share of firms' costs, and these costs are shaped by supply-chain relationships that are neither random nor frictionless. Production networks define which firms are exposed to which upstream disturbances, with what intensity, and with what timing. This networked structure implies that shocks do not transmit proportionally to aggregate output or the aggregate price level; rather, they propagate along specific paths, can be amplified by bottlenecks, and may generate cascading effects when constraints bind. In network economies, local disturbances can therefore produce macroeconomic outcomes that look ``aggregate'' in the data but are in fact the emergent consequence of decentralised interactions. The network view is by now well-established in macroeconomic research \citep{Acemoglu2012}, but its implications for inflation hinge on the micro foundations of pricing, the evolution of mark-ups and market power, and the endogenous adjustment of production and credit constraints.

A conscious complexity-oriented macroeconomic view is that crises and fluctuations may arise endogenously from the interaction of balance sheets, credit terms, and inter-firm linkages, generating avalanche-like dynamics. In evolving credit-network environments, the financial position of firms and banks co-determines production, investment, and bankruptcy risk; sufficiently large shocks or sufficiently tight financial conditions can trigger chains of defaults that feed back into credit availability and loan pricing, producing amplification and persistence. This mechanism is central in the network-based financial accelerator tradition, where borrowers' net worth affects both the quantity and the price of credit, and where defaults reduce lenders' balance-sheet capacity and induce further tightening. In this class of models, the macroeconomy is not only subject to exogenous shocks; it can generate endogenous regimes characterised by clustered failures and amplified downturns \citep{DelliGatti2006,DelliGatti2010b}. A key methodological implication, stressed in the AB-SFC tradition, is that these feedbacks cannot be treated as add-ons: they require a representation in which the inter-linkages between the real and financial sides of the economy are explicit, and in which money endogeneity is not abstracted away \citep{Battiston2007}. While much of the financial-network literature focuses on output and crises, the same logic points to an inflation mechanism: financing costs are part of firms' variable costs, and changes in spreads, rationing, and working-capital conditions can pass through to posted prices, especially when supply constraints and market power make cost-plus pricing operative. Indeed, the ABM literature emphasises that even simple micro behaviours can generate complex systemic properties because of feedbacks and structural effects arising from dispersed interactions; in that spirit, credit conditions and firms' finance become natural candidates for an emergent inflation mechanism \citep{Boissay2006,GreenwaldStiglitz1993}.

The connection between networks and inflation becomes particularly sharp when one focuses on the \emph{cascading} nature of price adjustments in a production system with intermediate inputs. When upstream prices rise, downstream firms face higher unit costs; if they pass these costs through, they raise their own prices, which in turn raises costs for their downstream buyers, and so on. The macro implication is that inflation can be generated and sustained by the topology of the production network and by the sequential structure of pass-through, even if each firm follows a relatively simple pricing rule. Recent work formalises this idea explicitly by modelling \emph{pricing cascades} in networked economies, showing how input-output linkages can generate amplification, persistence, and cross-sectoral comovement of inflation, and how the aggregate inflation process may inherit the network's centralities and bottlenecks \citep{GhassibeNakov2025}. In this framework, inflation is not merely the aggregation of independent idiosyncratic price changes; it is the outcome of a structured propagation process in which local price-setting decisions are coupled through upstream cost exposure. Importantly, cascades do not require that firms coordinate; they are the emergent result of decentralised responses to local cost conditions within a connected production structure.

A further crucial element is that price-setting behaviour itself is heterogeneous and state-dependent, and it changes with the economic environment. A growing empirical literature indicates that firms' choice between time-dependent and state-dependent price adjustment is correlated with inflation, with uncertainty, and with the composition of costs, suggesting that the micro adjustment regime is endogenous to the macro regime. In particular, higher uncertainty and higher inflation are associated with a higher likelihood of state-dependent pricing, while firm size and the share of non-labour costs also matter \citep{Bunn2026}. This matters for an emergent-inflation perspective because the network propagation of costs interacts nonlinearly with the distribution of adjustment hazards: if upstream firms react more frequently or more aggressively when inflation and uncertainty are high, they can accelerate cascades; conversely, if downstream firms remain time-dependent, pass-through may be delayed and inflation persistence may rise. Therefore, emergent inflation should be thought of as a regime-dependent outcome of the interplay between (i) a networked cost structure and (ii) an endogenous micro pricing regime.

Expectations provide an additional layer of emergence. Inflation is both a realised statistical object and a belief held by firms and households, influencing future pricing, wage bargaining, and consumption/saving decisions. Yet, expectations are not formed from a common information set in a frictionless way. They are heterogeneous, shaped by attention, experiences, media exposure, and the salience of specific prices. Work on subjective inflation expectations documents systematic patterns in forecast errors across demographic groups and socioeconomic status, recurring across countries and over time: for instance, gender differences, education and income gradients, and persistent heterogeneity in levels of expected inflation \citep{DAcuntoWeber2024}. The same literature stresses that survey design advances now allow one to measure a term structure of consumers' expectations; in many periods the term structure is flat, but during the post-2020 inflation episode short-term expectations jumped earlier and more strongly, while longer-horizon expectations moved later and less, remaining nevertheless elevated relative to pre-pandemic levels \citep{WeberEtAl2022}. These patterns are macro-relevant because they speak directly to the informational and behavioural foundations of ``anchoring'': if short- and long-horizon expectations co-move in sustained inflation episodes, the conventional narrative that long-term expectations remain insulated by central bank credibility becomes empirically non-trivial and policy-relevant \citep{DAcuntoMalmendierWeber2022}. In the presence of such heterogeneity, aggregate inflation dynamics interact with the distribution of beliefs: not only can inflation affect expectations, but expectations can affect demand allocation and firms' pricing incentives, potentially generating feedback loops.

This motivates models where agents learn adaptively from the price signals they observe, with finite and heterogeneous memory, rather than holding model-consistent expectations. A key insight from agent-based work on expectations is that the interaction between expected and realised inflation can be understood as a two-way feedback: expectations affect current price dynamics, and realised inflation shapes subsequent expectations \citep{Palestrini2024}. In standard macroeconomics this loop is often closed through rational expectations \`a la \citet{Muth1961}, but the same work stresses that in environments with high ``complexity'' agents may rely on heuristics because model-consistent expectations are too demanding to implement, a point consistent with limited attention and informational frictions \citep{Sims1998,Sims2003}. Moreover, laboratory evidence suggests that while subjects often use adaptive heuristics, their forecasting performance is not as systematically biased as canonical adaptive rules would imply; this motivates expectation mechanisms in which adaptive rules are augmented by belief-correction terms that respond to past trends and help mitigate persistent errors \citep{Anufriev2019,Colasante2017,Heemeijer2009}. In other words, bounded rationality need not mean naive extrapolation: adaptive expectations can embed structured adjustment mechanisms, such as anchoring-and-adjustment heuristics, that reshape the inflation--expectations feedback loop \citep{Heemeijer2009,HommesZhu2014,Kahneman2011}. In a networked economy, this is particularly consequential because the relevant signal is not merely the aggregate price index, but the market-level and input-level prices that firms and households directly face. The implication is that expectations are likely to be \emph{local} and \emph{network-mediated}: different agents see different prices, with different volatility, and update with different speeds. As a result, the macro inflation process can emerge from the co-evolution of prices and beliefs under decentralised and heterogeneous learning.

A natural methodological consequence of these considerations is that inflation should be studied in frameworks capable of representing heterogeneity, networks, discrete interactions, and out-of-equilibrium adjustment. Agent-based macroeconomics is specifically designed for this purpose, because it builds macro outcomes from micro behavioural rules and explicit interaction structures, while allowing for bounded rationality and evolving states. In addition, agent-based models can be integrated with rigorous accounting and financial consistency, making them suited for analysing monetary and credit channels in inflation dynamics. Stock-flow consistent macroeconomics, stemming from \citet{Godley1997} and \citet{GodleyLavoie2007}, provides a discipline in which every flow originates somewhere and goes somewhere and every stock is recorded as an asset for someone and a liability for someone, eliminating ``financial black holes'' and imposing a coherence that is naturally aligned with national accounts. This is particularly relevant for inflation as an emergent phenomenon because mechanisms involving credit creation, interest payments, cash-flow constraints, and balance-sheet feedbacks can otherwise be misrepresented or implicitly ruled out. At the same time, traditional stock-flow consistent (SFC) models are often highly aggregated; agent-based modeling (ABM) approaches help overcome this limitation by introducing heterogeneity, decentralised matching, and network-based balance sheet relationships, enabling the analysis of selection, self-organisation, contagion, and the propagation of shocks through financial linkages \citep{Nier2007}. In this regard, ABM contributions have highlighted how credit-network topology can spread financial fragility via contagion effects \citep{DelliGatti2005,DelliGatti2008,DelliGatti2010b}, while related work investigates links between business cycles and monetary aggregates \citep{Cincotti2010} and the role of regulation and capital requirements in debt dynamics and cycles \citep{Raberto2012}; these are precisely the ingredients that one expects to matter for cost-based pricing and, ultimately, for inflation regimes.

The aim of this paper is to develop an agent-based framework in which inflation arises endogenously from the interaction of (i) monopolistic competition with heterogeneous mark-ups, (ii) production networks linking intermediate inputs across goods, (iii) endogenous money creation and firm-specific financing costs, and (iv) heterogeneous expectation formation based on market-level price signals and an aggregate inflation anchor. The conceptual contribution is to treat inflation not as an imposed shock process nor as the reduced-form outcome of an aggregate Phillips curve, but as the emergent macro pattern generated by decentralised pricing and quantity decisions in a networked economy subject to financial constraints and heterogeneous learning. The methodological contribution is to provide a coherent mechanism through which micro-level cost changes propagate as pricing cascades along supply chains, while credit conditions and belief dynamics shape both the amplitude and persistence of the resulting inflation process.

Our results will show that the baseline economy remains active and non-inflationary and that inflation appears when specific sources of pressure are activated and when the structure of the economy allows those pressures to propagate. Mark-up pressure is the strongest inflationary source, but it is accompanied by a sharp contraction in output. Financial-cost shocks and policy-rate shocks generate more moderate cost-push inflation through firms' borrowing costs. Natural-capital shocks operate upstream and become consumer-price inflation only when downstream production is sufficiently exposed to intermediate inputs. Production networks amplify pass-through when downstream exposure is high, but dense interdependence within the intermediate sector mainly generates fragility. Expectations do not create inflation from nothing; they amplify and prolong price dynamics once an inflationary pressure is already present. These results have direct policy implications. If inflation is an emergent outcome of heterogeneous cost structures, network propagation, credit conditions, mark-up dynamics, and expectation feedbacks, then a purely aggregate reaction to the observed CPI is not sufficient. In particular, a uniform increase in the policy rate inappropriate and even counterproductive when inflation is driven by financing costs, upstream bottlenecks, natural-resource costs, or mark-up dynamics. The policy problem should be, therefore, how to identify the mechanism that generates inflation.

The perspective we propose has several implications. First, it highlights that the inflation process is intrinsically linked to market structure and mark-up dynamics: changes in competitive performance can translate into mark-up adjustments, which interact with cost shocks and network propagation, potentially generating persistent inflation even when primary disturbances dissipate. Second, it emphasises that the micro pricing regime is not fixed: inflation and uncertainty can shift firms toward state-dependent adjustment, altering pass-through speed and persistence. Third, it stresses that expectations are heterogeneous and experience-based: demographic and socioeconomic gradients, horizon-specific dynamics, and episodes in which short- and long-term expectations co-move in ways that challenge simple anchoring narratives all suggest that the distribution of beliefs matters for macro dynamics. In a complex economy, these belief distributions can feed back into pricing and demand decisions, reinforcing emergent dynamics, and they can do so even when agents rely on boundedly rational heuristics rather than model-consistent forecasting rules \citep{Palestrini2024,Sims1998,Sims2003}. Finally, it places the role of production and credit networks at the centre of inflation analysis: not all shocks matter equally, and disturbances to central suppliers or to financially constrained nodes can generate disproportionately large inflationary effects, consistent with the network role and with cascade mechanisms. 

In sum, the emerging picture is that inflation is best understood as a complex-system outcome: a macro variable generated by a web of uncoordinated interactions among firms, banks, and households, where network structure, financial conditions, and heterogeneous learning jointly determine propagation, amplification, and persistence. The model developed in this paper integrates micro frictions, network topology, and endogenous credit creation to study the evolution of inflationary regimes, including episodes generated by cascades and feedback loops rather than by a single aggregate driver.

The remainder of the paper is organised as follows. Section~2 presents the model. Section~3 describes the simulation design and discusses the main results and policy implications. Section~4 concludes.

\section{The Model}
The economy is populated by a set of banks $b\in\mathcal{B}$ and a set of firms $i\in\mathcal{I}$ operating in two productive sectors: a capital/intermediate-goods sector $K$ and a consumption-goods sector $C$. Money is endogenously created through bank lending: firms finance production costs via bank credit, which simultaneously creates deposits. Banks set firm-specific lending rates $r_{i,t}$ by adding an idiosyncratic spread (mark-up) to the policy rate $\rho_t$ fixed by the Central Bank. Hence, the credit channel determines the effective financing cost of production, while the Central Bank accommodates the reserve needs induced by credit creation at the policy rate. Consumers/workers $h\in\mathcal{H}$ earn their income by means of their employment in firms and can purchase consumption goods, according to their preferences.

\subsection{Goods, varieties, and market structure}
Each sector produces a \emph{finite} number of \emph{goods} (markets), while each good is supplied by multiple firms in monopolistic competition (interpretable as brand differentiation). Let $\mathcal{G}^C=\{1,\dots,G_C\}$ denote the set of consumption goods (c-goods) and $\mathcal{G}^K=\{1,\dots,G_K\}$ the set of capital/intermediate goods (k-goods). Each firm belongs to one sector only $s(i)\in\{C,K\}$ and produces one good $g(i)\in\mathcal{G}^{s(i)}$. For each sector--good pair $(s,g)$, define the set of firms operating in that market as
\begin{equation}
	\mathcal{I}_{g}^{s}\equiv\{\, i\in\mathcal{I}\;:\; s(i)=s,\; g(i)=g\,\} \qquad n_{g}^{s}\equiv\left|\mathcal{I}_{g}^{s}\right|
\end{equation}
Firms within the same market $g$ set heterogeneous prices $p_{i,t}$ and compete to match demand for that good. Let $\Phi_{i,t}$ denote realised sales of firm $i$ and $Q_{i,t}$ its realised output; market-level output and sales are simply: $\Omega_{g,t}^{s}\equiv\sum_{i\in\mathcal{I}_{g}^{s}}Q_{i,t}$ and $\Phi_{g,t}^{s}\equiv\sum_{i\in\mathcal{I}_{g}^{s}}\Phi_{i,t}$. 

\subsection{Production technologies and production networks}
Firms produce using labour and intermediate inputs purchased through a \emph{production network} defined at the \emph{good level}. The network is designed to be coherent with monopolistic competition: firms producing the same good share the same set of upstream input goods (and therefore the same set of potential supplier markets), though they may purchase from different individual suppliers (varieties) within those upstream markets. We represent input linkages through two directed adjacency matrices:
\begin{align}
	\mathbf{A}^{KC} &\in\{0,1\}^{G_K\times G_C}
	&
	A_{g_K,g_C}^{KC}=1 \;\Longleftrightarrow\; \text{k-good}g_K\text{ is an input for c-good }g_C
	\\
	\mathbf{A}^{KK} &\in\{0,1\}^{G_K\times G_K}
	&
	A_{g'_K,g_K}^{KK}=1 \;\Longleftrightarrow\; \text{k-good}g'_K\text{ is an input for k-good}g_K
\end{align}
Equivalently, for each $g_C\in\mathcal{G}^C$ define the upstream set of required k-goods markets
\begin{equation}
	\mathcal{S}_{g_C}^{C}\equiv\{\,g_K\in\mathcal{G}^K:\; A_{g_K,g_C}^{KC}=1\,\}
\end{equation}
and for each $g_K\in\mathcal{G}^K$ define its upstream k-good markets
\begin{equation}
	\mathcal{S}_{g_K}^{K}\equiv\{\,g'_K\in\mathcal{G}^K:\; A_{g'_K,g_K}^{KK}=1\,\}
\end{equation}
If $i,i'\in\mathcal{I}_{g}^{s}$ produce the same good, they share the same upstream sets $\mathcal{S}_{g}^{s}$.

Given the upstream set of required markets, each firm type, $s(i) \in \{C,K\}$, purchases inputs from a subset of \emph{firms} (varieties) operating in those markets. The corresponding feasible supplier set is
\begin{equation}
	\mathcal{U}_{i,t}^{s(i)}
	\equiv
	\bigcup_{\tilde g_K\in\mathcal{S}^{s(i)}_{g(i)}} \mathcal{I}_{\tilde g_K}^{K}
	\label{feasible_suppliers_unified}
\end{equation}
The feasibility constraint is always network-consistent because it is inherited from the good-to-good structure; conditional on feasibility, however, upstream suppliers are selected through a probabilistic choice mechanism that trades off \emph{convenience} (low input prices) against \emph{attractiveness} (brand differentiation proxied by mark-ups). In particular, when firm $i$ needs to purchase an intermediate input from an upstream market, it evaluates each feasible supplier $j\in\mathcal{U}^{s(i)}_{i,t}$ through an attractiveness index
\begin{equation}
	\mathcal{A}_{j,t}
	=
	\frac{\left(1+\mu_{j,t}\right)^{\psi}}{p_{j,t}^{\varphi}}
	\qquad \psi\ge 0,\;\varphi\ge 0
	\label{supplier_attractiveness}
\end{equation}
where $p_{j,t}$ is the price charged by supplier $j$ and $\mu_{j,t}$ is its idiosyncratic mark-up. The parameter $\psi$ captures the strength of brand/quality considerations (higher mark-ups proxy stronger differentiation and perceived reliability), while $\varphi$ captures the sensitivity to input prices. Supplier selection is then probabilistic: firm $i$ draws its supplier from the feasible set with probability proportional to attractiveness; purchases are executed with a sequential rationing rule. For each required upstream market $\tilde g_K$, firm $i$ draws candidate suppliers in $\mathcal{I}^{K}_{\tilde g_K}$ according to eq.\eqref{supplier_choice_prob}. If the selected supplier cannot satisfy the required quantity because its available stock is insufficient, firm $i$ purchases the available amount and draws another supplier from the same upstream market (without replacement) until either the required quantity is satisfied or no further suppliers are available. If the required quantity cannot be fully sourced in any upstream market, planned output is scaled down to the maximum feasible level implied by the scarcest required input, and all input requirements are recomputed consistently with the reduced output level.

Operationally, after attempting to source inputs from all available suppliers in each required upstream market $\tilde g_K\in\mathcal{S}^{s(i)}_{g(i)}$, we define the feasible output level as
\begin{equation}
	\bar Q_{i,t}
	=
	\min_{\tilde g_K\in\mathcal{S}^{s(i)}_{g(i)}}
	\left\{
	\frac{\sum_{j\in\mathcal{I}^{K}_{\tilde g_K}} \bar x_{i\leftarrow j,t}}{a_{X,i,\tilde g_K}}
	\right\},
	\label{Qbar_def}
\end{equation}
where $\bar x_{i\leftarrow j,t}$ denotes the quantity of the intermediate input actually obtained by firm $i$ from supplier $j$ in market $\tilde g_K$ in period $t$. Planned output is then updated as $Q^{\text{exp}}_{i,t}\leftarrow\min\{Q^{\text{exp}}_{i,t},\bar Q_{i,t}\}$, and all input requirements are recomputed consistently.
\begin{equation}
	\Pr\!\left(j\mid i,t,\tilde g_K\right)
	=
	\frac{\mathcal{A}_{j,t}}{\sum_{m\in\mathcal{I}^{K}_{\tilde g_K}}\mathcal{A}_{m,t}}
	\qquad j\in\mathcal{I}^{K}_{\tilde g_K},\ \ \tilde g_K\in\mathcal{S}^{s(i)}_{g(i)}
	\label{supplier_choice_prob}
\end{equation}
This stochastic choice rule generates heterogeneous and time-varying input-cost exposures even among firms producing the same good, while preserving structural coherence: firms producing the same good share the same upstream \emph{markets} (hence the same feasible supplier pools), yet may end up purchasing from different varieties within those pools. As a result, idiosyncratic pricing decisions by upstream suppliers propagate along production networks through a decentralised and uncoordinated pass-through mechanism, contributing to the emergence and persistence of inflation at the aggregate level. 

\subsubsection{Firm--firm interactions: production-network links}
Firm-to-firm interactions arise because each firm purchases intermediate inputs from upstream suppliers through the production network defined at the good level. As specified above, the technology layer determines, for each firm $i$, the set of feasible upstream supplier markets $\mathcal{S}^{s(i)}_{g(i)}$ and, consequently, the feasible supplier set of varieties $\mathcal{U}_{i,t}^{s(i)}$ as defined in eq.\eqref{feasible_suppliers_unified}.

Conditional on feasibility, actual purchases induce a time-varying firm--firm network. Let $\mathbf{G}^{FF}_t=(\mathcal{I},\mathcal{E}^{FF}_t)$ denote the directed firm--firm graph at time $t$, where a directed edge $(i\leftarrow j)\in\mathcal{E}^{FF}_t$ indicates that firm $i$ purchases at least one input from firm $j$ in period $t$. Equivalently, define the firm--firm adjacency matrix $\mathbf{M}^{FF}_t\in\{0,1\}^{|\mathcal{I}|\times|\mathcal{I}|}$ as
\begin{equation}
	M^{FF}_{i j,t}=1
	\;\Longleftrightarrow\;
	j\in\mathcal{U}_{i,t}^{s(i)} \ \text{and}\ x_{i\leftarrow j,t}>0
	\label{MFF_def}
\end{equation}
The network is dynamic because supplier selection is repeated each period and depends on suppliers' posted prices and mark-ups. In particular, when firm $i$ needs to source an input, it evaluates each feasible supplier $j\in\mathcal{U}^{s(i)}_{i,t}$ using the convenience--attractiveness index eq.\eqref{supplier_attractiveness}, and draws suppliers probabilistically according to eq.\eqref{supplier_choice_prob}. As a result, changes in upstream prices $p_{j,t}$ and mark-ups $\mu_{j,t}$ affect downstream input costs and downstream prices.

\subsubsection{Input requirements and production.}
To keep the supply-chain logic transparent, we assume fixed-coefficient (Leontief-type) input requirements at the firm level. For each firm $i$, planned output $Q_{i,t}^{\text{exp}}$ requires labour and intermediate inputs proportional to output. Thus, expected factor requirements are:
\begin{align}
	N_{i,t}^{\text{exp}} &= a_{N,i}\, Q_{i,t}^{\text{exp}},
	\\
	X_{i,\tilde g_K,t}^{\text{exp}} &= a_{X,i,\tilde g_K}\, Q_{i,t}^{\text{exp}},\qquad \tilde g_K\in\mathcal{S}^{s(i)}_{g(i)}
\end{align}
with coefficients $a_{N,i}>0$ and $a_{X,i,\tilde g_K}>0$ for all $\tilde g_K\in\mathcal{S}^{s(i)}_{g(i)}$. For each required upstream market $\tilde g_K\in\mathcal{S}^{s(i)}_{g(i)}$, firm $i$ selects a supplier $j\in\mathcal{I}^{K}_{\tilde g_K}$ according to eq.\eqref{supplier_choice_prob} (computed within that market) and attempts to purchase the required quantity $X^{\text{exp}}_{i,\tilde g_K,t}$ subject to suppliers' available stocks. For c-firms ($s(i)=C$), required intermediates are k-goods purchased from $\mathcal{U}_{i,t}^{C}$. For k-firms ($s(i)=K$), required intermediates are k-goods purchased from $\mathcal{U}_{i,t}^{K}$ and production additionally requires natural capital:
\begin{equation}
	NK_{i,t}^{\text{exp}} = a_{NK,i}\, Q_{i,t}^{\text{exp}}, \qquad s(i)=K,\;\; a_{NK,i}>0
\end{equation}

Realised production $Q_{i,t}$ may fall short of $Q_{i,t}^{\text{exp}}$ if inputs cannot be fully financed or obtained (credit rationing, upstream shortages, or both). This mechanism makes network bottlenecks and financial conditions jointly relevant for realised output and, therefore, for aggregate imbalances and inflation. Output expectations are correspondingly reduced to the feasible level implied by the prevailing input-availability and credit constraints, i.e.\ $Q_{i,t}=\min\{Q_{i,t}^{\text{exp}},\,\bar Q_{i,t}\}$, where $\bar Q_{i,t}$ denotes the maximum output attainable given realised access to required inputs and working-capital finance in period $t$.

Starting from an initial value $Q^{\text{exp}}_{i,0}>0$, at the end of period $t$ each firm records inventories $Q^{i}_{i,t}\ge 0$ (the quantity of produced units that remain unsold and are carried over) and unmet demand $Q^{u}_{i,t}\ge 0$, defined as the gap between total demand addressed to firm $i$ and realised unit sales:
\[
Q^{u}_{i,t}\equiv \max\{0,\; D_{i,t}-\Phi_{i,t}\}
\]
where $D_{i,t}$ denotes total demand (in units) directed to firm $i$ in period $t$ and $\Phi_{i,t}$ are realised sales. Analogously, household-level unmet consumption demand is defined as the difference between desired purchases and realised purchases at posted prices, and aggregates consistently into firms' unmet demand through the demand allocation mechanism. Expected production for the next period is then updated as
\begin{equation}
Q^{\text{exp}}_{i,t+1}
=
(1-\alpha_Q)Q^{\text{exp}}_{i,t}
+
\alpha_Q
\max\Big\{0,\; Q_{i,t}+Q^u_{i,t}-Q^i_{i,t}\Big\},
\qquad 0<\alpha_Q\le 1
	\label{Qexp_update}
\end{equation}
Accordingly, unmet demand increases planned output, while inventories reduce it, generating a minimal yet effective feedback loop between sales realisations, production plans, and downstream/upstream propagation through the production network.

\subsubsection{Unit costs}
Let the wage rate $w_t$ follow an exogenous AR(1) process, as in Palestrini et al (2024)
\begin{equation}
	w_t=\omega_w w_{t-1} + d + \epsilon_w
	\label{wageAR_repeated}
\end{equation}
where $d$ is the drift, $\omega_w \in (0,1)$ is the autoregressive coefficient and $\epsilon_w \sim \mathcal{N}(0,\sigma)$ is a Gaussian noise;  further, let $p_{j,t}$ be the price charged by upstream supplier $j$. Denote by $x_{i\leftarrow j,t}$ the quantity of intermediate input purchased by $i$ from $j$ in period $t$, and by $p_{NK,t}$ the unit price of natural capital (exogenous or policy-controlled in the baseline). We define total variable costs of firm $i$ as
\begin{equation}
	\mathcal{C}_{i,t}
	=
	w_t\,N_{i,t}
	+
	\sum_{j\in\mathcal{U}_{i,t}^{s(i)}} p_{j,t}\,x_{i\leftarrow j,t}
	+
	p_{NK,t}\, NK_{i,t}
	+
	\mathcal{F}_{i,t}
	\label{costs_def}
\end{equation}
where $\mathcal{F}_{i,t}$ captures financing costs associated with working capital and/or loans (see below). Consumption firms do not use natural capital, hence $NK_{i,t}=0$ for $s(i)=C$. The corresponding unit cost is
\begin{equation}
	\textsc{uc}_{i,t}
	\equiv
	\frac{\mathcal{C}_{i,t}}{Q_{i,t}}
	\label{unit_cost}
\end{equation}
When $Q_{i,t}=0$, unit costs are not defined. In this case we impose the convention that the posted price does not change, $p_{i,t}=p_{i,t-1}$, and mark-ups are updated only if realised unit sales are strictly positive, $\Phi_{i,t}>0$.

\subsubsection{Mark-ups}
Each firm has an idiosyncratic mark-up parameter $\mu_{i,t}>0$ that evolves endogenously over time. Mark-ups are initialised at $\mu_{i,0}>0$ and then updated as a function of the firm's \emph{quantity-based} competitive performance within its sector, captured by its sales share in physical units (rather than revenues). Let $\Phi_{i,t}$ denote realised sales (units) of firm $i$ in period $t$ and $\mathcal{I}^{s}$ the set of firms in sector $s\in\{C,K\}$. The sectoral sales share of firm $i$ is
\begin{equation}
	\xi_{i,t}
	\equiv
	\frac{\Phi_{i,t}}{\sum_{j\in\mathcal{I}^{s(i)}}\Phi_{j,t}}
	\label{sales_share_units}
\end{equation}
Mark-ups follow an adaptive rule that combines competitive performance, success-contingent pricing power, excess-demand pressure, and inventory discipline. In particular, firms whose unit sales share increases may expand their mark-up, while firms losing market share face pressure to compress it. At the same time, firms that successfully sell most of the quantity brought to the market may perceive stronger pricing power, whereas unsold production disciplines mark-up adjustment. A parsimonious specification is
\begin{equation}
	\mu_{i,t+1}
	=
	\begin{cases}
		\max\Big\{
		\underline{\mu},\;
		\mu_{i,t}
		+
		\zeta_{\mu}\left(\xi_{i,t}-\xi_{i,t-1}\right)
		+
		\zeta_{g}\left[\vartheta_{i,t}-\bar{\vartheta}\right]_{+}
		+
		\zeta_{u}\,u_{i,t}
		-
		\zeta_{I}\,\iota_{i,t}
		\Big\},
		& \text{if } \Phi_{i,t}>0,\\[6pt]
		\mu_{i,t},
		& \text{if } \Phi_{i,t}=0,
	\end{cases}
	\label{markup_update}
\end{equation}
where $[x]_{+}\equiv\max\{x,0\}$. The first term inside the adjustment rule captures the change in firm $i$'s quantity-based competitive performance. The variable
\[
u_{i,t}\equiv \frac{Q^{u}_{i,t}}{D_{i,t}}
\]
measures unmet-demand pressure, with $D_{i,t}>0$ denoting total demand directed to firm $i$. The sell-through indicator
\[
\vartheta_{i,t}
\equiv
\frac{\Phi_{i,t}}{\Phi_{i,t}+Q^{i}_{i,t}}
\]
measures the extent to which firm $i$ succeeds in selling the quantity effectively brought to the market, while
\[
\iota_{i,t}
\equiv
\frac{Q^{i}_{i,t}}{Q_{i,t}}
\]
captures the share of current production that remains unsold and therefore disciplines mark-up adjustment. The parameter $\bar{\vartheta}\in[0,1]$ is the sell-through threshold above which success-contingent mark-up pressure becomes active. The coefficients $\zeta_{\mu}\ge0$, $\zeta_g\ge0$, $\zeta_u\ge0$, and $\zeta_I\ge0$ govern, respectively, the responsiveness of mark-ups to changes in sales shares, success-contingent pricing power, unmet demand, and inventories. The lower bound $\underline{\mu}>0$ prevents non-positive mark-ups.

This specification nests the simpler adaptive mark-up rule as a limiting case. When $\zeta_g=\zeta_u=\zeta_I=0$, mark-ups respond only to changes in sectoral unit-sales shares. When $\zeta_g>0$, firms with high sell-through rates can raise mark-ups in response to successful market performance; however, this mechanism remains disciplined by demand because higher prices reduce firms' attractiveness in the probabilistic choice rules governing households' purchases and firms' input sourcing. When $\zeta_u>0$, local excess demand raises perceived pricing power, whereas $\zeta_I>0$ makes unsold production exert downward pressure on mark-ups.

\subsubsection{Expected price setting}
Expectations are formed using information from two distinct levels: the firm's own market and the aggregate price dynamics of the economy. The expectation mechanism builds on the heterogeneous adaptive-learning algorithm proposed by Palestrini \textit{et al.} (2024), extended here to market-specific prices and to richer belief-correction dynamics based on finite and heterogeneous memory, while also allowing a macro-level inflation signal to enter expectation formation. 

Let $p_{i,t}$ denote the price set by firm $i$ in period $t$ and let $\Phi_{i,t}$ denote its realised sales (in units). For each sector $s\in\{C,K\}$ and each good $g$, define the corresponding market as the set of firms $\mathcal{I}_{g}^{s}\equiv\{\, i:\; s(i)=s,\; g(i)=g\,\}$. The market price of good $g$ in sector $s$ is the sales-weighted average of firm-level prices within that market:
\begin{equation}
	p_{g,t}^{s}
	=
	\sum_{i\in\mathcal{I}_{g}^{s}}
	\left(
	\frac{\Phi_{i,t}}{\sum_{j\in\mathcal{I}_{g}^{s}}\Phi_{j,t}}
	\right)
	p_{i,t}
	\label{pg_def}
\end{equation}
This definition ensures that firms with higher realised sales carry greater weight in determining the price signal observed by other firms operating in the same market. In case  $\sum_{j\in\mathcal{I}^{s}_{g}}\Phi_{j,t}=0$, we set $p^{s}_{g,t}=p^{s}_{g,t-1}$.

Firm $i$ observes the past market price $p_{g(i),t-1}^{s(i)}$ and sets its expected market price $p^{e}_{i,t}$ as
\begin{equation}
	p^{e}_{i,t}
	=
	\lambda_i\, p^{s(i)}_{g(i),t-1}
	+
	(1-\lambda_i)\,p^{e}_{i,t-1}
	+
	\Delta^{e}_{i,t}
	+
	\chi_{\pi}\,\pi_{t-1} 
	\qquad 0<\lambda_i<1
	\label{expect_rule_simplified}
\end{equation}
where $\lambda_i$ is a firm-specific gain parameter, $\Delta^{e}_{i,t}$ is a belief-correction term capturing the expected one-period change in the market price of the good produced by $i$, and $\pi_{t-1}$ is the aggregate inflation rate of the economy (defined below). The parameter $\chi_{\pi}\ge 0$ controls the strength with which the macro inflation signal enters market-level expectations.

Define the one-period change in the relevant market price as
\begin{equation}
	\Delta p_{g(i),t-h}^{s(i)} \equiv p_{g(i),t-h}^{s(i)} - p_{g(i),t-h-1}^{s(i)}
	\label{deltap_def}
\end{equation}
To allow for richer expectation dynamics than na\"{\i}ve extrapolation, belief correction is formed using a finite memory window of heterogeneous length $\varsigma_i\in\mathbb{N}$:
\begin{equation}
	\Delta^{e}_{i,t}
	=
	\sum_{h=1}^{\varsigma_i}\omega_{i,h,t}\,\Delta p_{g(i),t-h}^{s(i)}
	\qquad 
	\omega_{i,h,t}\ge 0
	\qquad
	\sum_{h=1}^{\varsigma_i}\omega_{i,h,t}=1
	\label{belief_correction_simplified}
\end{equation}

We consider alternative specifications for the weighting scheme $\{\omega_{i,h,t}\}_{h=1}^{\varsigma_i}$, which govern how firms process past information within the window and therefore how local price dynamics are propagated into expectations:
\begin{itemize}
	\item[-] Equal weights
	\begin{equation}
		\omega_{i,h,t}=\frac{1}{\varsigma_i}
		\qquad h=1,\dots,\varsigma_i
		\label{weights_equal}
	\end{equation}
	\item[-] Geometric decay
	\begin{equation}
		\omega_{i,h,t}
		=
		\frac{(1-\theta)\theta^{h-1}}{\sum_{m=1}^{\varsigma_i}(1-\theta)\theta^{m-1}}
		\qquad 0<\theta<1
		\qquad h=1,\dots,\varsigma_i
		\label{weights_geom}
	\end{equation}
	\item[-] Magnitude-based weighting
	\begin{equation}
		\omega_{i,h,t}
		=
		\frac{\left|\Delta p^{s(i)}_{g(i),t-h}\right|^{\gamma}}{\sum_{m=1}^{\varsigma_i}\left|\Delta p^{s(i)}_{g(i),t-m}\right|^{\gamma}}
		\qquad \gamma\ge 0
		\qquad h=1,\dots,\varsigma_i
		\label{weights_mag}
	\end{equation}
	where $\gamma=0$ collapses to equal weights.
	\item[-] Combined timing--magnitude weighting
	\begin{equation}
		\omega_{i,h,t}
		=
		\frac{\theta^{h-1}\left|\Delta p^{s(i)}_{g(i),t-h}\right|^{\gamma}}
		{\sum_{m=1}^{\varsigma_i}\theta^{m-1}\left|\Delta p^{s(i)}_{g(i),t-m}\right|^{\gamma}}
		\qquad 0<\theta<1,\;\gamma\ge 0
		\qquad h=1,\dots,\varsigma_i
		\label{weights_combined}
	\end{equation}
\end{itemize}
In this framework, heterogeneity in $\lambda_i$, $\varsigma_i$, and the weighting parameters governs how quickly firms react to market-specific price signals and how strongly local shocks propagate into expectations. At the same time, the common macro signal $\pi_{t-1}$ provides a potential feedback channel through which aggregate inflation can influence decentralised beliefs.
\subsubsection{Pricing rule}
Given unit costs $\textsc{uc}_{i,t}$, the endogenous mark-up $\mu_{i,t}$, and market-specific price expectations $p^{e}_{i,t}$ defined in eq.\eqref{expect_rule_simplified}, each firm sets its price as:
\begin{equation}
	p_{i,t}
	=
	(1+\mu_{i,t})\,\textsc{uc}_{i,t}\left(1+\kappa\,\pi^{\,e}_{i,t}\right)
	\qquad \kappa\ge 0
	\label{pricing_rule}
\end{equation}
where $\pi^{\,e}_{i,t}$ is the expected rate of price change of firm $i$, that we define alternatively:
\begin{equation}
	\pi^{\,e}_{i,t}
	\equiv
	\begin{cases}
		\quad \dfrac{p^{e}_{i,t}-p^{e}_{i,t-1}}{p^{e}_{i,t-1}}
		& \text{expectations-on-expectations anchoring}\\[18pt]
		\quad \dfrac{p^{e}_{i,t}-p_{i,t-1}}{p_{i,t-1}}
		& \text{price anchoring}
	\end{cases}
	\label{pi_e_expect_anchor}
\end{equation}
When $\kappa=0$, firms follow pure cost-plus pricing; as $\kappa$ increases, expectations amplify price adjustments and can generate persistence in market-level and aggregate inflation dynamics even when unit costs stabilise. To assess the quantitative contribution of this channel, Appendix~\ref{app:kappa_zero} compares the pure cost-plus case, $\kappa=0$, with the expectation-augmented specification, $\kappa=0.35$, under no independent price pressure and under mark-up, bank-cost, policy-rate, and natural-capital pressures. This exercise isolates the incremental effect of expectations on inflation and output dynamics.

\subsection{Banking, endogenous money, and credit conditions}
Each firm finances production through credit obtained from one bank. The policy rate is $\rho_t$. Each bank $b$ is characterized by a time-invariant mark-up $m_b$ drawn at $t=0$ from a distribution $\mathcal{D}_m(\cdot)$ whose dispersion is controlled by a parameter. The loan rate applied to firm $i$ is firm-specific and given by $r_{i,t}=\rho_t+m_{b(i,t)}+\varepsilon_{i,t}$, where $\varepsilon_{i,t}$ is an idiosyncratic component (or, under endogenous scoring, a score-implied spread adjustment). In the endogenous-money setting, the central bank accommodates the banking system's demand for reserves at the policy rate, while the quantity of reserves adjusts endogenously to credit creation and settlement needs. Bank profits are defined as
\begin{equation}
	\pi_{b,t}
	=
	\sum_{i\in\mathcal{I}_{b,t-1}} r_{i,t}\,L_{i,t-1}
	-
	\text{LL}_{b,t}
	-
	w_t N_{b,t}
	-
	\rho_{t}\,B_{b,t-1}
\end{equation}
where $\mathcal{I}_{b,t-1}$ denotes the set of firms with an outstanding loan from bank $b$ at time $t-1$; $\sum_{i\in\mathcal{I}_{b,t-1}} r_{i,t}\,L_{i,t-1}$ is the interest revenue on outstanding loans; $LL_{b,t}$ is the value of loan losses; $w_t N_{b,t}$ is the wage bill for bank personnel; and $\rho_{t}\,B_{b,t-1}$ is the interest cost on central-bank funding. Credit losses are modelled as a probabilistic event driven by an exogenous default-risk parameter $\delta\in[0,1]$: for any requested loan, bank approval occurs with probability $1-\delta$ in the baseline probabilistic regime. In an alternative endogenous-scoring regime, approval and spreads depend on a bank-specific score aggregating an expected profitability component, a risk component (e.g.\ bank--firm trait mismatch or borrower fragility), and a financial-need/capacity component, thereby endogenising both credit rationing and heterogeneous financing costs.

To connect finance to pricing and production costs, we model working-capital finance in reduced form by letting a fraction $\chi\in[0,1]$ of variable costs require credit each period. Thus, firm $i$ borrows the amount $L_{i,t}$, defined as
\begin{equation}
	L_{i,t}
	=
	\chi\,\Bigg(
	w_t\,N_{i,t}
	+
	\sum_{j\in\mathcal{U}_{i,t}^{s(i)}} p_{j,t}\,x_{i\leftarrow j,t}
	+
	p_{NK,t}\,NK_{i,t}
	\Bigg)
	\label{loan_request}
\end{equation}
consumption firms do not use natural capital, hence $NK_{i,t}=0$ for $s(i)=C$. Obtaining funds at rate $r_{i,t}$, financing costs are
\begin{equation}
	\mathcal{F}_{i,t} = r_{i,t}\,L_{i,t}
	\label{finance_cost}
\end{equation}
Tighter credit conditions (higher spreads or rationing) directly increase unit costs eq.\eqref{unit_cost} and therefore firms' posted prices via the pricing rule eq.\eqref{pricing_rule}. This channel is central to the emergence of inflation in an endogenous-money economy: credit allocation and network concentration shape the distribution of financing costs and propagate cost shocks through production networks into the price dynamics of multiple markets.

\subsubsection{Bank--firm interactions: credit-network links}
Financial interactions are, thus, organised through a temporal bipartite credit network linking banks $b\in\mathcal{B}$ and firms $i\in\mathcal{I}$. Let $\mathbf{G}^{BF}_t=(\mathcal{B}\cup\mathcal{I},\mathcal{E}^{BF}_t)$ denote the bank--firm bipartite graph at time $t$, where a link $(i\leftarrow b)\in\mathcal{E}^{BF}_t$ indicates that bank $b$ provides credit to firm $i$ in period $t$. Define the corresponding bipartite adjacency matrix $\mathbf{M}^{BF}_t\in\{0,1\}^{|\mathcal{I}|\times|\mathcal{B}|}$ as
\begin{equation}
	M^{BF}_{i b,t}=1
	\;\Longleftrightarrow\;
	L_{i,t}>0 \ \text{and}\ b=b(i,t)
	\label{MBF_def}
\end{equation}
where $b(i,t)\in\mathcal{B}$ denotes the lending bank selected by firm $i$ in period $t$ and $L_{i,t}$ is the granted loan. The network is dynamic because firms may revise their lending relationships over time and because credit approval is not guaranteed.

In the \emph{exogenous probabilistic} regime, for any requested loan, bank $b$ approves with probability $1-\delta$ ($\delta\in[0,1]$ exogenous), implying that credit expansion is subject to an externally imposed approval/default risk. In the \emph{endogenous scoring} regime, approval and the applied spread depend on a bank-specific score aggregating an expected profitability component (e.g.\ expected profit or repayment capacity), a risk component (e.g.\ bank--firm mismatch in behavioural traits or borrower fragility), and a financial-need/capacity component (requested amount relative to available lending capacity). Hence, the realised firm-specific lending rate $r_{i,t}$ is heterogeneous and time-varying, and affects firms' financing costs eq.\eqref{finance_cost} and, through unit costs eq.\eqref{unit_cost}, the pricing rule eq.\eqref{pricing_rule}.

\subsection{Household consumption and demand allocation}

Households consume out of current income and purchase consumption goods from c-firms. Consistently with the supplier-selection mechanism used for firms' input purchases, households choose among differentiated varieties within each c-good market using a probabilistic rule based on an attractiveness--convenience indicator that combines brand appeal (proxied by mark-ups) and prices. The aggregate marginal propensity to consume, $c$, is considered as an average of a higher one for workers, $c_W$, and a lower one for profit recipients (entrepreneurs and bankers), $c_X$. Let $y_{h,t}$ denote household income in period $t$. Individual consumption expenditure is determined by means of an individual parameter $c_z\in(0,1)$, with $z=1$ for workers and $z=2$ for entrepreneurs and bankers, which is exogenous and drawn at $t=0$ from group-specific distributions:  
\begin{equation}
	c_z \sim \mathcal{D}_{z(h)} \quad \text{with} \quad \mathbb{E}[c_z\mid z(h)=1]=c_W,\;\;\mathbb{E}[c_z\mid z(h)=2]=c_X
	\label{mpc_draw}
\end{equation}
where $c_W$ and $c_X$ are group-level mean propensities (with $c_W>c_X$ by assumption). Given $c_z$, the household consumption budget is
\begin{equation}
	C_{h,t}=c_z\,y_{h,t}
	\label{Ch_def}
\end{equation}
and residual income $(1-c_z)y_{h,t}$ is correspondingly saved.

Household consumption expenditure is allocated across consumption-good markets according to fixed budget shares $\{\eta_g\}_{g\in\mathcal{G}^C}$ (the same weights used in the CPI). Hence, the market-specific consumption budget is
\begin{equation}
	C_{h,g,t}=\eta_g\,C_{h,t}
	\qquad g\in\mathcal{G}^C
	\label{Chg_def}
\end{equation}

Within each consumption-good market $g$, households choose among the set of available c-firms $\mathcal{I}^{C}_g$. For any firm $i\in\mathcal{I}^{C}_g$, define a convenience--attractiveness indicator
\begin{equation}
	\mathcal{V}_{i,t}
	\equiv
	\frac{(1+\mu_{i,t})^{\psi}}{p_{i,t}}
	\qquad \psi\ge 0
	\label{V_cons_def}
\end{equation}
which increases with the firm's mark-up (interpreted as brand differentiation/appeal) and decreases with its price. Variety choice is modelled as a simple probability proportional to this indicator:
\begin{equation}
	\Pr\!\left(i\mid g,t\right)
	=
	\frac{\mathcal{V}_{i,t}}{\sum_{j\in\mathcal{I}^{C}_g}\mathcal{V}_{j,t}}
	\qquad i\in\mathcal{I}^{C}_g
	\label{choice_prob_consumers}
\end{equation}
so that firms with higher $\mathcal{V}_{i,t}$ attract a larger share of demand, while choice remains stochastic and dispersed across varieties.

Given its market-specific consumption budget $C_{h,g,t}$ defined in eq.\eqref{Chg_def}, each household $h$ attempts to purchase consumption goods by selecting c-firms within the relevant markets according to the probabilistic choice rule eq.\eqref{choice_prob_consumers}. Purchases are executed subject to goods availability: households buy \emph{what they find} at posted prices until either their budget is exhausted or the selected varieties are out of stock. If the initially selected variety is out of stock, the household may attempt alternative varieties within the same market $g$ according to \eqref{choice_prob_consumers}. If all available varieties in market $g$ are out of stock, any remaining unspent amount of the market-specific budget $C_{h,g,t}$ is recorded as forced saving (and contributes to unmet consumption demand).

\subsection{Aggregate income, expenditure, and macro variables}
Aggregate variables follow from accounting identities. Let $W_t = w_t\sum_{i\in\mathcal{I}}N_{i,t}$ denote the aggregate wage bill and let $X_t=\sum_{i\in\mathcal{I}}X^f_{i,t}+\sum_{b\in\mathcal{B}}X^b_{b,t}$ denote total profit income accruing to entrepreneurs and bankers. Aggregate income is then given by the sum of labour income and profit income, computed as:
\begin{equation}
	Y_t= W_t+X_t
\end{equation}
Aggregate realised consumption is obtained by summing households' \emph{effective} expenditures in the goods market. Let $\widetilde{C}_{h,t}\le C_{h,t}$ denote the amount of the consumption budget actually spent by household $h$ in period $t$ (with the gap $C_{h,t}-\widetilde{C}_{h,t}$ capturing forced reductions due to shortages). Aggregate realised consumption expenditure is then
\begin{equation}
	\widetilde{C}_t \equiv \sum_{h\in\mathcal{H}} \widetilde{C}_{h,t}
	\label{C_realised_from_households}
\end{equation}
Finally, we define the consumer price index (CPI) as an expenditure-weighted aggregate of market prices of all consumption goods:
\begin{equation}
	P_t
	\equiv
	\sum_{g\in\mathcal{G}^C}\eta_g\, p^{C}_{g,t}
	\label{CPI_def}
\end{equation}
Aggregate inflation is simply defined as the rate of change in the CPI,
\begin{equation}
	\pi_t = \frac{P_t - P_{t-1}}{P_{t-1}}
	\label{infl_def}
\end{equation}
Analogously, a producer/intermediate-goods price index can be defined over $g\in\mathcal{G}^K$ to track upstream inflation and quantify network pass-through into consumer prices. Because market prices $p^{s}_{g,t}$ are constructed from sales-weighted firm prices (see eq.\eqref{pg_def}), the CPI and aggregate inflation inherit the underlying heterogeneity in unit costs, mark-ups, credit conditions, and supply-chain constraints, thereby allowing inflation to arise as an emergent macroeconomic outcome of decentralised pricing and production decisions in a networked, endogenous-money economy.

\section{Simulations}
\label{sec:simulations}

This section studies the inflationary regimes generated by the model, with the aim to identify the conditions under which decentralised pricing decisions produce aggregate inflation as an emergent outcome. The central distinction is between \emph{sources of pressure} and \emph{propagation mechanisms}. Mark-ups, financial costs, and natural-capital costs act as sources of pressure on firms' pricing decisions. Production networks and expectations, instead, determine whether these pressures remain local, dissipate through quantity adjustment, or become persistent aggregate inflation.

This distinction is important because it separates our approach from top-down representations in which inflation is mainly explained by a small number of aggregate drivers. In our model, inflation is measured ex post as the evolution of the consumer price index generated by heterogeneous firms that face different input costs, credit conditions, mark-ups, sales outcomes, and expectation rules. This bottom-up perspective is consistent with the production-network literature, which shows that input-output linkages can transform microeconomic shocks into aggregate fluctuations \citep{Acemoglu2012, CarvalhoTahbazSalehi2019, BaqaeeFarhi2019}, and with empirical work documenting that producer-price inflation can propagate through input linkages \citep{Auer2019}. It is also consistent with micro evidence on price-setting heterogeneity \citep{NakamuraSteinsson2008} and with agent-based stock-flow consistent macroeconomics, where macroeconomic regularities emerge from decentralised interactions under accounting and financial consistency \citep{Caiani2016, Dosi2010}.

All simulations are run on the same baseline population: households, banks, consumption-good firms, and capital/intermediate-good firms interact through goods, credit, and production networks. The baseline setting, configured as reported in Appendix, is deliberately non-inflationary. It uses a production-network structure in which consumption-good producers depend on upstream intermediate-good markets, while the intermediate sector is treated as the upstream productive base. All other configurations, then, activate specific channels one at a time.

\subsection{Baseline dynamics}

\begin{figure}[htbp]
	\centering
	\includegraphics[width=\textwidth]{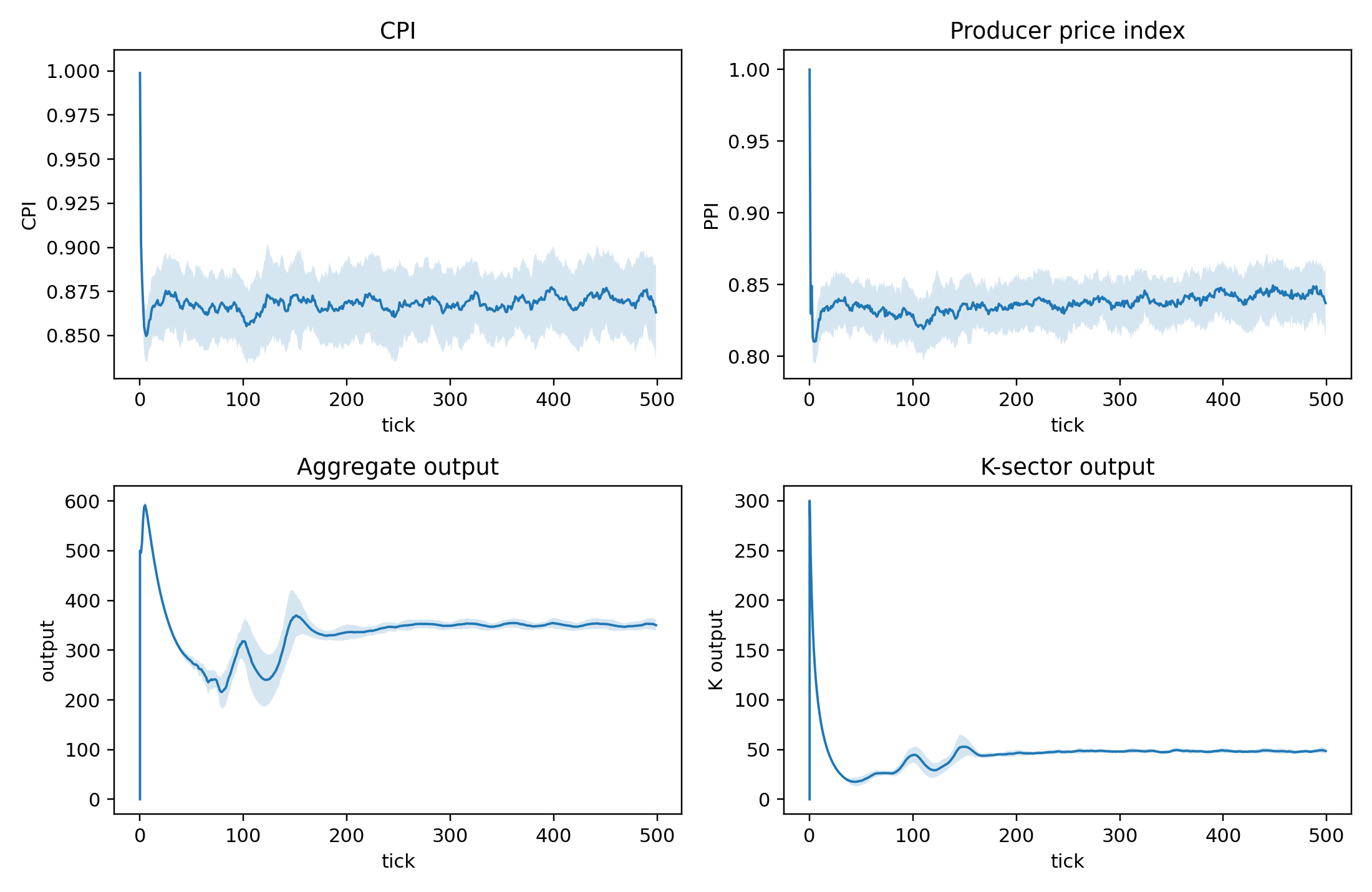}
	\caption{Baseline validation: CPI, producer price index, aggregate output, and intermediate-sector output. Lines report Monte Carlo averages; shaded bands report cross-seed standard deviations.}
	\label{fig:baseline_validation}
\end{figure}

Figure~\ref{fig:baseline_validation} reports the baseline dynamics. The figure plots Monte Carlo averages; the surrounding bands measure cross-seed variability. The baseline economy remains active throughout the simulation: aggregate output and intermediate-sector output stay positive, the firm--firm production network remains operative, and credit relations continuously finance production. At the same time, neither the CPI nor the producer price index displays a persistent inflationary pattern. This result is central for identification: inflation is not mechanically built into the model. The baseline is a stable, non-inflationary reference regime generated by decentralised production, credit, and pricing interactions.

The baseline also plays a methodological role. It shows that heterogeneity, endogenous credit, production-network links, and adaptive expectations do not, by themselves, generate sustained inflation. Hence, when inflation arises in the following experiments, it can be attributed to specific sources of pressure and to the mechanisms that propagate or amplify them. This is the first sense in which the model differs from purely top-down treatments of inflation: aggregate price dynamics are not imposed as an autonomous macroeconomic process, but emerge from the interaction of heterogeneous microeconomic units. This bottom-up orientation is consistent with agent-based and stock-flow consistent approaches to macroeconomic dynamics \citep{Dosi2010,Caiani2016}, and with micro-price evidence showing that aggregate price indices conceal substantial heterogeneity in price-setting behaviour \citep{NakamuraSteinsson2008}.

\subsection{Cost-pressure channels}

\begin{figure}[htbp]
	\centering
	\includegraphics[width=\textwidth]{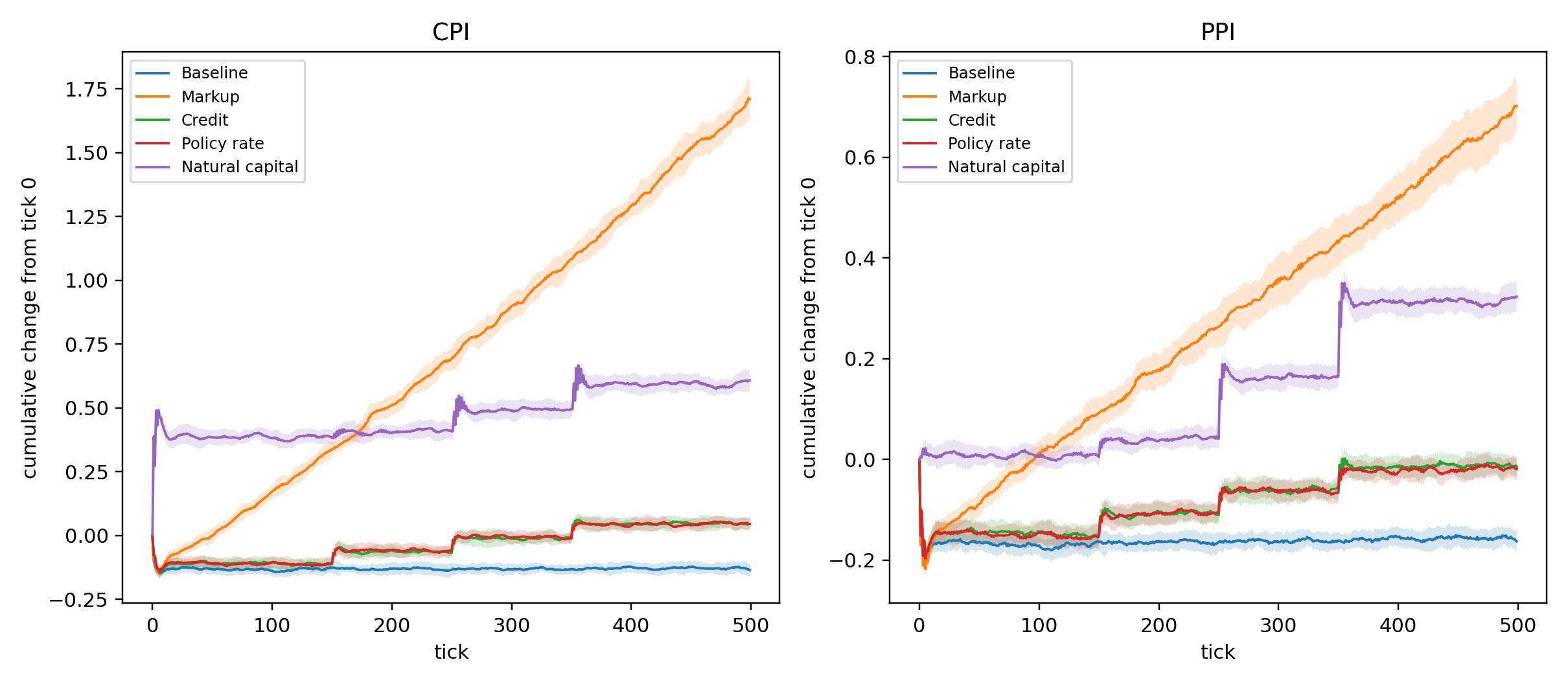}
	\caption{Cost-pressure channels: cumulative CPI and producer-price dynamics.}
	\label{fig:cost_channels}
\end{figure}

Figure~\ref{fig:cost_channels} compares the main cost-pressure channels. The model distinguishes between sources of pressure and mechanisms of propagation. Mark-up pressure, financial costs, policy-rate changes, and natural-capital costs are sources of pressure. Production networks and expectations determine whether those pressures remain local, propagate downstream, or become persistent aggregate inflation.

The mark-up channel generates the strongest inflationary response. This follows directly from the pricing rule: when successful firms increase mark-ups, posted prices rise immediately. The mechanism is nevertheless not purely mechanical, because firms that raise prices too aggressively may lose demand through the probabilistic choice rules. The result is a mark-up-driven inflation regime accompanied by a substantial contraction of real activity. This is consistent with the empirical literature showing that mark-up heterogeneity and market power can have macroeconomic implications \citep{DeLoecker2020}. It is also consistent with firm-level evidence that variable mark-ups affect the transmission of shocks into domestic prices \citep{Amiti2019}. In the present model, however, mark-up inflation is not an aggregate wedge imposed from above: it emerges from decentralised pricing decisions by heterogeneous firms whose market power is endogenous to realised sales.

Financial-cost shocks generate more moderate inflation. When banks raise lending mark-ups stepwise, firms face higher financing costs, unit costs increase, and prices adjust upward. The effect is visible but less explosive than in the mark-up scenario. The policy-rate experiment belongs to the same financial-cost family: the origin of the cost increase is monetary policy rather than bank-specific spreads. This is consistent with the cost-channel view of monetary transmission, according to which interest rates enter firms' marginal costs \citep{RavennaWalsh2006}. It is also consistent with empirical evidence that financial frictions shape firms' pricing behaviour: liquidity-constrained firms may raise prices in adverse financial conditions rather than cut them, because they face a stronger need to generate internal funds \citep{Gilchrist2017}. The simulations reproduce this logic in a decentralised setting: the financial-cost channel raises prices and compresses real activity.

Natural-capital cost shocks generate an upstream cost-push mechanism. The direct effect falls on intermediate-good producers through the cost of natural capital. The impact on consumer prices depends on downstream exposure to intermediate inputs. When this exposure is weak, the shock remains mostly upstream; when it is stronger, the producer-price increase propagates to the CPI and real activity declines. This result is consistent with empirical evidence on energy and commodity price shocks. Oil-price movements affect domestic inflation, but the magnitude and persistence of pass-through depend on the nature of the shock and on the structure of production and expenditure \citep{Kilian2009,Choi2018,BaumeisterKilian2016}. In the model, the natural-capital channel therefore operates as a resource-cost pressure whose final inflationary effect is conditional on input intensity and network pass-through.

\subsection{Stepwise transmission mechanisms}

\begin{figure}[htbp]
	\centering
	\includegraphics[width=\textwidth]{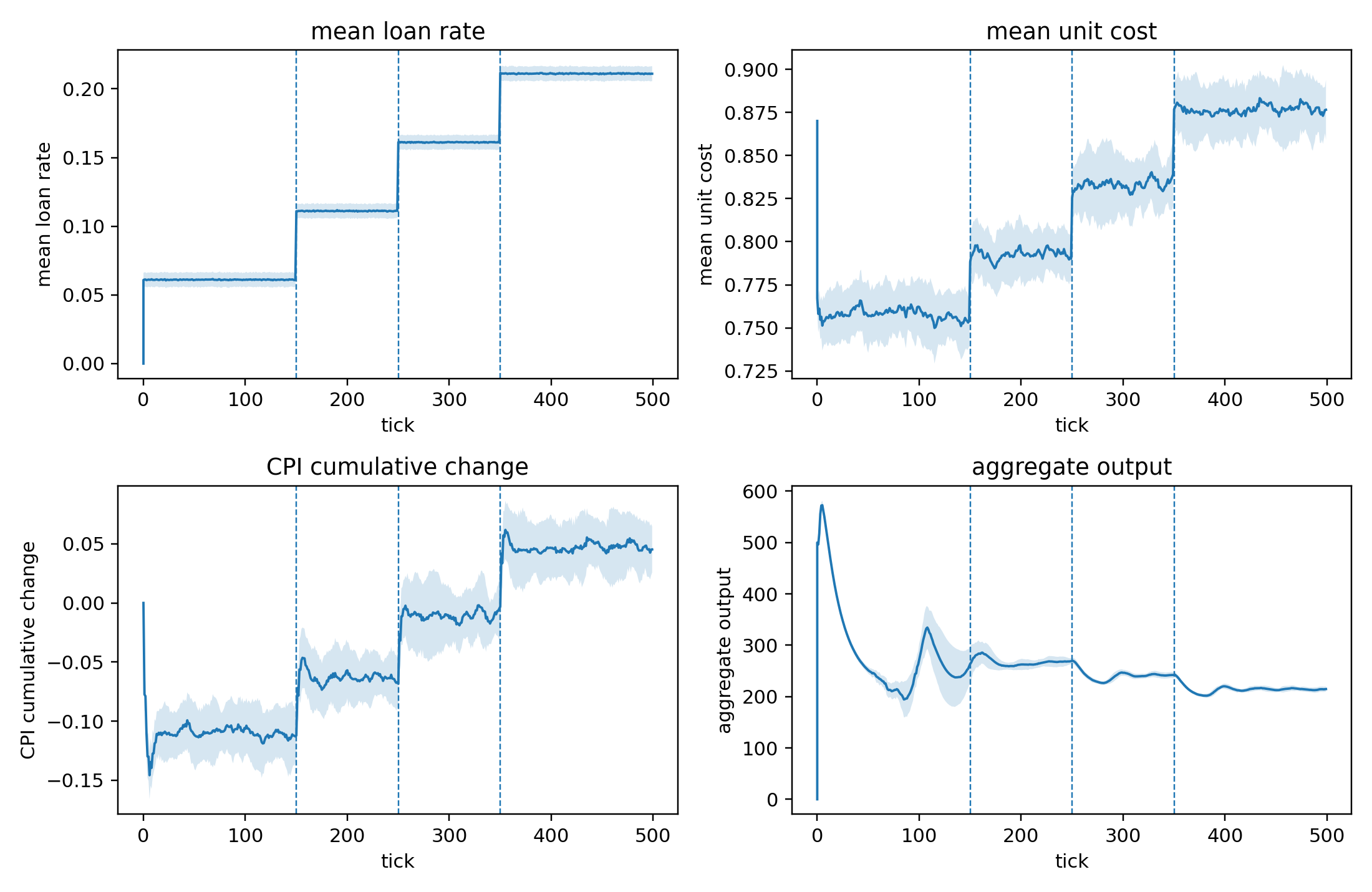}
	\caption{Bank-cost steps: lending rates, unit costs, CPI, and aggregate output.}
	\label{fig:bank_cost_steps}
\end{figure}

\begin{figure}[htbp]
	\centering
	\includegraphics[width=\textwidth]{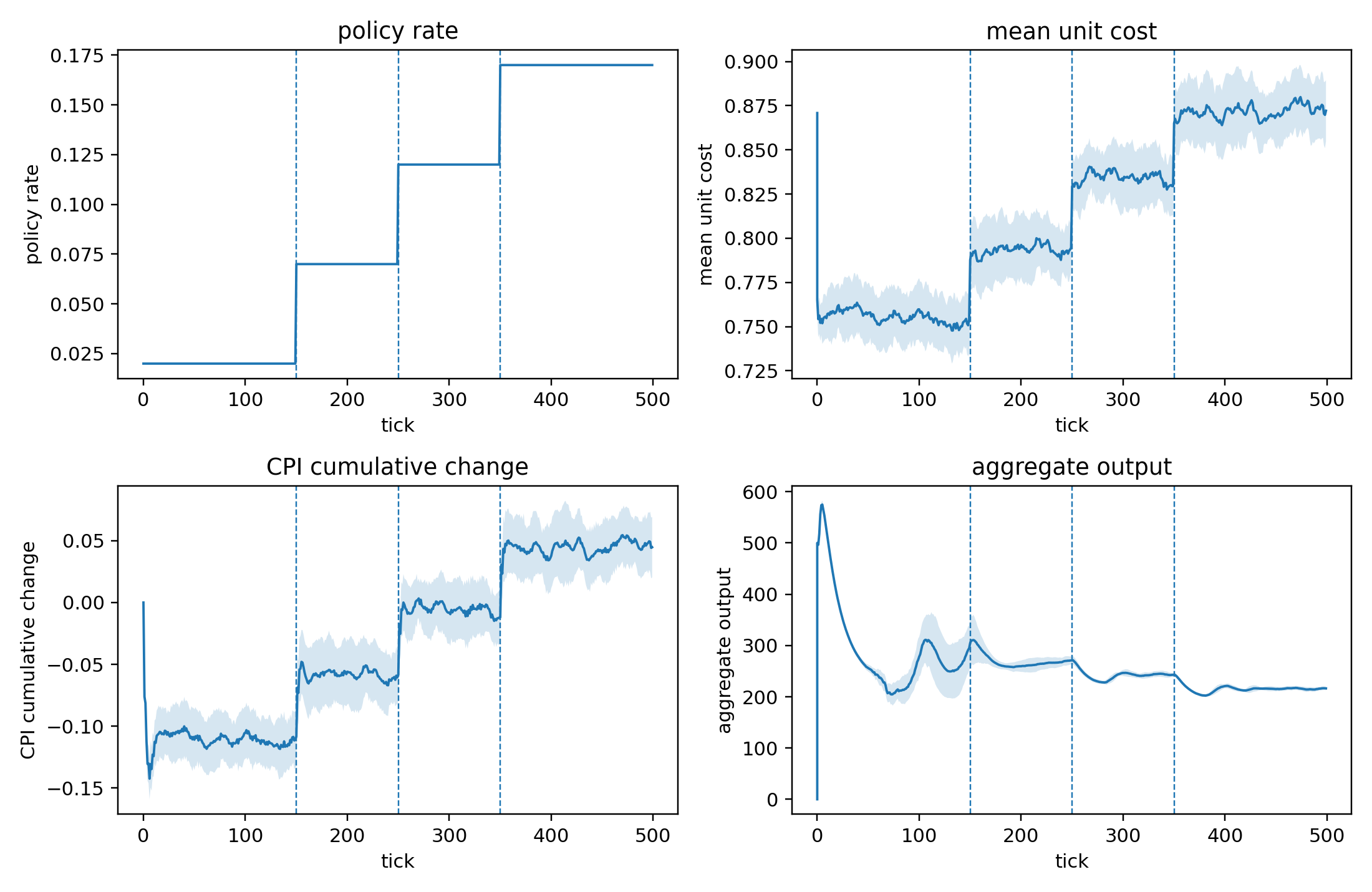}
	\caption{Policy-rate steps: policy rate, unit costs, CPI, and aggregate output.}
	\label{fig:policy_rate_steps}
\end{figure}

\begin{figure}[htbp]
	\centering
	\includegraphics[width=\textwidth]{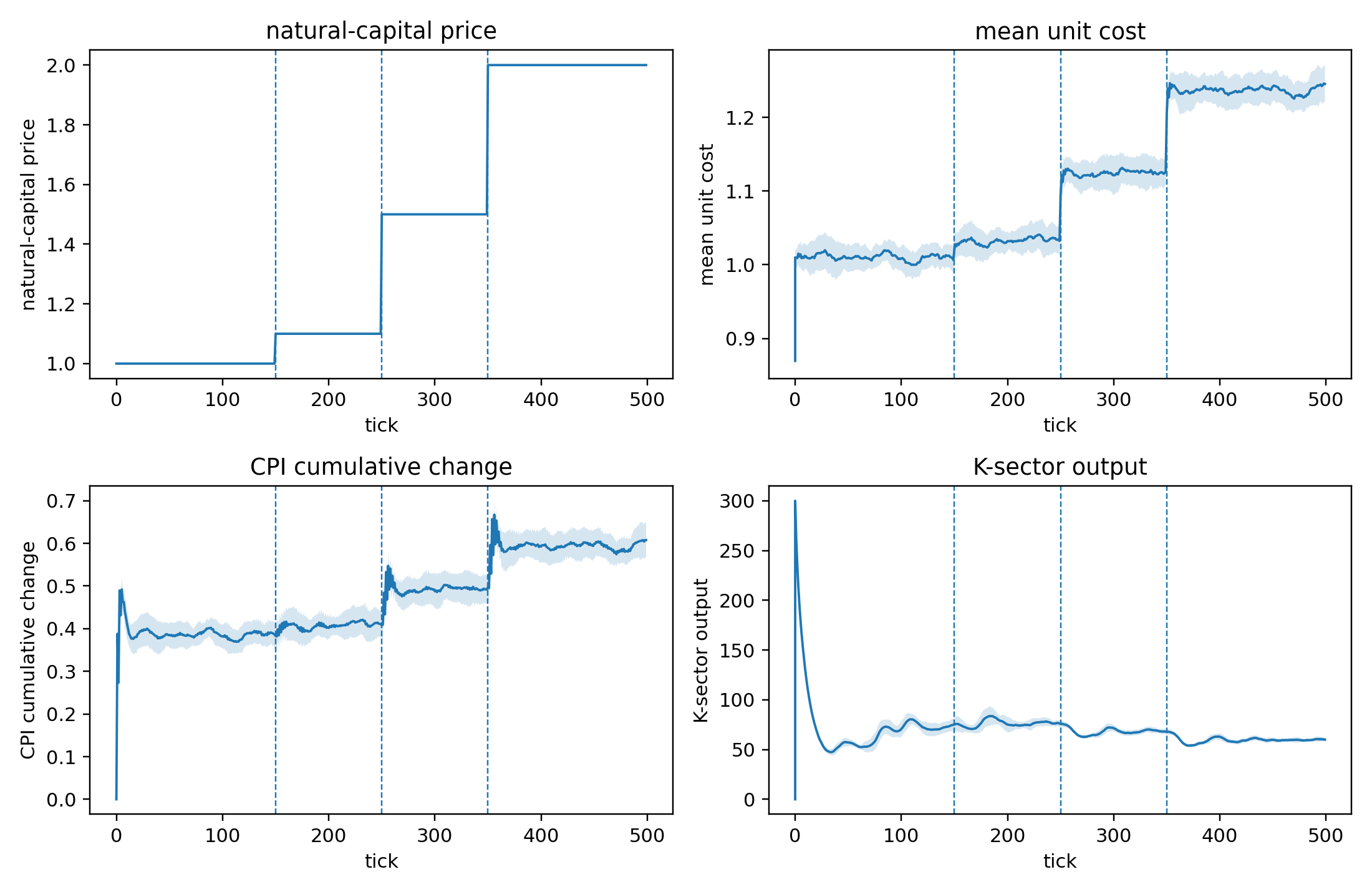}
	\caption{Natural-capital cost steps: natural-capital price, unit costs, CPI, and intermediate-sector output.}
	\label{fig:natural_capital_steps}
\end{figure}

Figures~\ref{fig:bank_cost_steps}, \ref{fig:policy_rate_steps}, and \ref{fig:natural_capital_steps} report the stepwise transmission experiments. In each case, the relevant cost variable changes discretely during the simulation. This design makes the timing of pass-through visible.

The bank-cost experiment shows a clean financial-cost pass-through. Stepwise increases in bank lending mark-ups raise the average lending rate. This increases firms' financing costs, raises unit costs, and produces upward movements in prices. The inflationary response is concentrated around the adjustment windows: each increase in bank costs shifts the economy toward a higher nominal-cost regime. This result is consistent with evidence that financial constraints affect price-setting at the firm level \citep{Gilchrist2017}, and with models in which working-capital finance creates a cost channel from interest rates to inflation \citep{RavennaWalsh2006}. In our framework, the channel is explicitly bottom-up: banks impose heterogeneous lending conditions, firms incorporate financing costs into unit costs, and aggregate inflation is observed only after these firm-level adjustments have propagated through markets.

The policy-rate experiment produces the same type of financial-cost mechanism. A higher policy rate raises the lending rate faced by firms and thereby increases the financial component of unit costs. The simulation therefore highlights two effects: tighter monetary conditions can depress activity, but they can also raise firms' costs when production relies on working-capital finance. In the model, the net outcome is determined by the interaction between credit dependence, production feasibility, and cost-plus pricing. The result supports a cost-channel interpretation of monetary transmission: a policy-rate increase is not only a demand-management tool, but can also affect the supply side by raising financing costs for firms \citep{RavennaWalsh2006}.

The natural-capital experiment highlights the upstream nature of resource-cost inflation. The shock first raises costs in the intermediate-goods sector. Producer prices respond directly, while the CPI responds according to the strength of downstream input exposure. Thus, the model does not imply one-to-one pass-through from natural-resource costs to consumer prices. Pass-through depends on technological input coefficients and on the network structure linking upstream and downstream markets. This is consistent with empirical work showing that energy-price shocks pass through to inflation with heterogeneous timing and intensity across economies and sectors \citep{Choi2018,BaumeisterKilian2016}. In our model, resource-cost pressure becomes consumer-price inflation only when downstream production is sufficiently exposed to the affected upstream sector.

\subsection{Production-network topology}

The production network is studied through comparative statics over alternative technological architectures. The good-level network is fixed at setup; it is not changed during a simulation. Across simulations, however, we vary the number of upstream intermediate-good markets required by consumption-good and intermediate-good producers. This allows us to distinguish downstream exposure from intermediate-sector interdependence.

\begin{figure}[htbp]
	\centering
	\includegraphics[width=\textwidth]{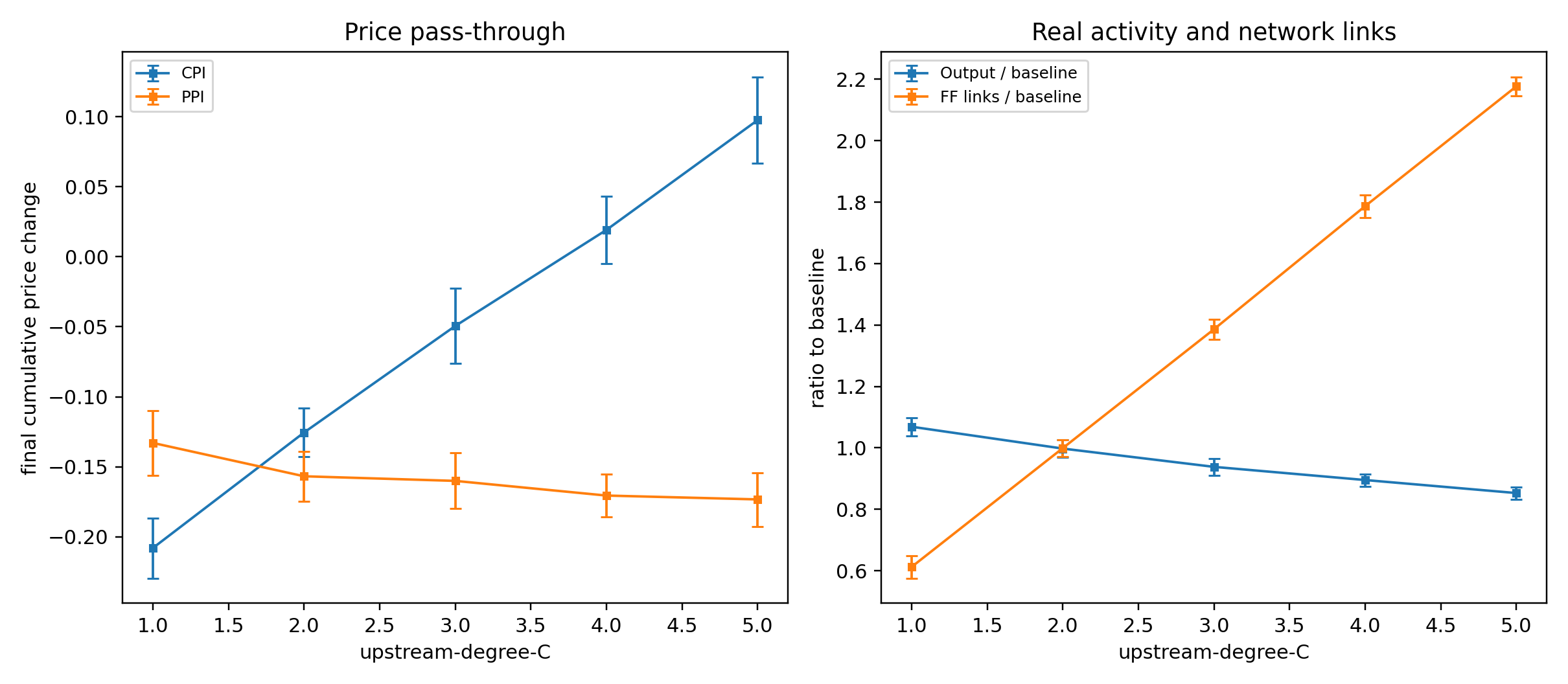}
	\caption{Downstream network exposure: pass-through as upstream-degree-C increases, with upstream-degree-K fixed at zero.}
	\label{fig:network_c_degree}
\end{figure}

Figure~\ref{fig:network_c_degree} reports the effect of increasing downstream exposure to intermediate inputs while keeping the intermediate sector as the upstream productive base. Higher upstream-degree-C increases the number of firm--firm links and strengthens pass-through from upstream conditions to consumer prices. The network therefore acts as a propagation mechanism: it does not create inflation by itself, but it determines how strongly local cost pressures affect aggregate prices.

This result is consistent with the production-network literature. Input-output linkages can transform microeconomic shocks into aggregate fluctuations \citep{Acemoglu2012,BaqaeeFarhi2019}, and international input linkages can propagate producer-price inflation across countries and sectors \citep{Auer2019}. In the present simulations, the same mechanism operates internally: the more consumption-good firms depend on intermediate inputs, the more upstream pressures can be transmitted to final consumer prices.

\begin{figure}[htbp]
	\centering
	\includegraphics[width=\textwidth]{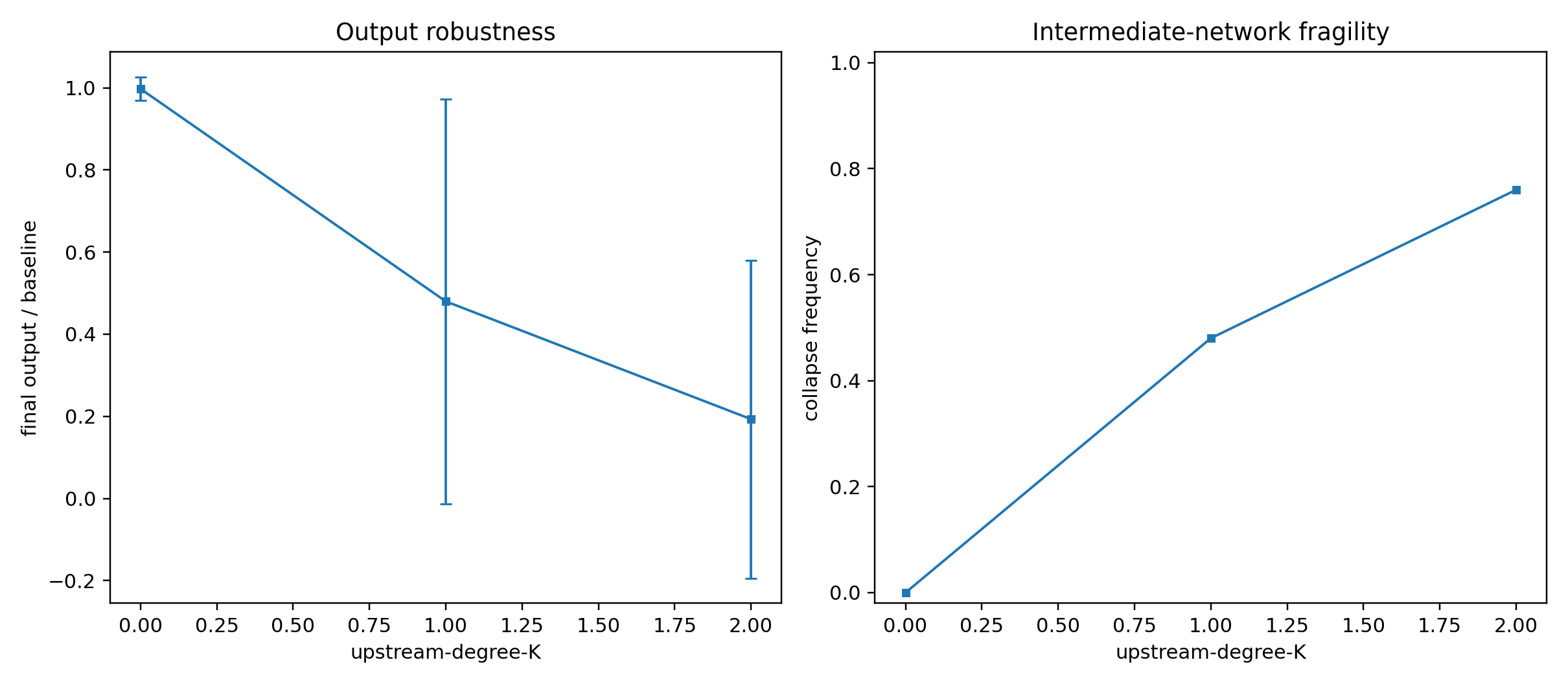}
	\caption{Intermediate-sector interdependence: final output and collapse frequency as upstream-degree-K increases.}
	\label{fig:network_k_degree}
\end{figure}

Figure~\ref{fig:network_k_degree} focuses on the intermediate sector. Increasing interdependence among intermediate-good markets has a different effect. Rather than producing smooth inflationary propagation, higher upstream-degree-K increases fragility. Dense intermediate-sector dependence can generate bottlenecks and deadlocks because intermediate producers require other intermediate inputs that may themselves become unavailable. This result is consistent with the production-network literature: input-output linkages transmit shocks, but they can also amplify disruptions when constraints bind \citep{Acemoglu2012,CarvalhoTahbazSalehi2019}.

\begin{figure}[htbp]
	\centering
	\includegraphics[width=0.6\textwidth]{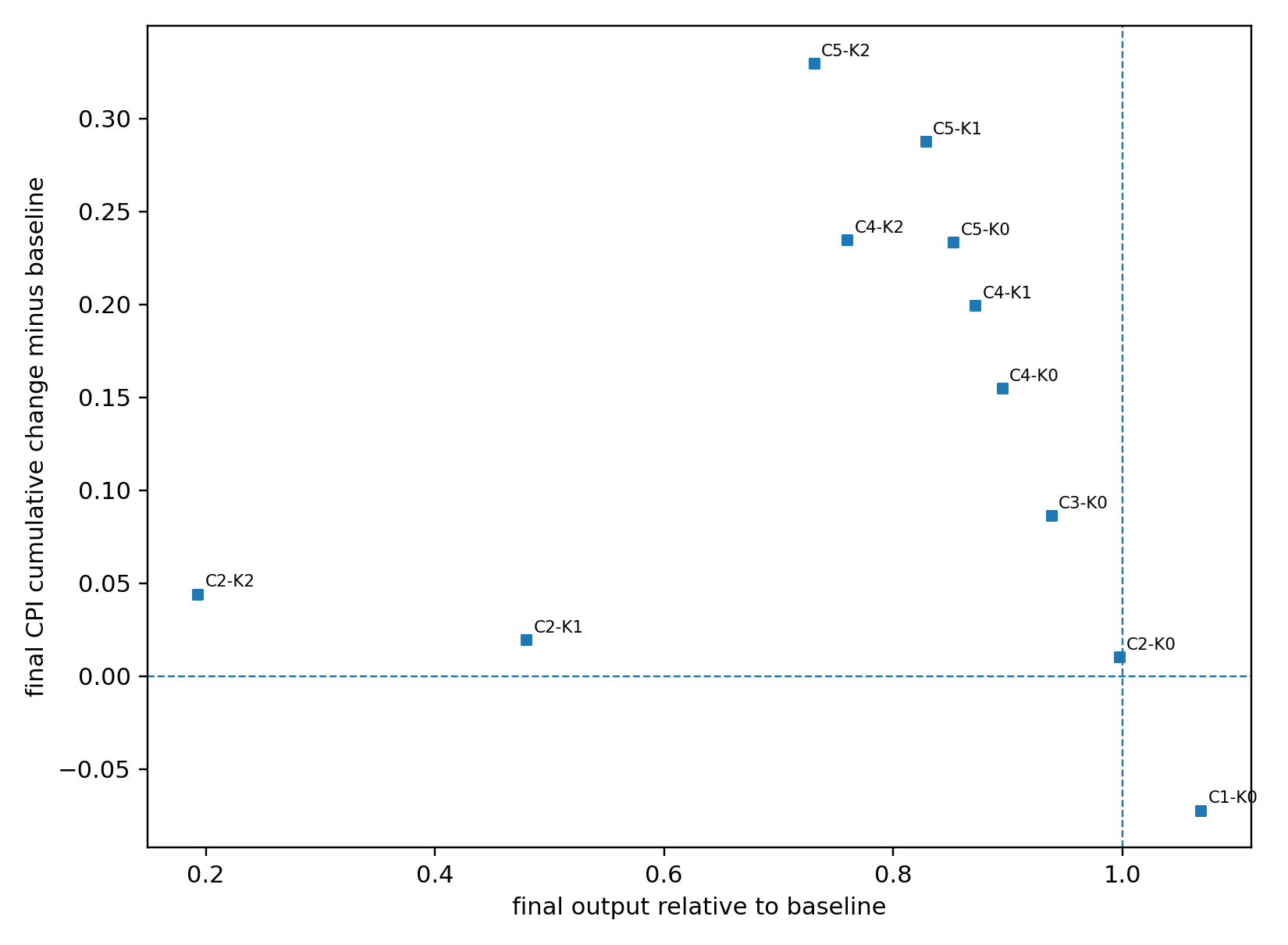}
	\caption{Network architectures and the inflation-output trade-off.}
	\label{fig:network_scatter}
\end{figure}

Figure~\ref{fig:network_scatter} summarises the inflation-output trade-off generated by alternative network architectures. Each point is a network configuration. The horizontal axis reports final output relative to the baseline; the vertical axis reports cumulative CPI growth relative to the baseline. The figure shows that network topology affects both price propagation and real robustness. Some architectures increase pass-through without destroying production; others primarily generate fragility.

Additional network diagnostics under alternative pressure contexts and anchoring rules are reported in Appendix~\ref{app:network_diagnostics}. They confirm the same qualitative distinction: downstream exposure amplifies pass-through, whereas dense intermediate-sector interdependence mainly raises fragility. Thus, production networks do not simply ``add inflation''. They shape the conditions under which local pressures become aggregate inflation or, alternatively, real disruption.

\subsection{Expectations}

\begin{figure}[htbp]
	\centering
	\includegraphics[width=\textwidth]{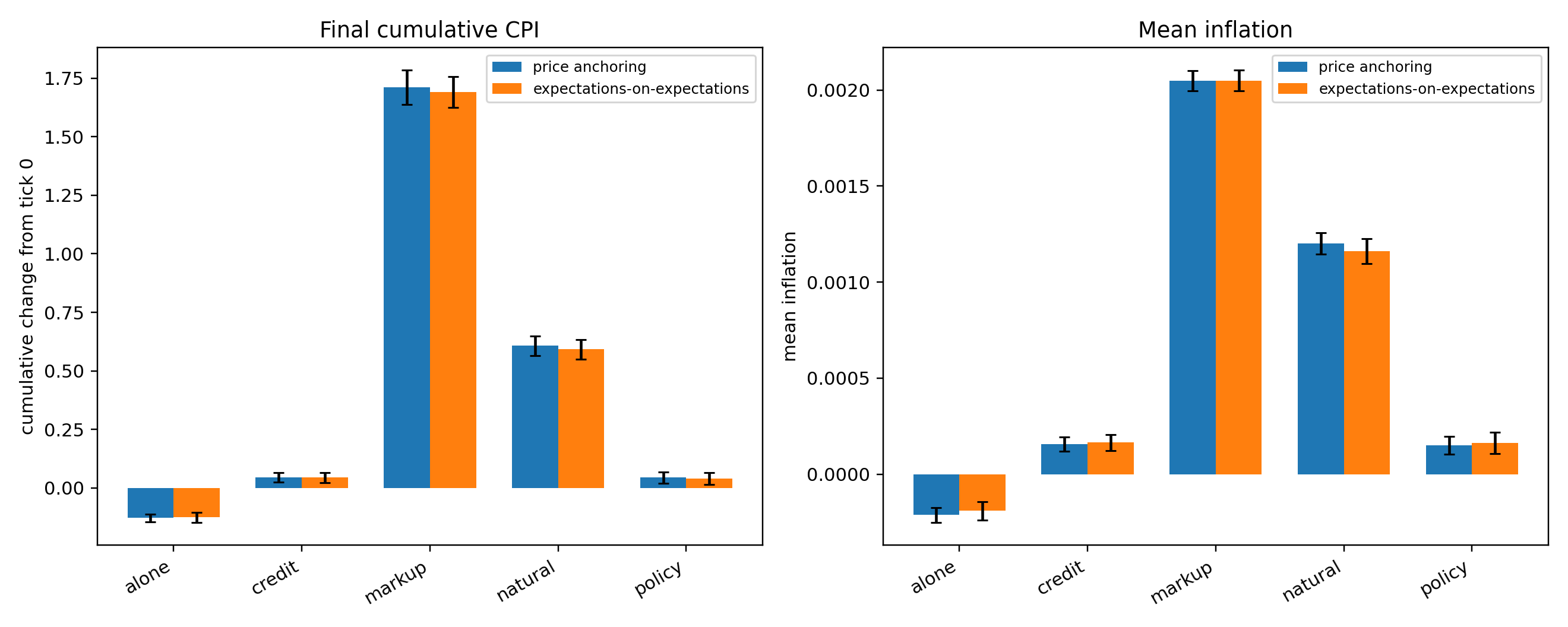}
	\caption{Expectation rules: anchoring mechanisms alone and under cost or mark-up pressure.}
	\label{fig:expectations}
\end{figure}

Figure~\ref{fig:expectations} shows the role of expectations. When no independent cost or mark-up pressure is active, neither price anchoring nor expectations-on-expectations generates inflation. Expectations are therefore not an arbitrary source of inflation in the model. They are adaptive responses to realised price signals.
A second result is that expectations matter once an inflationary pressure exists. Under mark-up, financial-cost, policy-rate, or natural-capital pressure, expectation rules can affect the persistence and amplitude of price movements: even an average per-tick inflation rate of $0.1\%$ would compound, if sustained over $500$ ticks, into a cumulative price-level increase of about $65\%$.

The difference between price anchoring and expectations-on-expectations is not dominant in every scenario, but expectations-on-expectations is more prone to persistence when realised price changes are already present. This is consistent with empirical evidence that expectations are neither fully rational nor homogeneous. Households and firms process inflation information through limited attention, personal experience, and salient local prices \citep{CoibionGorodnichenko2015,MalmendierNagel2016,DAcunto2021}. Therefore, the role that expectations play in inflationary terms is to amplify price pressures once decentralised price increases have already emerged. The combined weighting scheme $\{\omega_{i,h,t}\}_{h=1}^{\varsigma_i}$ used here is compared with all other ones in the Appendix.

\subsection{Mark-up dynamics}

Figure~\ref{fig:markup_mechanism} isolates the mark-up channel. In this scenario, firms that successfully sell most of their available output can increase mark-ups. The resulting inflation is strong and persistent, but it is also associated with a sharp contraction in real activity. This is a useful result because it shows that mark-up inflation is possible in the model but is not costless: demand reallocation and output losses discipline the process.

\begin{figure}[htbp]
	\centering
	\includegraphics[width=\textwidth]{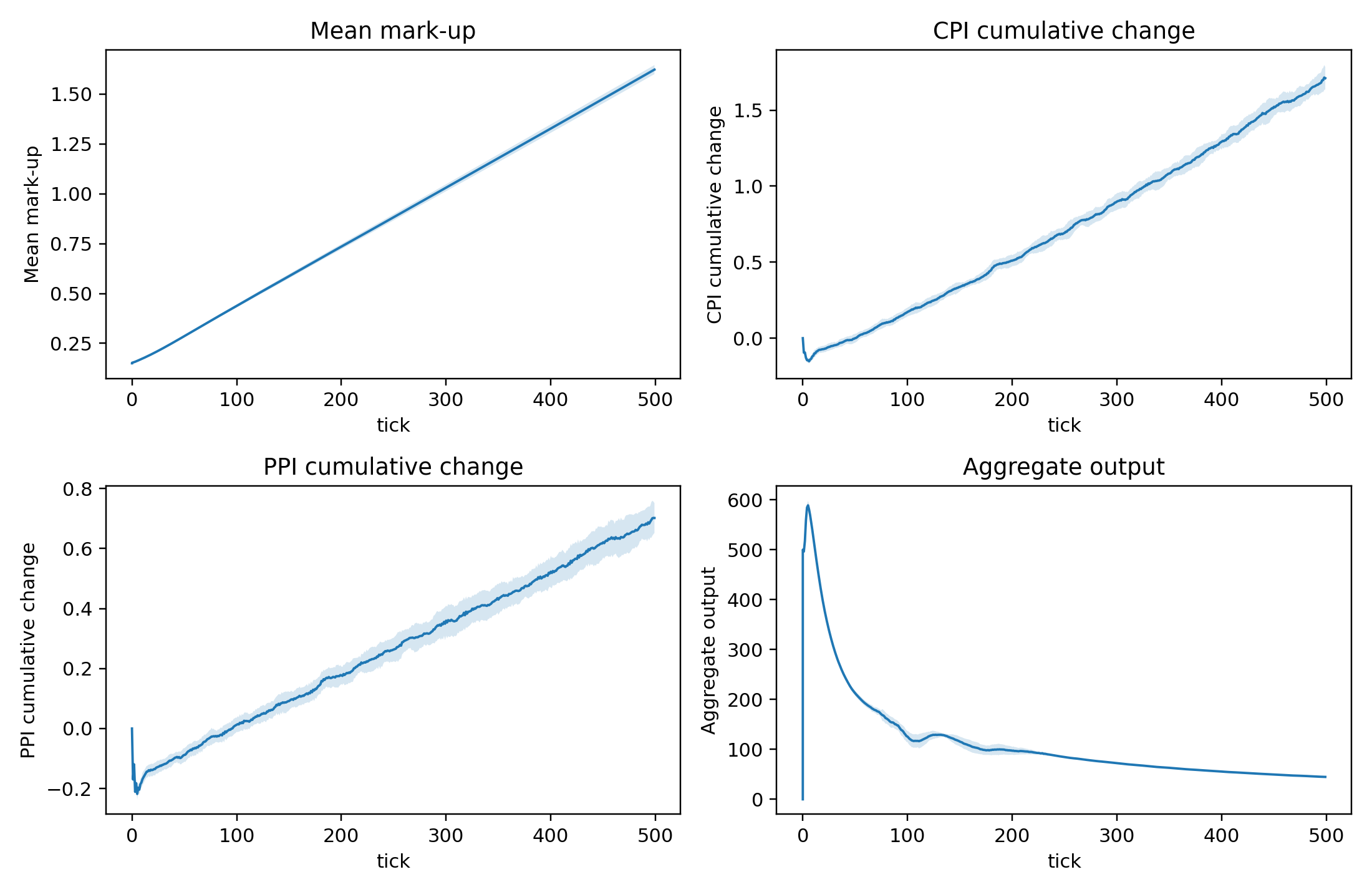}
	\caption{Mark-up mechanism: mark-ups, CPI, PPI, and aggregate output.}
	\label{fig:markup_mechanism}
\end{figure}

The mark-up experiment should therefore be read as a mark-up-driven or greedflation scenario. It differs from financial-cost and natural-capital inflation because the primary force is not a rise in input costs but the pricing behaviour of successful firms. The aggregate effect is nevertheless emergent: it results from heterogeneous firms adjusting mark-ups locally and from consumers and downstream firms reallocating demand across varieties. The mechanism is consistent with the empirical evidence that market power and mark-up heterogeneity matter for aggregate outcomes \citep{DeLoecker2020}, and with firm-level evidence that mark-ups shape domestic price responses to shocks \citep{Amiti2019}.

\subsection{Inflation-output trade-off}

\begin{figure}[htbp]
	\centering
	\includegraphics[width=0.6\textwidth]{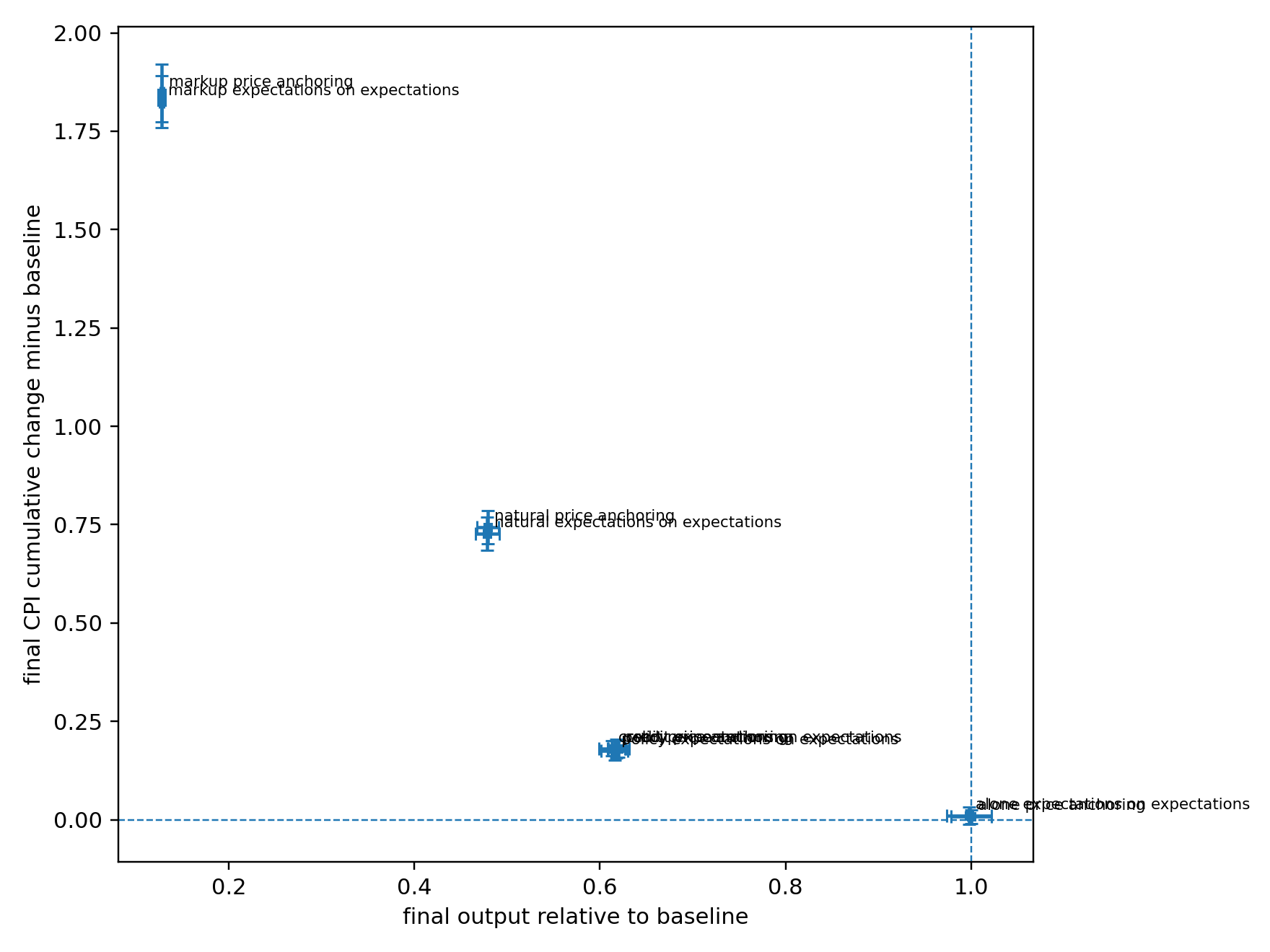}
	\caption{Scenario-level inflation-output trade-off.}
	\label{fig:inflation_output_tradeoff}
\end{figure}

Figure~\ref{fig:inflation_output_tradeoff} summarises the main regimes. Each point represents a scenario average. The horizontal axis measures final output relative to the baseline; the vertical axis measures cumulative CPI growth relative to the baseline. The figure shows that inflation is not a single-mechanism outcome. Mark-up pressure, financial-cost shocks, policy-rate shocks, natural-capital shocks, and network architectures generate different combinations of price increases and output losses.

\section{Policy implications}
\label{sec:policy_implications}

If inflation is understood as an emergent macroeconomic pattern, the policy response cannot be confined to interest-rate management. Indeed, the same CPI movement may originate from very different mechanisms. It may reflect demand pressure, upstream resource costs, mark-up increases, credit-cost shocks, expectation amplification, or network bottlenecks. Different real effects require different policy interpretations.

This point is particularly relevant for monetary policy in the euro area. The ECB's strategy defines price stability as a symmetric $2\%$ inflation target over the medium term and identifies the set of policy rates as the primary monetary-policy instrument \citep{ECB2021Strategy}. Although this institutional framework is considered operationally important, our results suggest that the policy rate is not a neutral instrument acting only through aggregate demand. In an economy in which firms finance production through working-capital credit, an increase in the policy rate also raises firms' borrowing costs. Through the cost channel, higher interest rates may increase unit costs and therefore generate cost-push pressure on prices, at least over the adjustment horizon \citep{BarthRamey2002,RavennaWalsh2006}. In this case, a policy-rate increase may simultaneously reduce demand and raise production costs. The final effect depends on firms' credit dependence, production feasibility, pricing power, and network exposure.

The policy-rate simulations capture precisely this ambiguity. A higher policy rate raises loan rates, increases the financial component of unit costs, and can produce an upward movement in prices before or together with output contraction. In a hypothetical scenario of a full employment, a demand-driven inflation can be faced by trying to restrict expenditure decisions; contrarily, when full employment is far from being reached, inflation is cost-driven and firms depend on credit to finance production, a rate increase can reinforce the cost pressure that monetary policy should help containing. Hence, the model supports the cost-channel view of monetary transmission, according to which nominal interest rates can enter firms' marginal costs \citep{RavennaWalsh2006}. It also aligns with credit-channel models in which lending spreads and financial frictions are themselves relevant for optimal policy \citep{DeFioreTristani2013}.

This result raises a more general issue about fixed policy rules. Taylor-type rules are considered useful as transparent benchmarks because they summarise monetary-policy responses to inflation and real activity in a simple reaction function \citep{Taylor1993}. But they are not sufficient as decision rules in a complex, networked economy. A rule that reacts only to aggregate inflation and an aggregate output gap cannot distinguish among the sources of inflation. The model shows that this distinction is essential, since otherwise policy risks confusing symptoms with causes. Our results provide evidence in this direction. Expectations alone do not generate inflation, but they amplify existing pressures. Networks alone do not generate inflation, but they determine the strength of pass-through and the probability of real disruption. Natural-capital shocks affect the CPI only when downstream firms are sufficiently exposed to intermediate inputs. Financial shocks affect prices through borrowing costs and credit dependence. Mark-up pressure generates the strongest inflationary response, but it also produces substantial output losses. 

A policy approach consistent with these results should therefore begin from source identification. Inflation management should distinguish at least five cases. First, if inflation is driven by demand pressure, conventional monetary tightening may be effective. Second, if inflation is driven by financial-cost pressure, higher policy rates can worsen firms' cost conditions and should be assessed together with credit spreads, working-capital dependence, and bank lending conditions. Third, if inflation is driven by upstream resource costs, policy should focus on pass-through, bottlenecks, and the exposure of downstream sectors rather than treating the CPI increase as homogeneous demand inflation. Fourth, if inflation is driven by mark-up dynamics, competition policy, market monitoring, and sector-specific interventions may be more relevant than a purely monetary response. Fifth, if inflation is amplified by expectations, communication and credibility matter, but expectations should be treated as an amplifier of realised price pressures rather than as an autonomous source of inflation.

The production-network results imply that central banks and policy institutions should monitor sectoral and network indicators, not only aggregate inflation. Measures of producer-price inflation, input-output centrality, sectoral bottlenecks, credit spreads, mark-up dispersion, and firm-level financing conditions are informative about future CPI dynamics. In this respect, stabilising inflation requires information about the underlying propagation structure. The literature on relative-price changes similarly shows that optimal monetary policy may depend on the sectoral composition of inflation and on whether price movements originate in flexible-price or sticky-price sectors \citep{Aoki2001}. Our framework extends this logic to production networks: the policy relevance of a price shock depends on where it occurs in the network and how it propagates downstream.

The endogenous-money dimension further strengthens this point. If bank lending creates purchasing power and finances production, monetary policy affects the operating conditions of firms and banks. The policy rate enters bank funding costs, loan rates, credit rationing, and firms' unit costs. Therefore, the practical conduct of central banking must account for the fact that credit creation, production costs, and price setting are jointly determined. This is precisely why balance-sheet policies, credit facilities, targeted refinancing operations, and other non-rate instruments have become central to the modern toolkit of central banking \citep{BorioDisyatat2010}. For the ECB, the implication is that the target should be pursued with a richer diagnostic framework. 

The broader policy lesson is that inflation management in a complex economy requires adaptive central banking. While rules may appear useful as disciplines and communication devices, they cannot replace structural diagnosis. The model shows that a policy framework adequate to emergent inflation should combine aggregate targets with disaggregated diagnostics, monetary instruments with credit and macroprudential tools, and inflation monitoring with network-based analysis of cost propagation.

\section{Conclusive remarks}
\label{sec:conclusions}

This paper has developed an agent-based model in which inflation is treated as an emergent macroeconomic regularity rather than as a primitive aggregate process. The motivation is that measured inflation is the outcome of heterogeneous price-setting decisions made by firms operating in different markets, connected through production networks, facing heterogeneous cost structures, different credit conditions, and different expectation rules. From this perspective, the relevant question is how decentralised cost pressures, local pricing decisions, input-output linkages, financial conditions, and expectations interact to generate persistent aggregate price dynamics.

The simulation results support this bottom-up interpretation. The model separates sources of inflationary pressure from propagation and amplification mechanisms. Mark-up dynamics, financial costs, policy-rate changes, and natural-capital costs are sources of pressure. Production networks and expectations, by contrast, determine whether these pressures remain local or become aggregate and persistent. 

Mark-up pressure generates the strongest inflationary response because mark-ups enter the pricing rule directly. The effect, however, is not costless: the simulations show a sharp contraction of real activity, indicating that mark-up-driven inflation is disciplined by demand reallocation and output losses.

Financial-cost shocks generate a more moderate but clear inflationary mechanism. Stepwise increases in banks' lending mark-ups raise firms' borrowing costs, increase unit costs, and lead to higher prices. The policy-rate experiment belongs to the same family: when the policy rate rises, firms face higher lending rates and the financial component of unit costs increases. The model therefore captures a cost-channel interpretation of monetary transmission. A tighter monetary stance may reduce activity, but in a production economy that relies on working-capital finance it can also raise firms' production costs and generate cost-push price pressure. The final outcome depends on credit dependence, production feasibility, and firms' pricing rules.

Natural-capital shocks operate through an upstream cost-push channel. When the price of natural capital rises, intermediate-good producers are affected first. Producer prices respond directly, while consumer prices respond only to the extent that downstream firms are exposed to intermediate inputs. The simulations therefore show that pass-through is conditional on input intensity and production-network exposure. This result is particularly relevant for interpreting energy and resource-price shocks: their inflationary consequences depend on the architecture of production and on the ability of downstream firms to transmit upstream costs.

The production-network experiments show that topology matters in two distinct ways. Greater downstream exposure of consumption-good producers to intermediate inputs strengthens pass-through from upstream costs to consumer prices. In this sense, the network is a propagation mechanism. By contrast, dense interdependence within the intermediate sector increases fragility. When intermediate producers require too many intermediate inputs from one another, bottlenecks become more likely. Thus, production networks do not simply amplify inflation; they can also transform local pressures into real disruption. The relevant distinction is between network structures that transmit costs downstream and network structures that make production itself fragile.

Expectations play an amplifying role, but they do not create inflation from nothing. When no independent cost or mark-up pressure is active, neither price anchoring nor expectations-on-expectations generates sustained inflation. Once price pressures are present, however, expectation rules can affect persistence and amplitude. This result is consistent with the view that expectations are adaptive responses to realised price signals rather than autonomous inflation engines. The model therefore supports a disciplined interpretation of expectations: they matter because they feed back into pricing behaviour, but only after decentralised price pressures have already appeared.

The main contribution of the paper is therefore conceptual and methodological. Conceptually, it provides a framework in which inflation is understood as a complex-system outcome: a macroeconomic pattern generated by decentralised interactions among heterogeneous firms, banks, and households. Methodologically, it shows how an agent-based model with endogenous credit, cost-based pricing, monopolistic competition, adaptive expectations, and production networks can distinguish between sources of pressure, propagation mechanisms, and amplification channels. This distinction would be difficult to recover in a purely aggregate framework, where inflation is typically represented as the reduced-form response of a representative price index to aggregate variables.

The results also clarify the policy interpretation. Policies that raise financing costs, in addition to resource costs, or firms' mark-ups have consequences on the structure of production, the degree of credit dependence, the distribution of market power, and the way expectations respond to realised prices. Aggregate inflation reveals to be the outcome of a sequence of decentralised adjustments: costs change locally, firms revise prices, buyers reallocate demand, production constraints bind or relax, and expectations update in response to observed price movements. Therefore, the attitude of policy-makers to face inflation, without properly distinguishing its sources, by simply rising the policy rate and worsening a situation of already raising costs, seems completely out of scope. 

\clearpage
\section*{Appendix: Additional network-topology diagnostics}
\label{app:network_diagnostics}

This appendix reports additional network diagnostics for different pressure contexts and anchoring rules. The figures complement the main text by showing how the same production-network architecture behaves when it is combined with different sources of price pressure. The main distinction remains unchanged: downstream exposure amplifies pass-through, while intermediate-sector interdependence raises fragility.

\subsection*{A.1 Network diagnostics under expectations-on-expectations}

When no independent source of inflationary pressure is active, the figures show that expectations alone do not transform network density into sustained inflation.

\begin{figure}[htbp]
	\centering
	\includegraphics[width=0.85\textwidth]{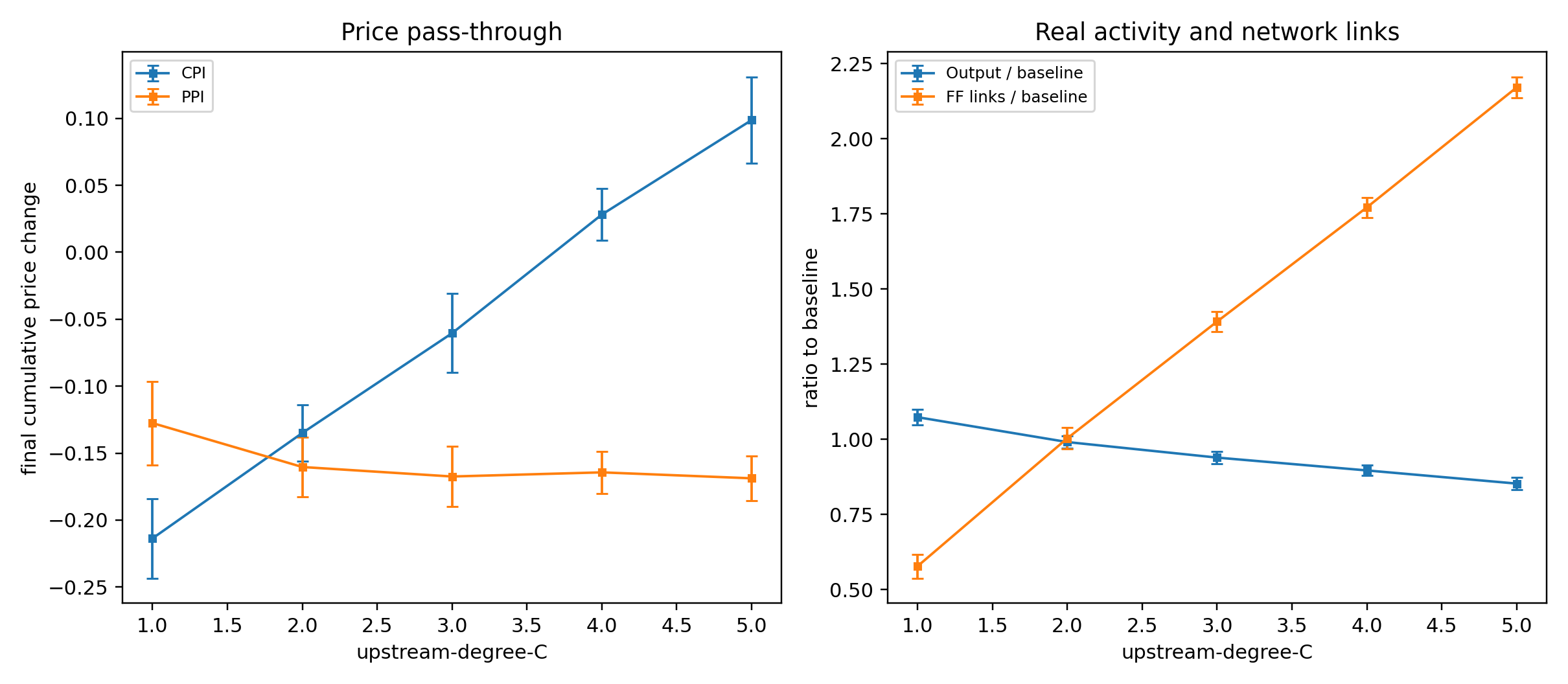}
	\caption{Downstream network exposure under expectations-on-expectations, no additional pressure.}
\end{figure}

The figure shows that increasing downstream exposure raises the density of firm--firm interactions, but it does not generate strong price dynamics in the absence of an underlying cost or mark-up pressure. Network exposure alone mainly changes propagation capacity.

\begin{figure}[htbp]
	\centering
	\includegraphics[width=0.85\textwidth]{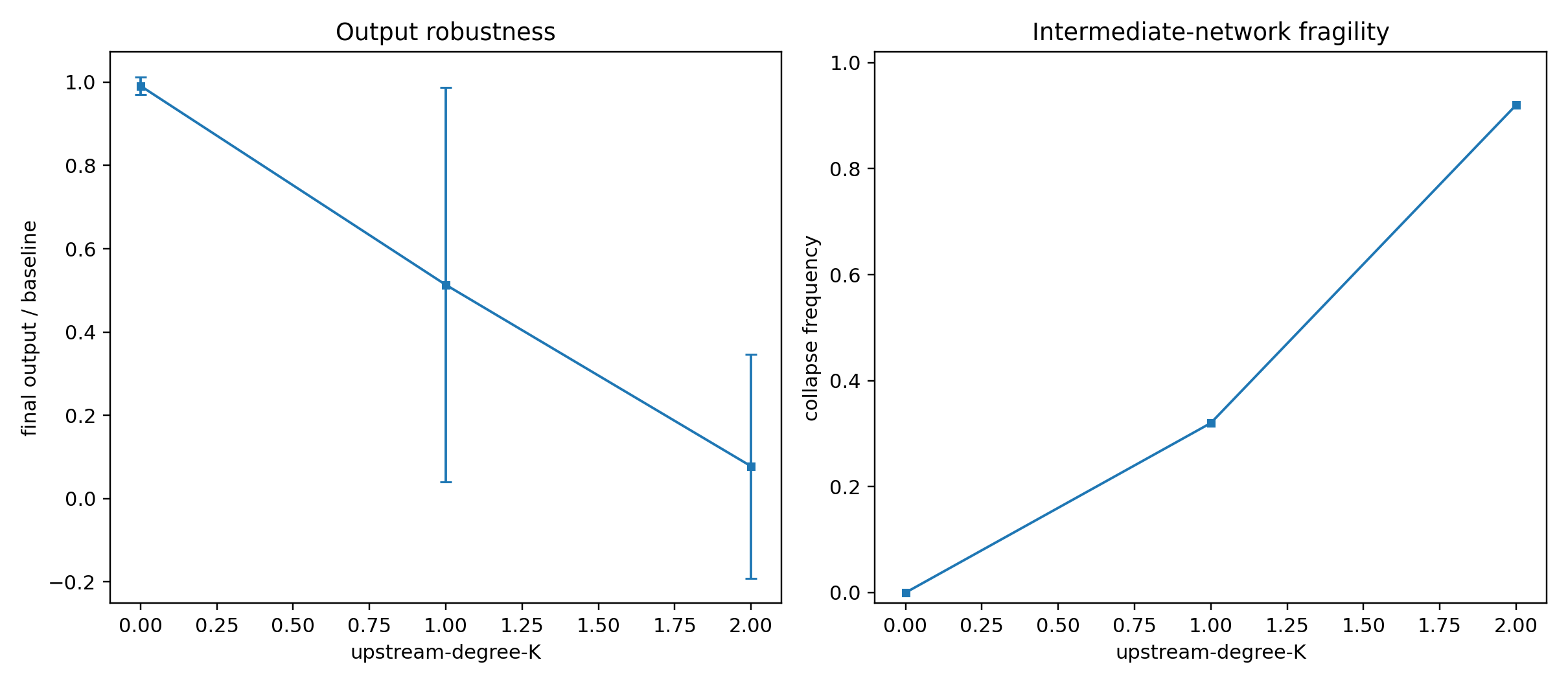}
	\caption{Intermediate-sector interdependence under expectations-on-expectations, no additional pressure.}
\end{figure}

The figure shows that increasing interdependence inside the intermediate sector primarily affects real robustness. Higher upstream-degree-K tends to reduce output and increase the probability of collapse, indicating that dense $K\to K$ linkages generate fragility rather than smooth inflationary propagation.

\begin{figure}[htbp]
	\centering
	\includegraphics[width=0.6\textwidth]{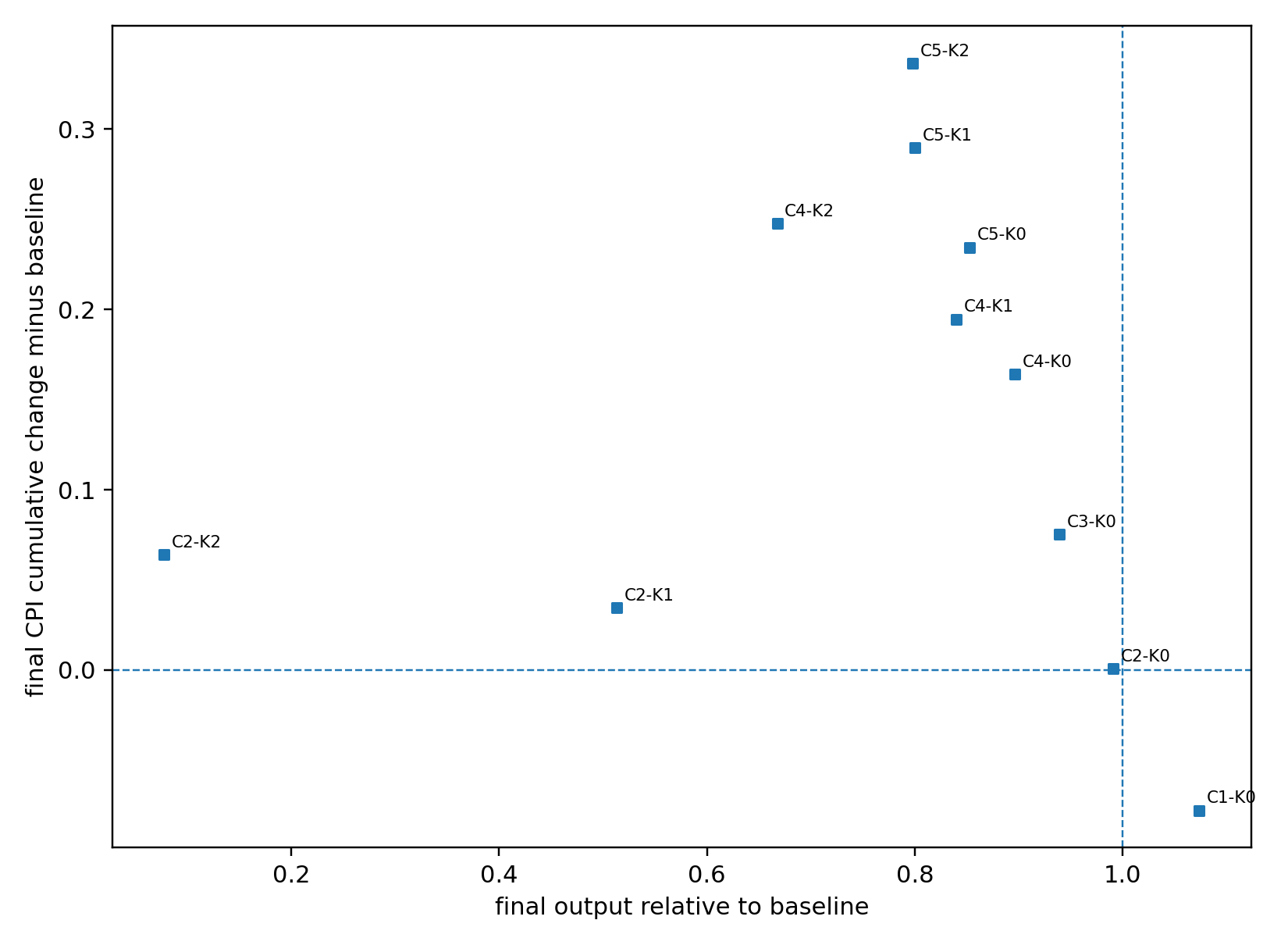}
	\caption{Network inflation-output trade-off under expectations-on-expectations, no additional pressure.}
\end{figure}

The scatter plot confirms that, without an independent source of price pressure, alternative network architectures mainly redistribute outcomes along the output dimension. Inflationary effects remain limited, while fragile architectures are identified by lower output.

\subsection*{A.2 Network diagnostics under mark-up pressure}

The figures show that mark-up dynamics are a genuine source of inflation, while the network determines the degree of propagation and the associated real contraction.

\begin{figure}[htbp]
	\centering
	\includegraphics[width=0.85\textwidth]{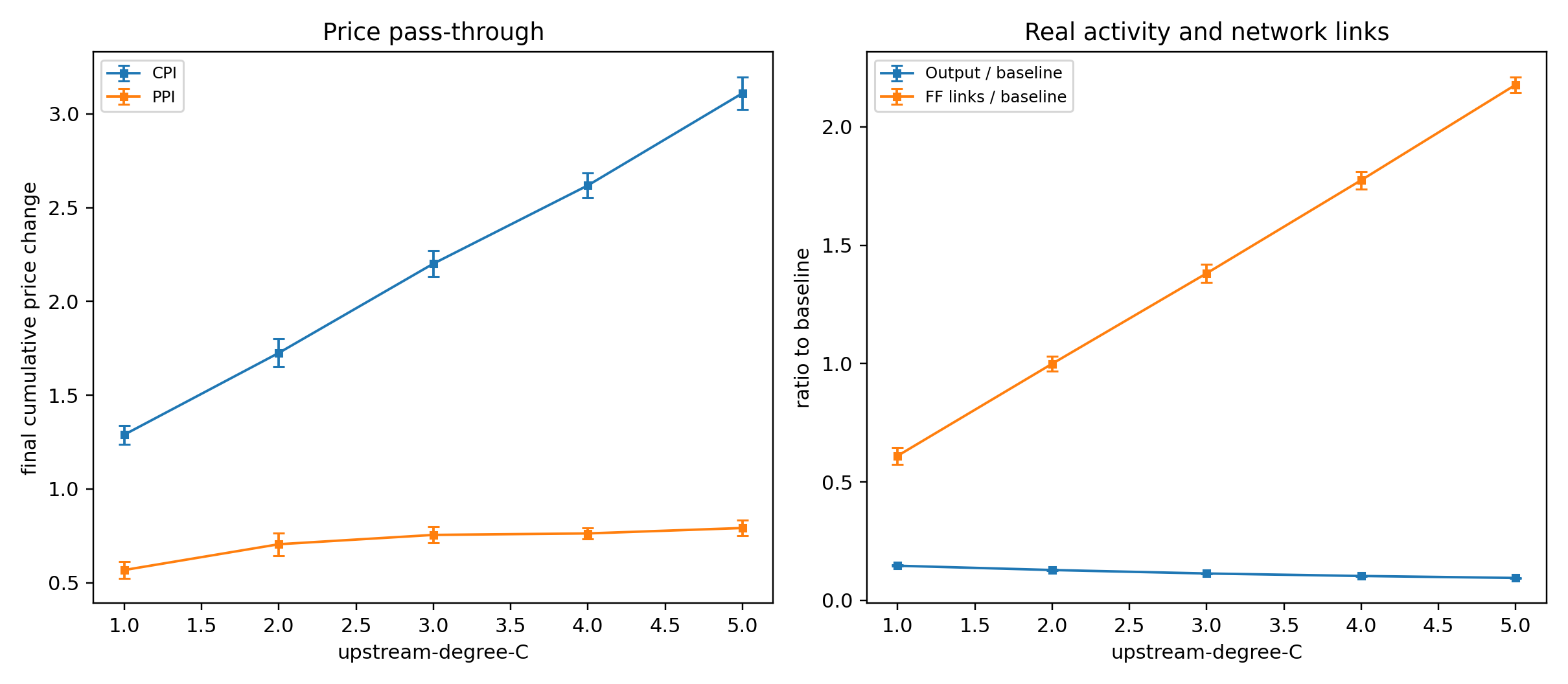}
	\caption{Downstream network exposure under mark-up pressure and price anchoring.}
\end{figure}

The figure shows that, under price anchoring, mark-up pressure produces stronger price increases as downstream exposure rises. Higher upstream-degree-C strengthens the transmission of firm-level pricing decisions through the production network.

\begin{figure}[htbp]
	\centering
	\includegraphics[width=0.85\textwidth]{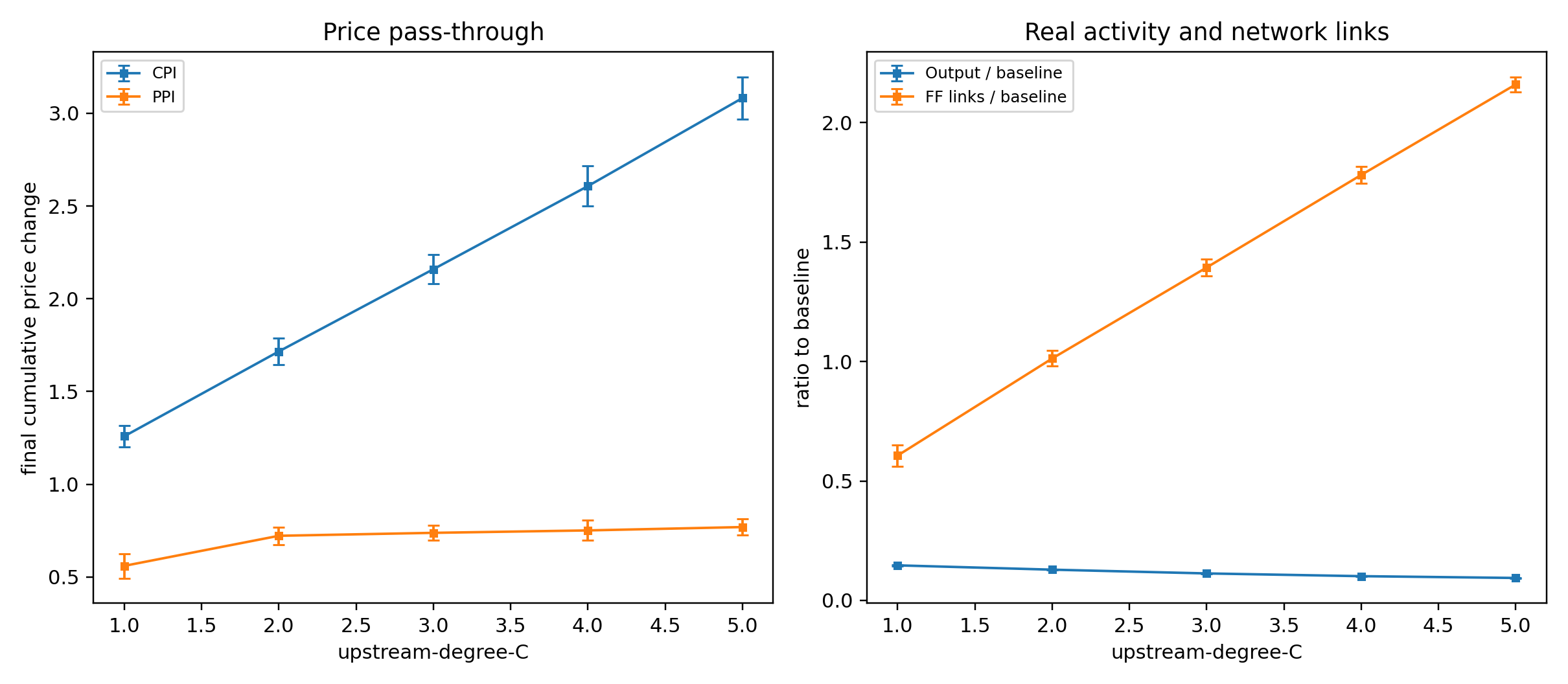}
	\caption{Downstream network exposure under mark-up pressure and expectations-on-expectations.}
\end{figure}

The same experiment, under expectations-on-expectations, shows a similar qualitative pattern: the mark-up channel drives inflation, while the network modulates its strength. The anchoring rule changes persistence more than the basic direction of the effect.

\begin{figure}[htbp]
	\centering
	\includegraphics[width=0.6\textwidth]{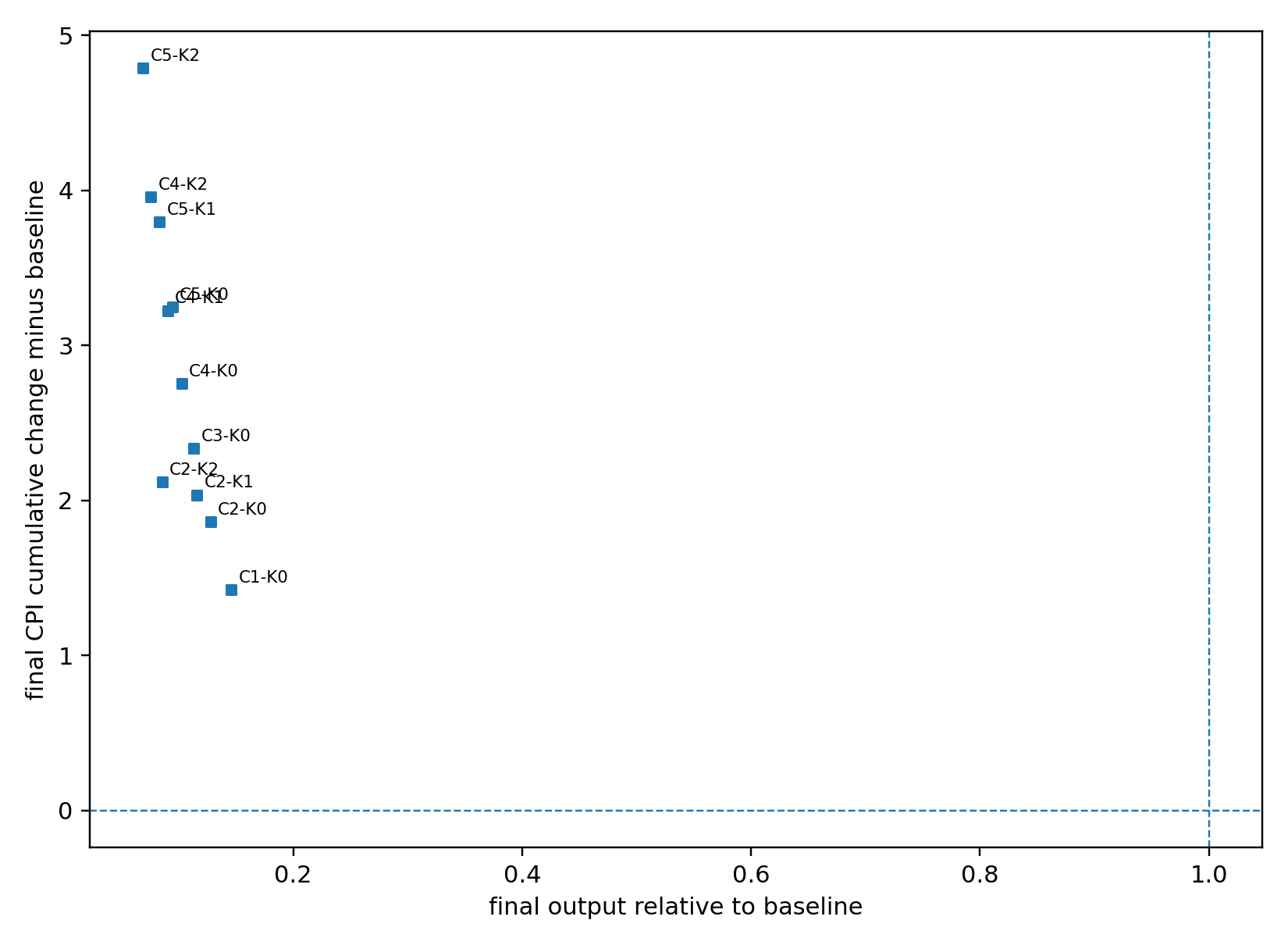}
	\caption{Network inflation-output trade-off under mark-up pressure and price anchoring.}
\end{figure}

The scatter plot shows that mark-up pressure shifts network configurations toward higher inflation and lower output. This confirms that mark-up inflation is not neutral on the real side: stronger price dynamics are associated with demand destruction and lower production.

\begin{figure}[htbp]
	\centering
	\includegraphics[width=0.6\textwidth]{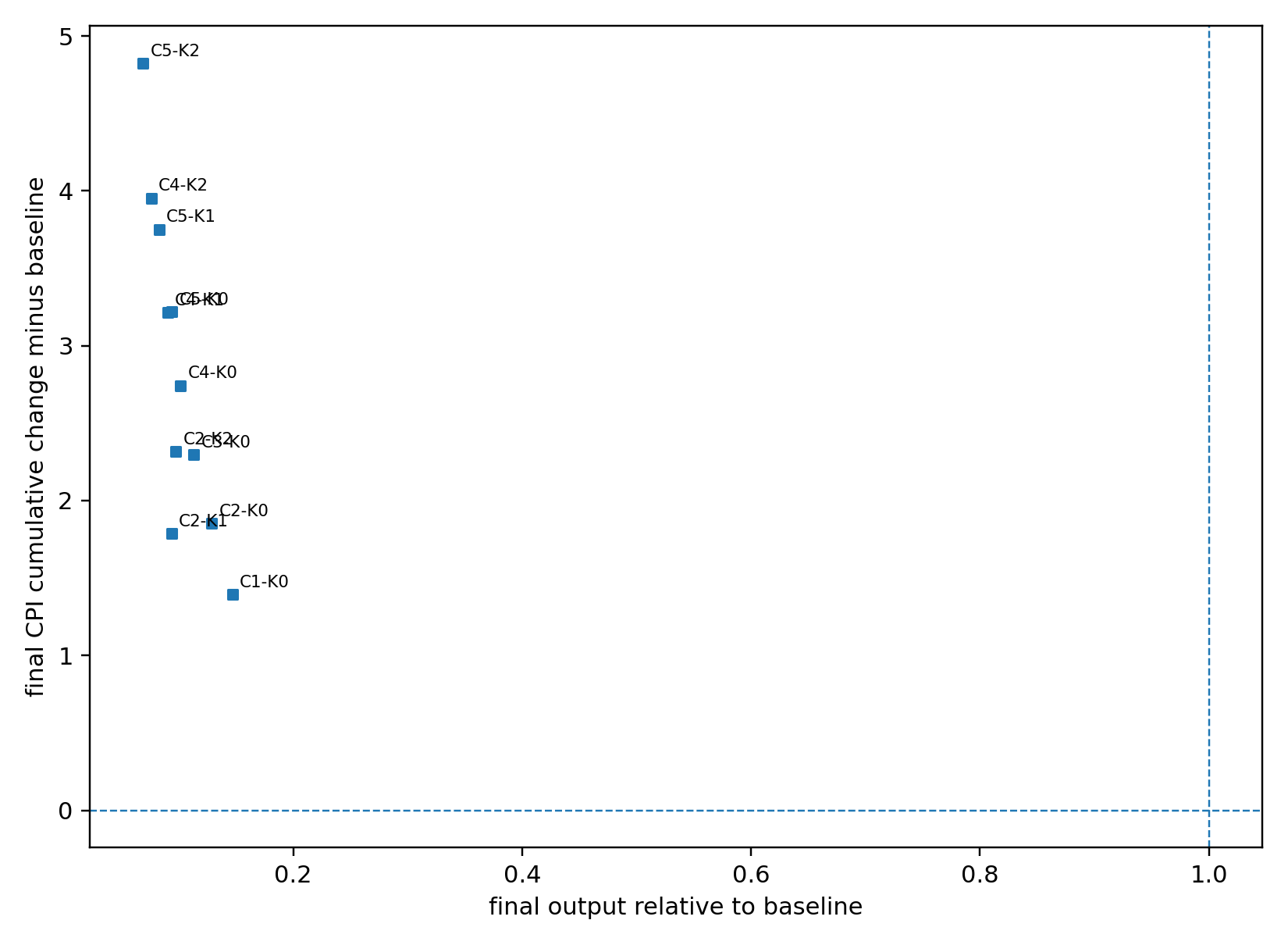}
	\caption{Network inflation-output trade-off under mark-up pressure and expectations-on-expectations.}
\end{figure}

The corresponding expectations-on-expectations scatter plot confirms the same trade-off. Relative to price anchoring, the main effect is not the emergence of a different regime, but a change in the persistence and amplitude of inflationary adjustment once mark-up pressure is present.

\subsection*{A.3 Network diagnostics under financial-cost pressure}

Network diagnostics, under financial-cost pressure, reveals that bank-cost and policy-rate shocks both operate through firms' financing costs, but the source of the increase differs. The figures show that downstream exposure strengthens cost pass-through, while fragile network architectures convert financial pressure into real contraction.

\begin{figure}[htbp]
	\centering
	\includegraphics[width=0.85\textwidth]{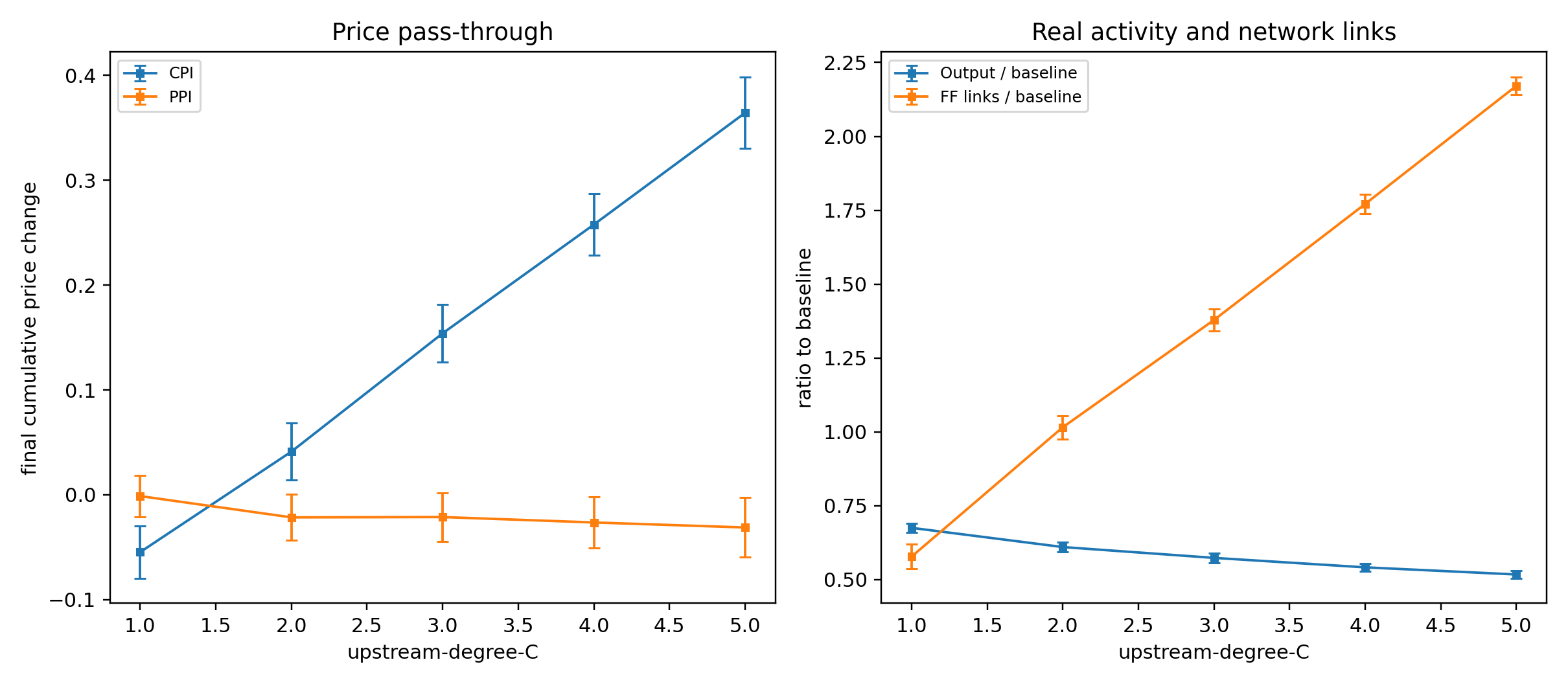}
	\caption{Downstream network exposure under bank-cost pressure and price anchoring.}
\end{figure}

The figure shows that bank-cost pressure becomes more visible when consumption-good firms are more exposed to intermediate inputs. Higher downstream network exposure increases the pass-through from financial costs to unit costs and then to consumer prices.

\begin{figure}[htbp]
	\centering
	\includegraphics[width=0.85\textwidth]{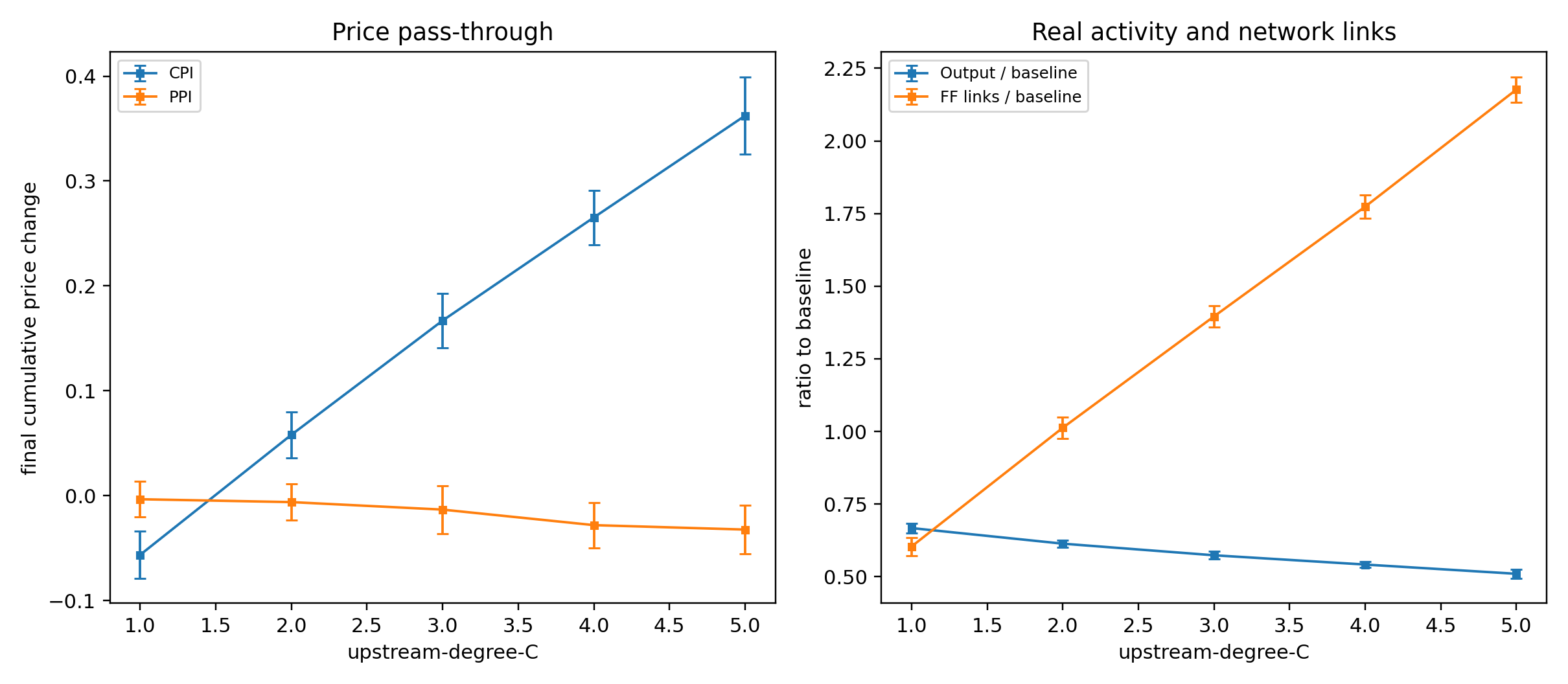}
	\caption{Downstream network exposure under policy-rate pressure and price anchoring.}
\end{figure}

Policy-rate pressure behaves as a financial-cost channel: as the policy raises lending rates, the effect on consumer prices is stronger when the production network transmits cost increases more intensely downstream.

\begin{figure}[htbp]
	\centering
	\includegraphics[width=0.5\textwidth]{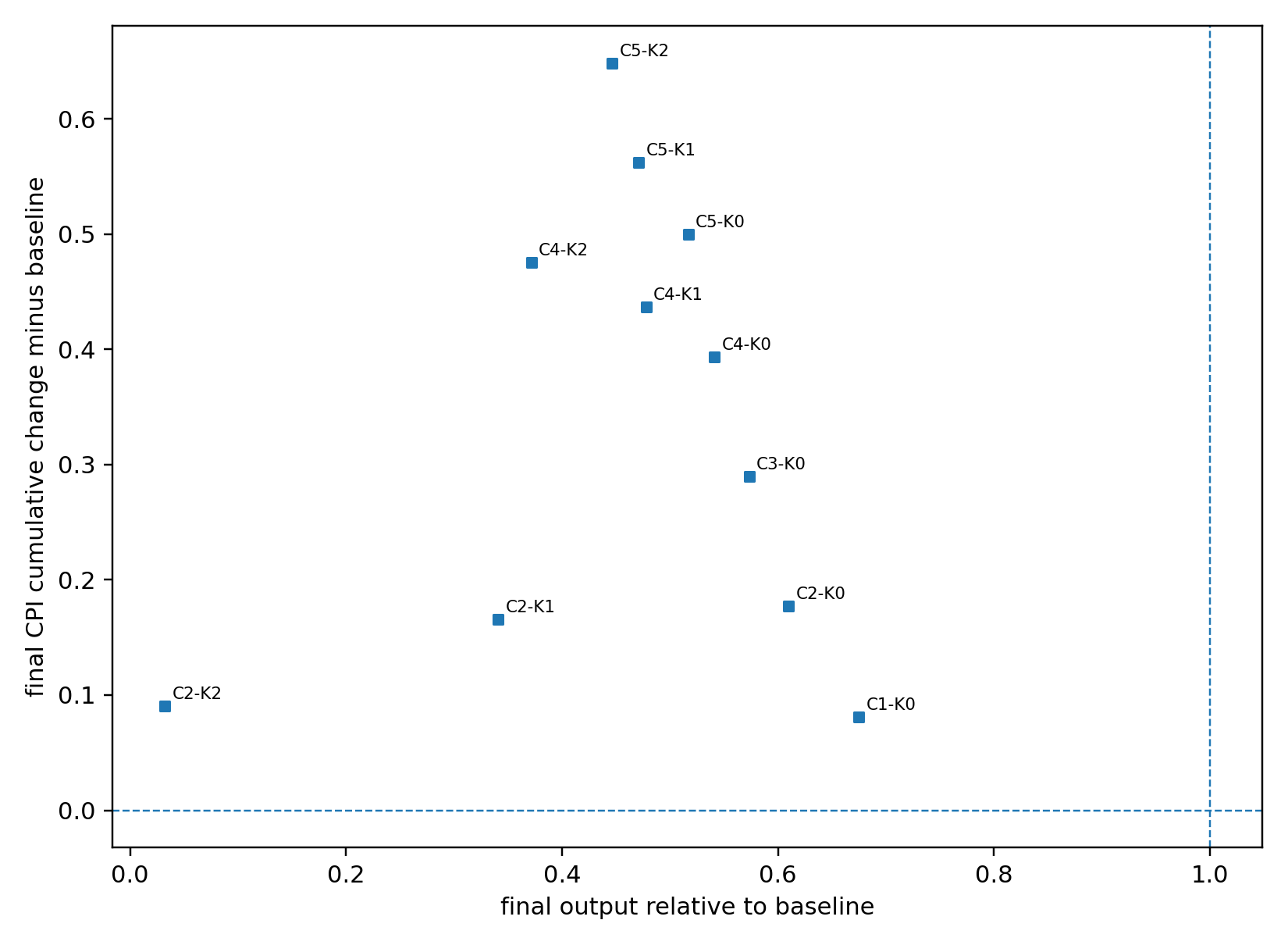}
	\caption{Network inflation-output trade-off under bank-cost pressure and price anchoring.}
\end{figure}

The scatter plot shows that bank-cost pressure produces a moderate inflation-output trade-off. Some network architectures transmit financial costs into prices while preserving output, whereas others mainly convert financial pressure into lower production.

\begin{figure}[htbp]
	\centering
	\includegraphics[width=0.5\textwidth]{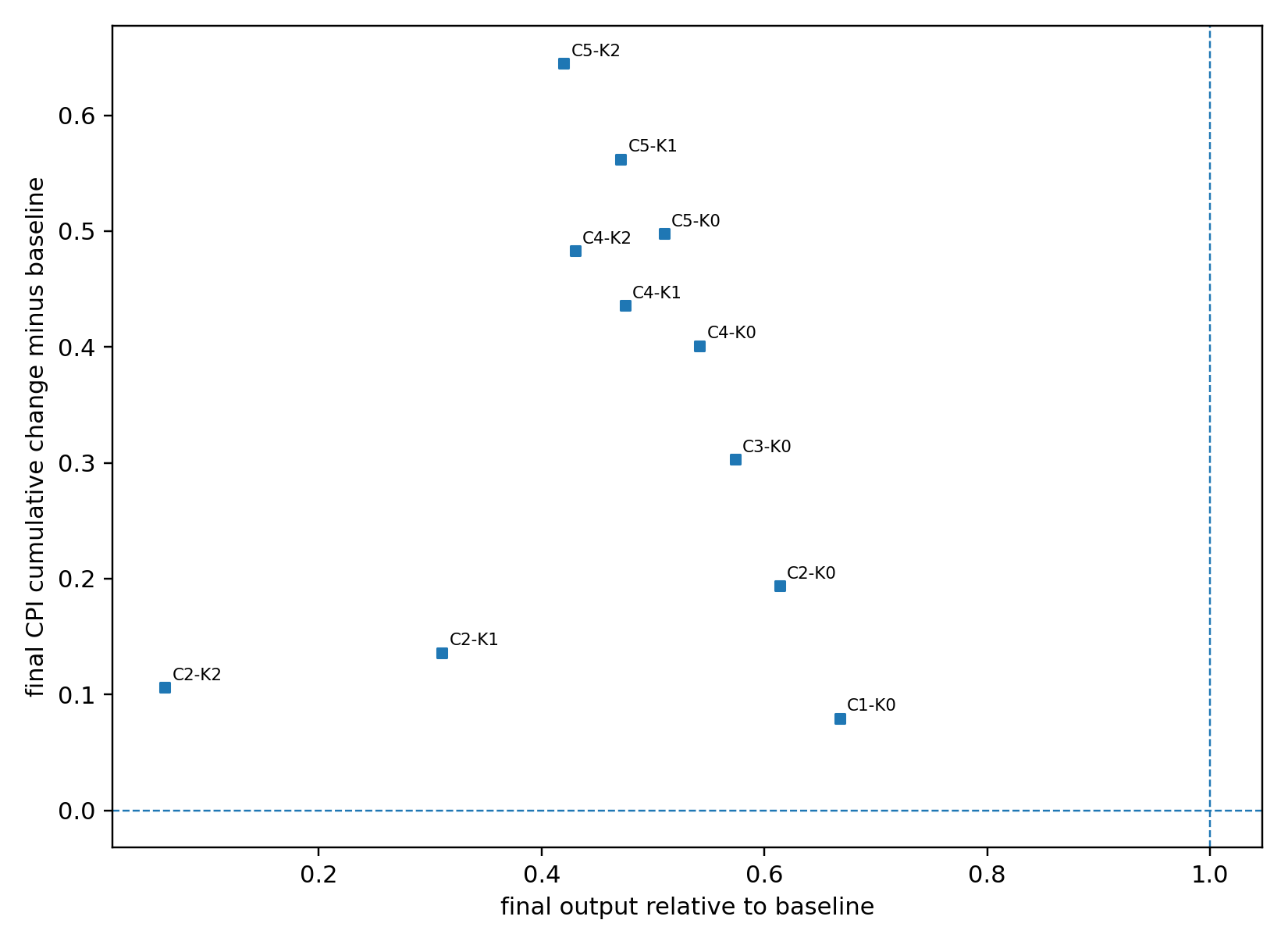}
	\caption{Network inflation-output trade-off under policy-rate pressure and price anchoring.}
\end{figure}

The policy-rate scatter plot shows the same mechanism from the policy side. The increase in the policy rate raises the financial component of costs; whether this appears mainly as inflation or as output loss depends on the production-network architecture.

\subsection*{A.4 Network diagnostics under natural-capital pressure}

Natural-capital shocks are transmitted first to producer prices and then to consumer prices when downstream exposure is sufficiently strong.

\begin{figure}[htbp]
	\centering
	\includegraphics[width=0.85\textwidth]{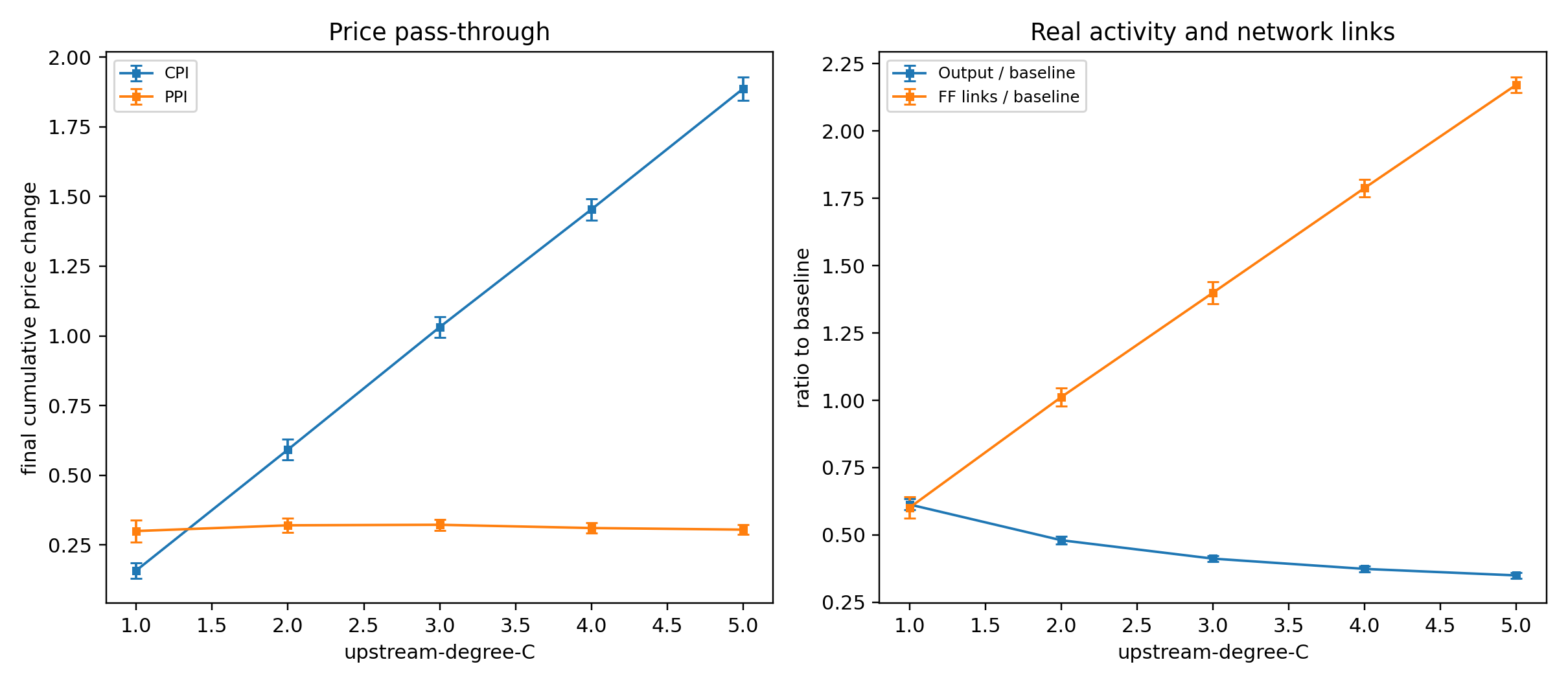}
	\caption{Downstream network exposure under natural-capital pressure and price anchoring.}
\end{figure}

The figure shows that higher downstream exposure strengthens the transmission of natural-capital cost shocks from intermediate-good producers to final consumption prices. The CPI response is therefore conditional on the technological dependence of consumption-good firms on upstream inputs.

\begin{figure}[htbp]
	\centering
	\includegraphics[width=0.85\textwidth]{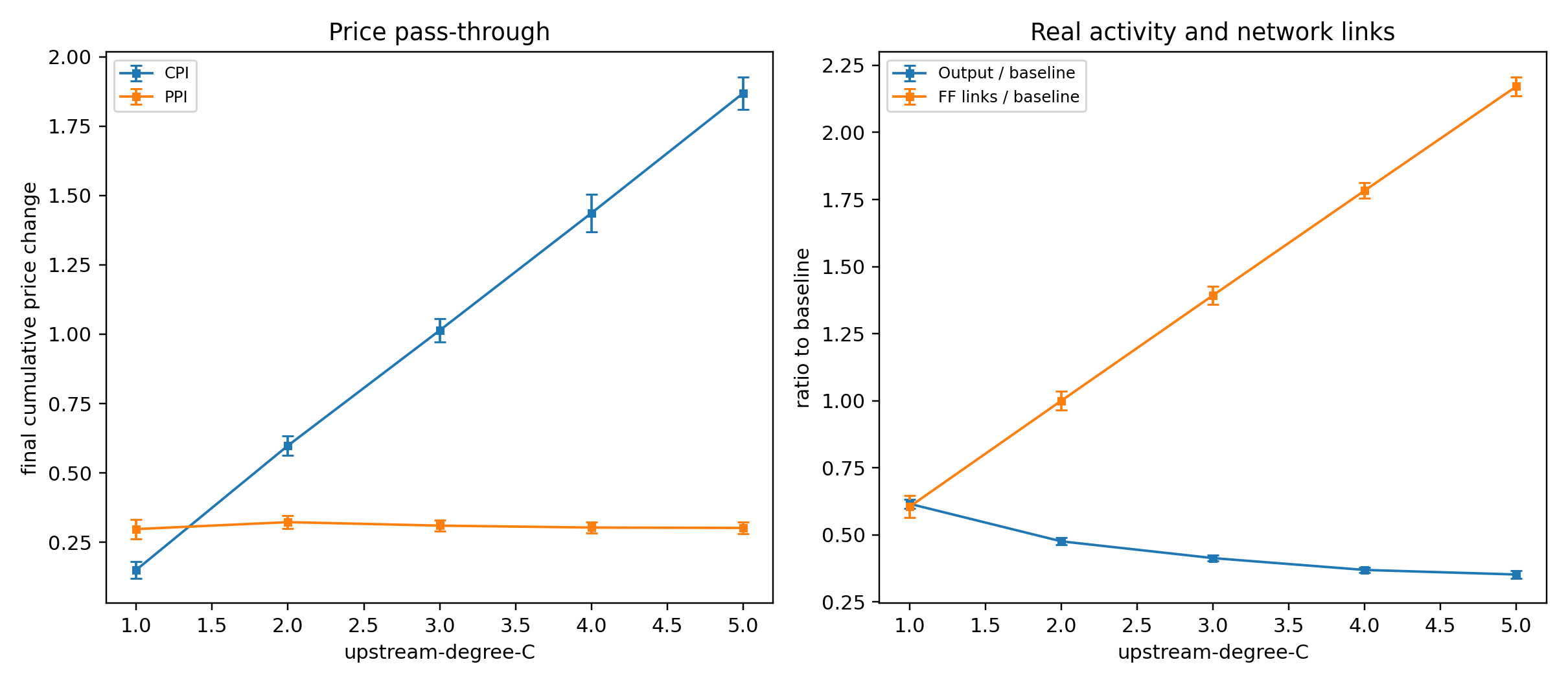}
	\caption{Downstream network exposure under natural-capital pressure and expectations-on-expectations.}
\end{figure}

The figure shows the same natural-capital experiment under expectations-on-expectations. The qualitative pattern remains: the shock is upstream, and downstream pass-through depends on the production network. Expectations can amplify the response once the price signal is present, but they do not replace the cost channel.

\begin{figure}[htbp]
	\centering
	\includegraphics[width=0.6\textwidth]{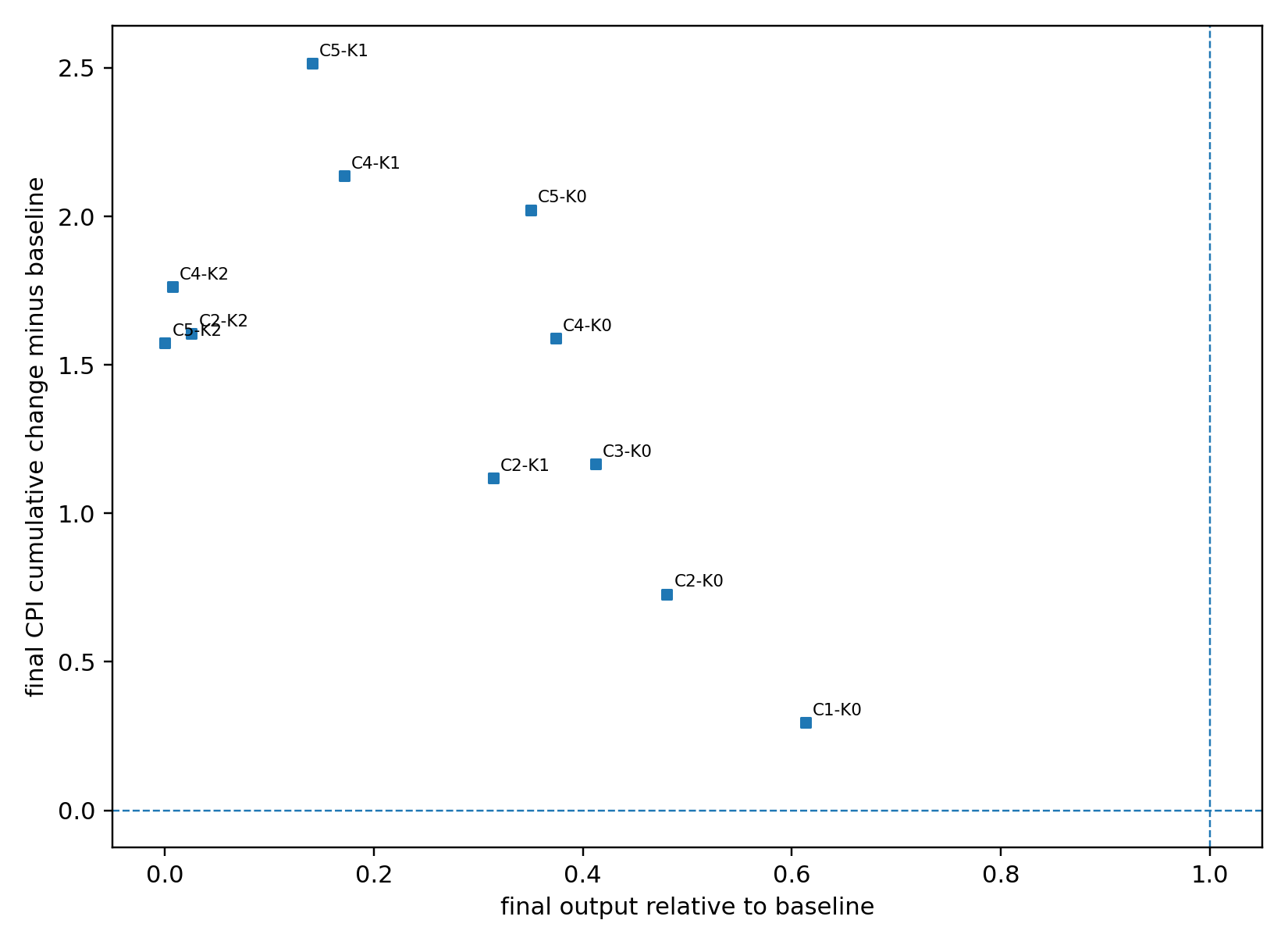}
	\caption{Network inflation-output trade-off under natural-capital pressure and price anchoring.}
\end{figure}

The scatter plot shows that natural-capital pressure generates stronger inflation when the network allows upstream cost increases to reach consumption-good markets. At the same time, configurations with stronger pass-through tend to be associated with lower output, reflecting the real cost of resource-price pressure.

\begin{figure}[htbp]
	\centering
	\includegraphics[width=0.6\textwidth]{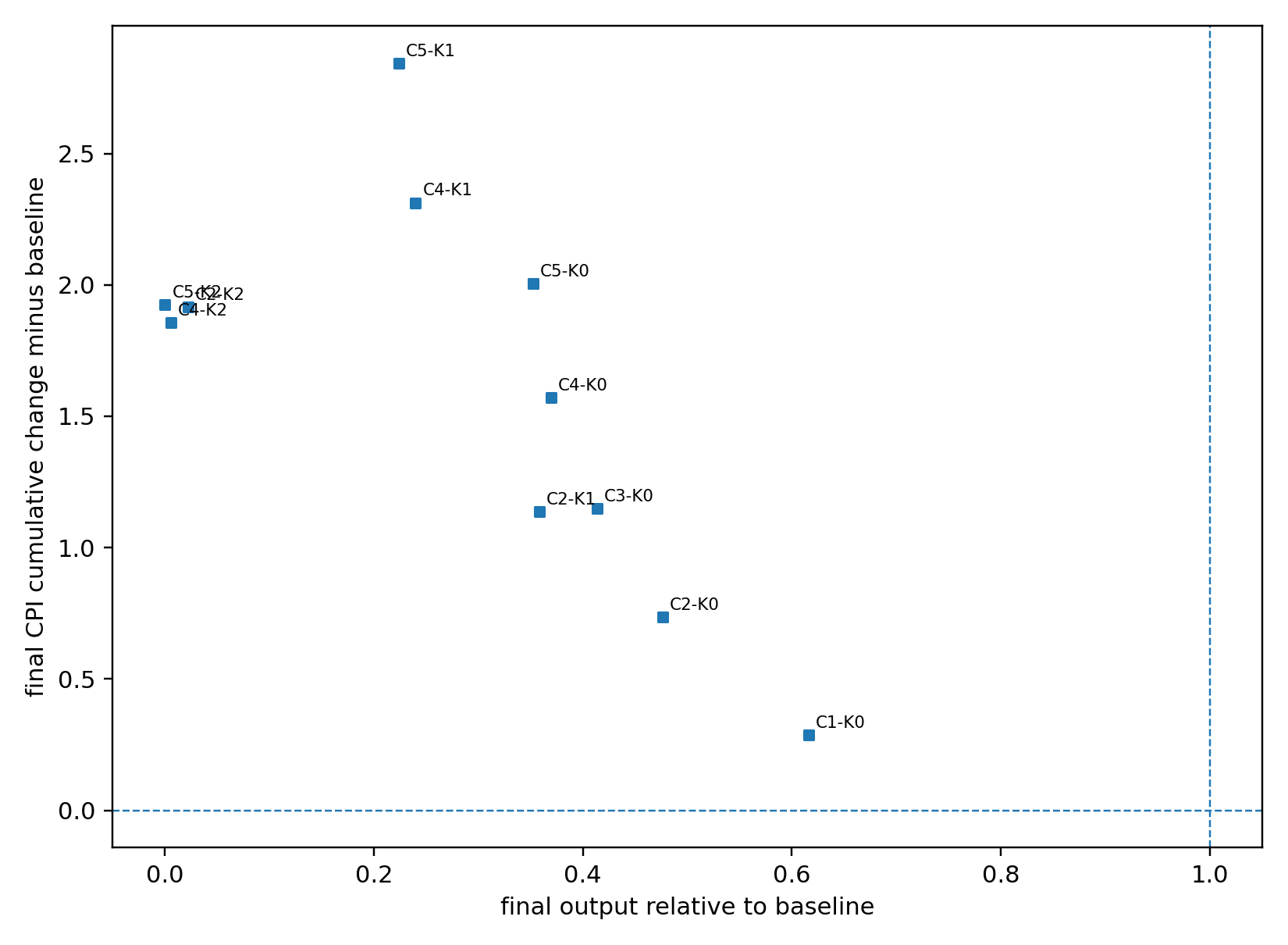}
	\caption{Network inflation-output trade-off under natural-capital pressure and expectations-on-expectations.}
\end{figure}

The corresponding expectations-on-expectations scatter plot confirms that natural-capital inflation is not purely an expectation phenomenon. Expectations affect the propagation of an existing upstream cost signal, while the strength of the inflation-output trade-off remains governed by network exposure.

\subsection*{A.5 Expectations-only robustness across belief-correction weights}
\label{app:expectations_weighting_robustness}

\begin{figure}[htbp]
	\centering
	\includegraphics[width=0.95\textwidth]{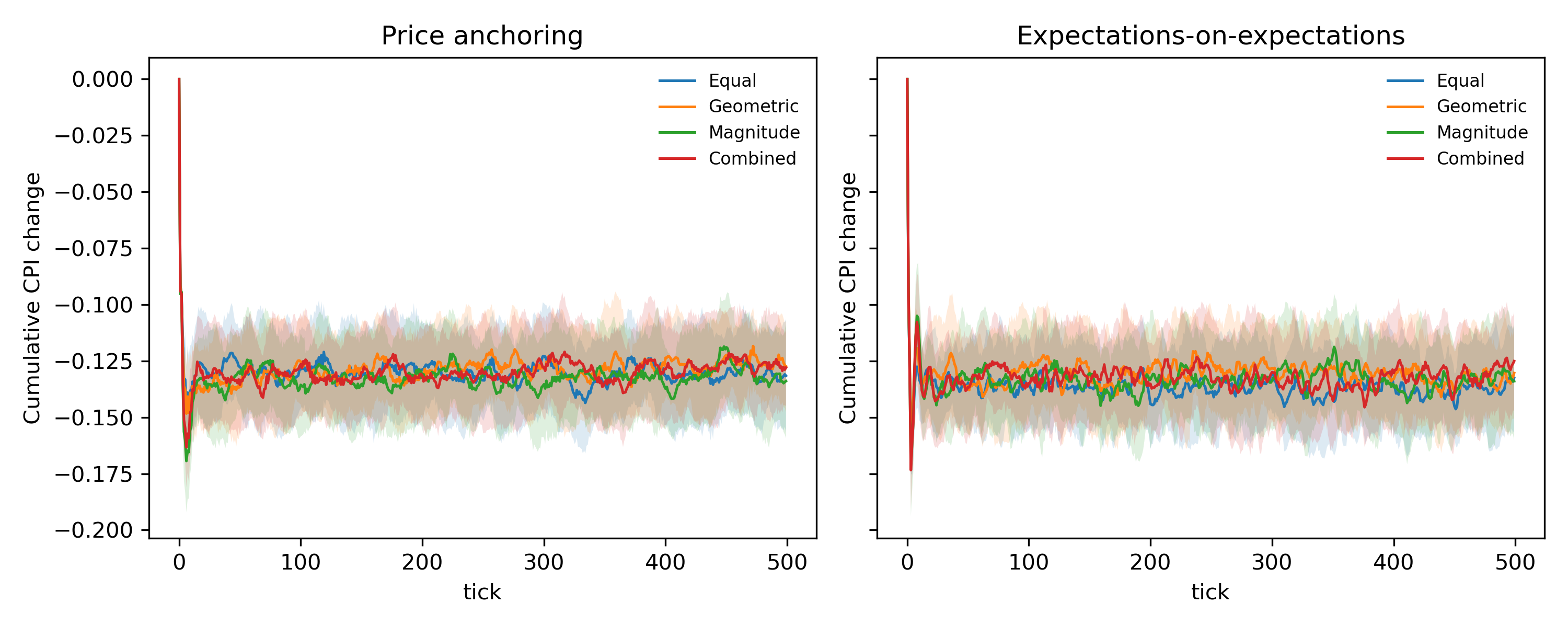}
	\caption{Expectations-only robustness across belief-correction weighting schemes. Lines report Monte Carlo averages; shaded bands report cross-seed standard deviations.}
	\label{fig:expectations_weighting_robustness}
\end{figure}

We compare the baseline combined timing--magnitude scheme with equal, geometric, and magnitude-based weights, under both price anchoring and expectations-on-expectations. Figure~\ref{fig:expectations_weighting_robustness} shows that the result is robust: in the absence of an independent cost or mark-up pressure, changing the memory-weighting rule does not generate sustained inflation. The alternative schemes affect small transitory movements and cross-seed variability, but they do not turn expectations into an autonomous inflationary source.

\subsection*{A.6 Robustness to the exclusion of the expectations channel}
\label{app:kappa_zero}

The pricing rule in Eq.~\eqref{pricing_rule} allows expected inflation to affect posted prices through the parameter $\kappa$. To isolate the quantitative contribution of this mechanism, we compare the expectation-augmented specification used in the main simulations, $\kappa=0.35$, with the limiting case $\kappa=0$. In the latter case, the pricing rule reduces to pure cost-plus pricing,
\[
p_{i,t}=(1+\mu_{i,t})\,\textsc{uc}_{i,t},
\]
so that expected inflation continues to be computed but no longer affects firms' posted prices.

The comparison is performed under five environments: no independent inflationary pressure, mark-up pressure, bank-cost pressure, policy-rate pressure, and natural-capital pressure. For $\kappa=0.35$, both price anchoring and expectations-on-expectations anchoring are considered. Each specification is evaluated over $25$ Monte Carlo replications. Lines and bars report Monte Carlo means, while shaded bands and error bars report cross-seed standard deviations.

\begin{figure}[htbp]
	\centering
	\includegraphics[width=0.85\textwidth]
	{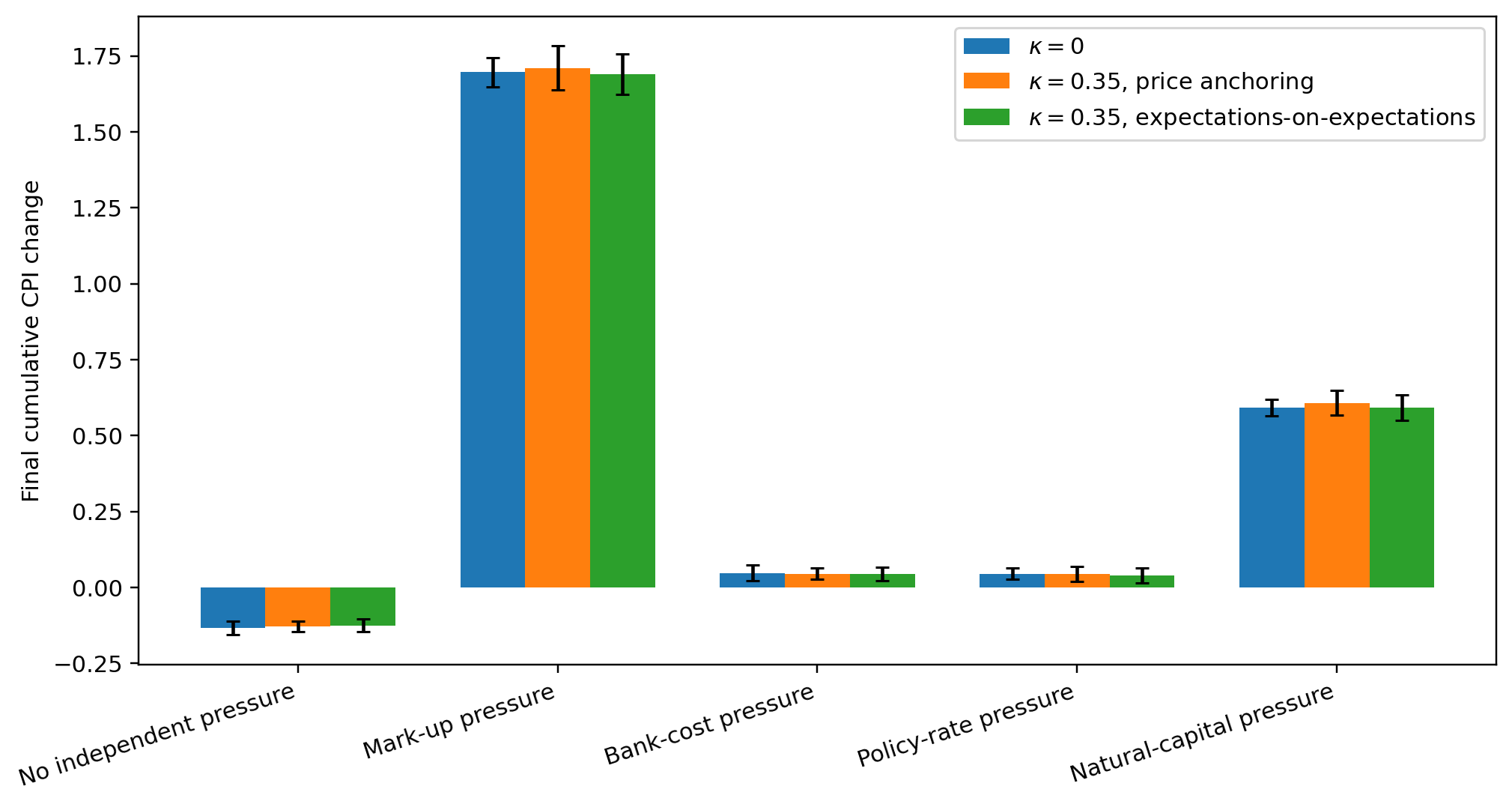}
	\caption{Final cumulative CPI change under pure cost-plus pricing, $\kappa=0$, and expectation-augmented pricing, $\kappa=0.35$, under the two anchoring rules. Error bars report cross-seed standard deviations.}
	\label{fig:kappa_cpi_summary}
\end{figure}

Figure~\ref{fig:kappa_cpi_summary} compares the final cumulative CPI response across the five environments. The distance between the $\kappa=0$ outcome and the two $\kappa=0.35$ outcomes measures the incremental contribution of the expectations channel. In the absence of an independent source of price pressure, the comparison tests whether expectations can generate inflation autonomously. In the remaining environments, it measures the extent to which expectations amplify inflation generated by mark-ups, financing costs, monetary-policy costs, or natural-capital costs.

\begin{figure}[htbp]
	\centering
	\includegraphics[width=0.85\textwidth]
	{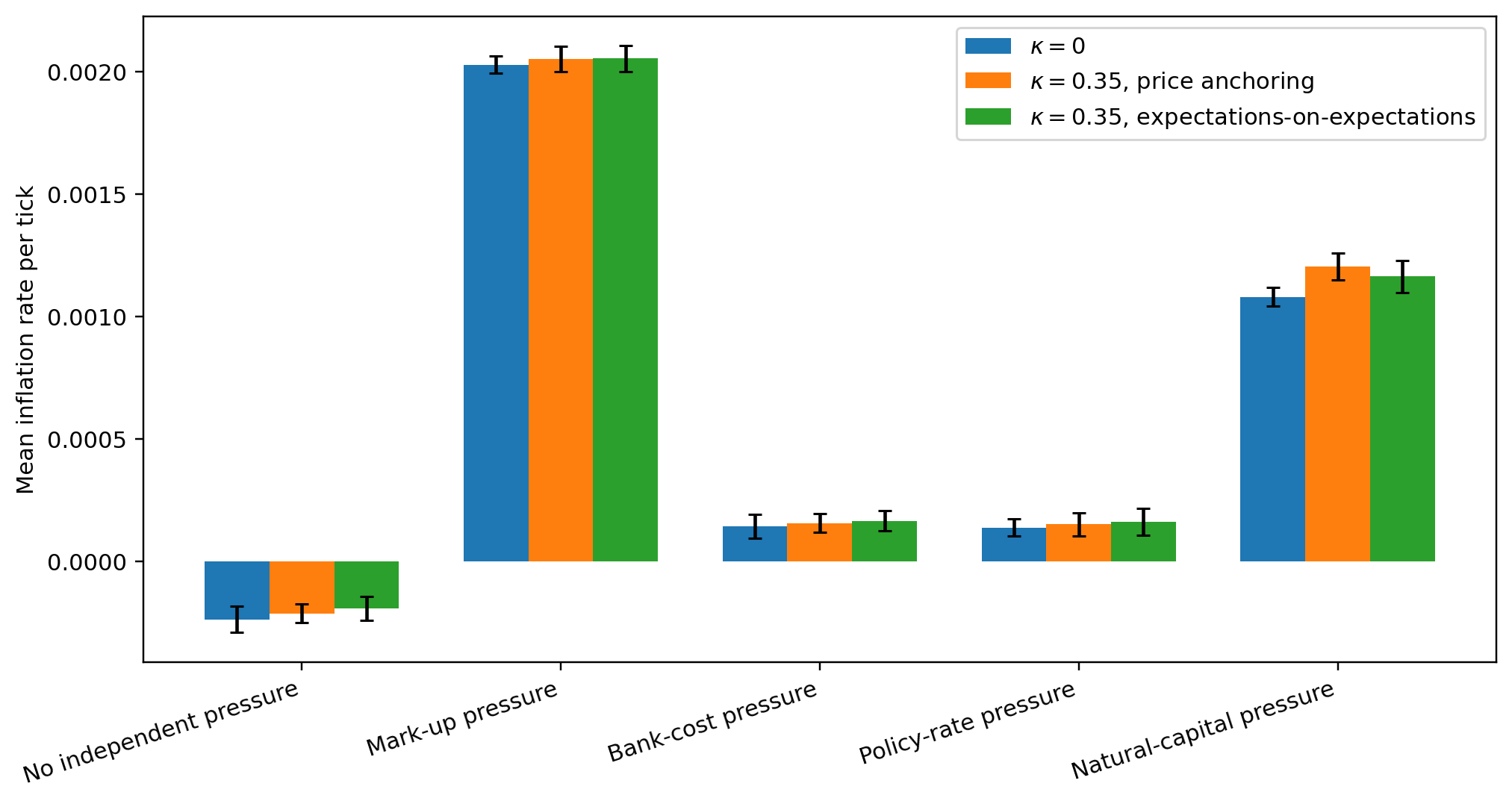}
	\caption{Mean per-tick inflation under pure cost-plus and expectation-augmented pricing. Error bars report cross-seed standard deviations.}
	\label{fig:kappa_inflation_summary}
\end{figure}

Figure~\ref{fig:kappa_inflation_summary} reports the corresponding average per-tick inflation rates. This measure complements cumulative CPI growth by showing whether expectations alter the average speed of price adjustment over the simulation horizon.

\begin{figure}[htbp]
	\centering
	\includegraphics[width=0.85\textwidth]
	{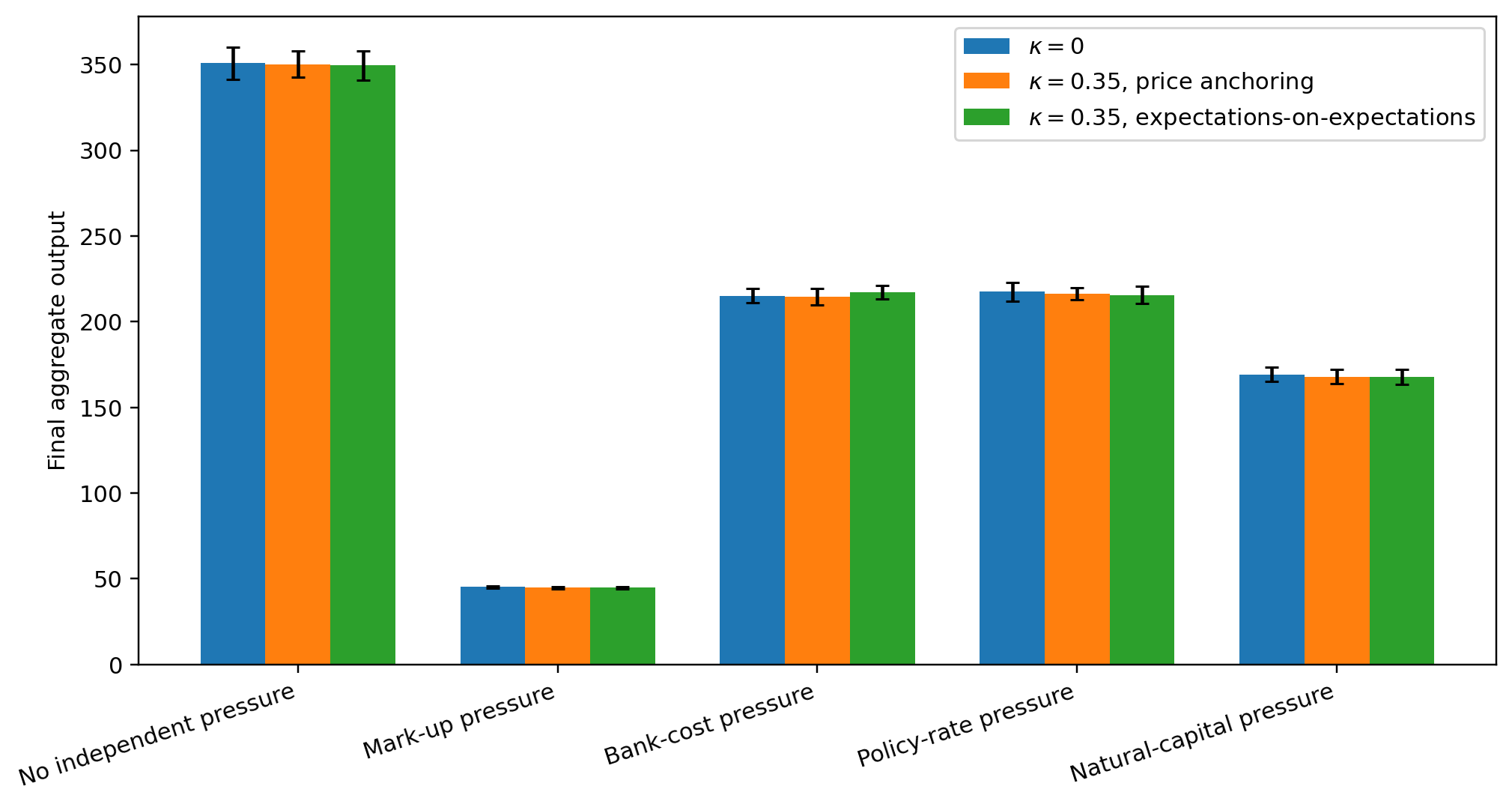}
	\caption{Final aggregate output under pure cost-plus and expectation-augmented pricing. Error bars report cross-seed standard deviations.}
	\label{fig:kappa_output_summary}
\end{figure}

Figure~\ref{fig:kappa_output_summary} shows the associated real effects. Comparing final output across the three pricing specifications makes it possible to determine whether stronger nominal amplification is accompanied by additional contraction in economic activity.

\begin{figure}[htbp]
	\centering
	\includegraphics[width=0.78\textwidth]
	{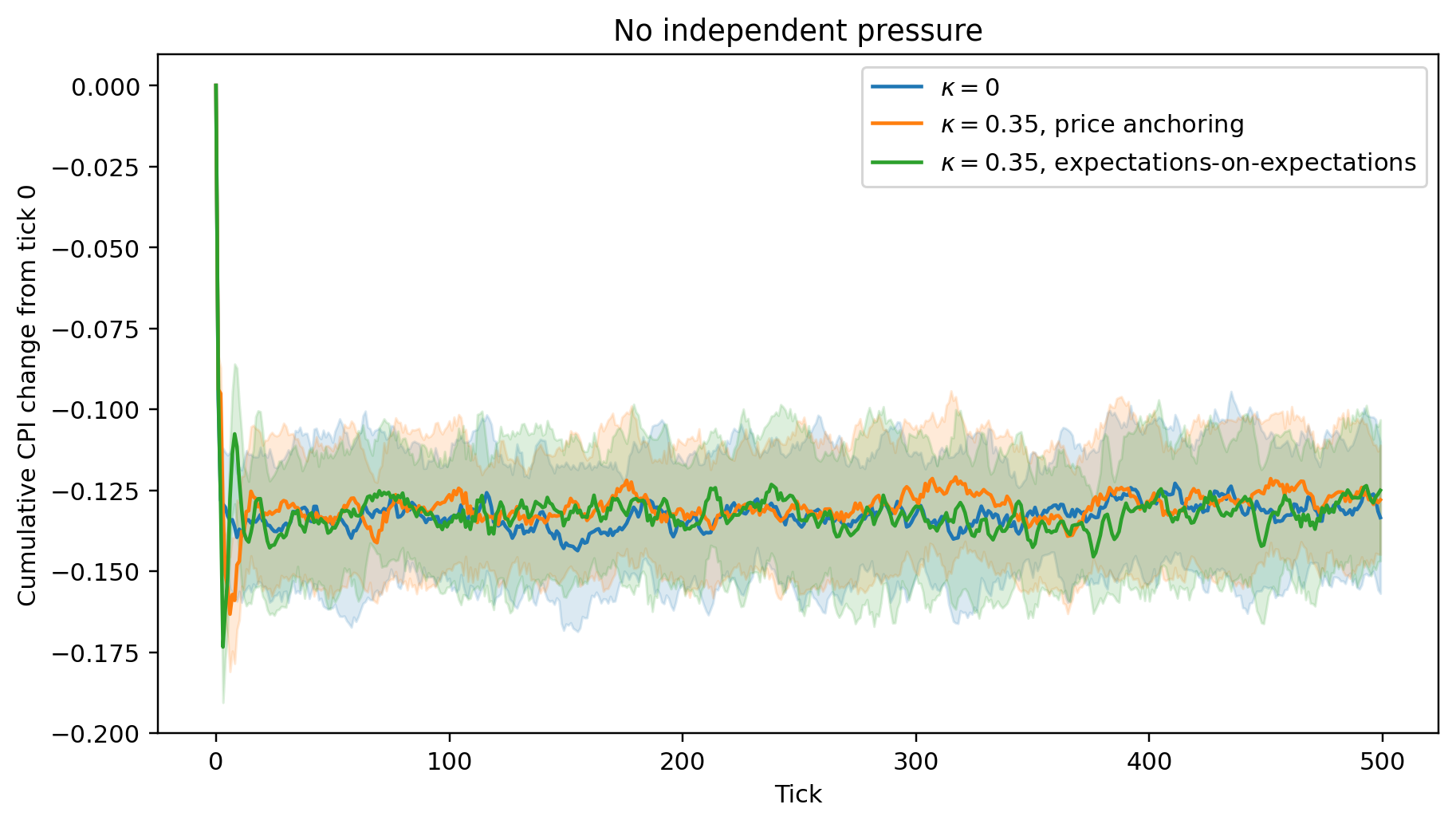}
	\caption{Cumulative CPI dynamics with no independent inflationary pressure.}
	\label{fig:kappa_cpi_alone}
\end{figure}

\begin{figure}[htbp]
	\centering
	\includegraphics[width=0.78\textwidth]
	{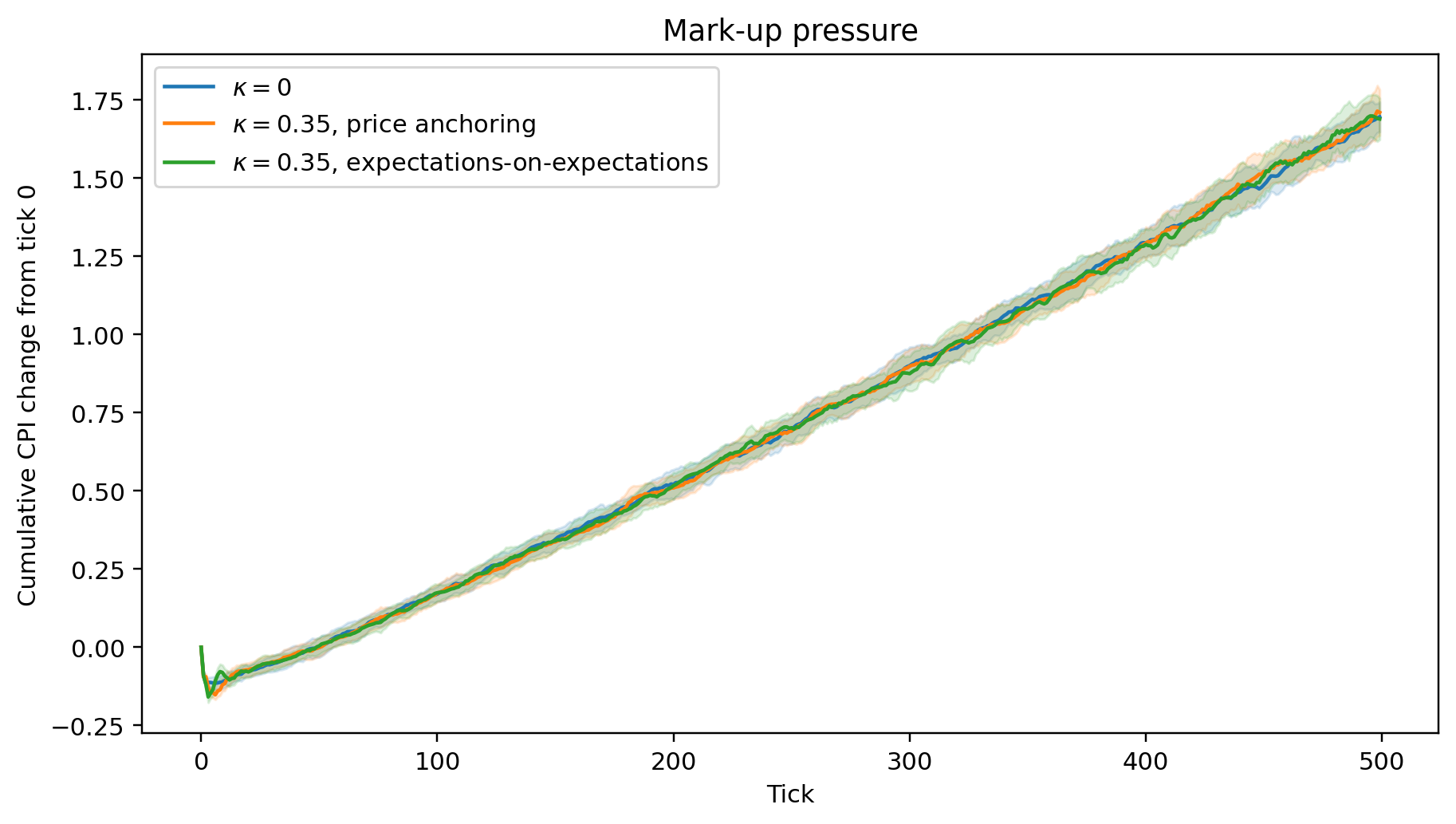}
	\caption{Cumulative CPI dynamics under mark-up pressure.}
	\label{fig:kappa_cpi_markup}
\end{figure}

\begin{figure}[htbp]
	\centering
	\includegraphics[width=0.78\textwidth]
	{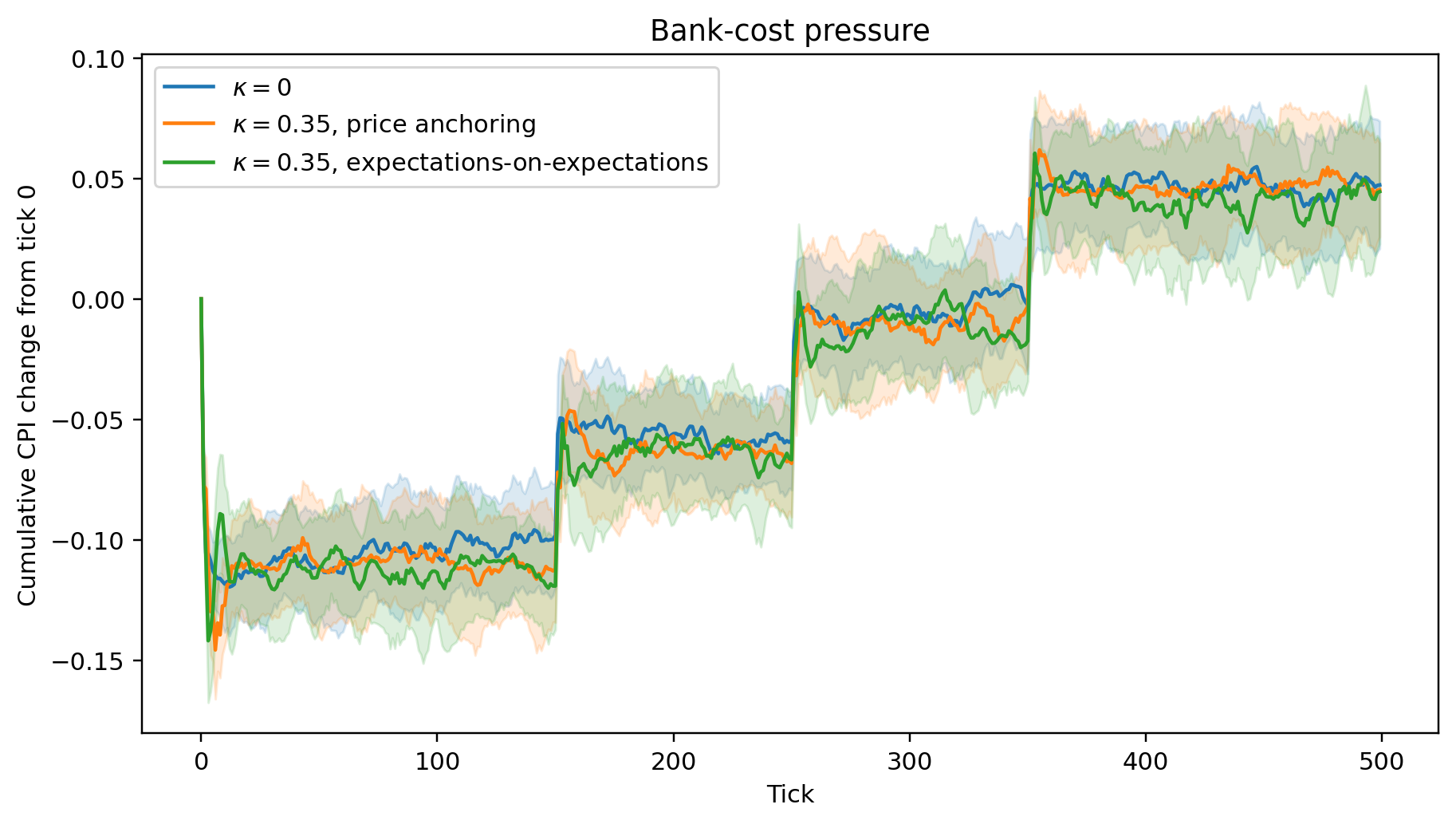}
	\caption{Cumulative CPI dynamics under bank-cost pressure.}
	\label{fig:kappa_cpi_credit}
\end{figure}

\begin{figure}[htbp]
	\centering
	\includegraphics[width=0.78\textwidth]
	{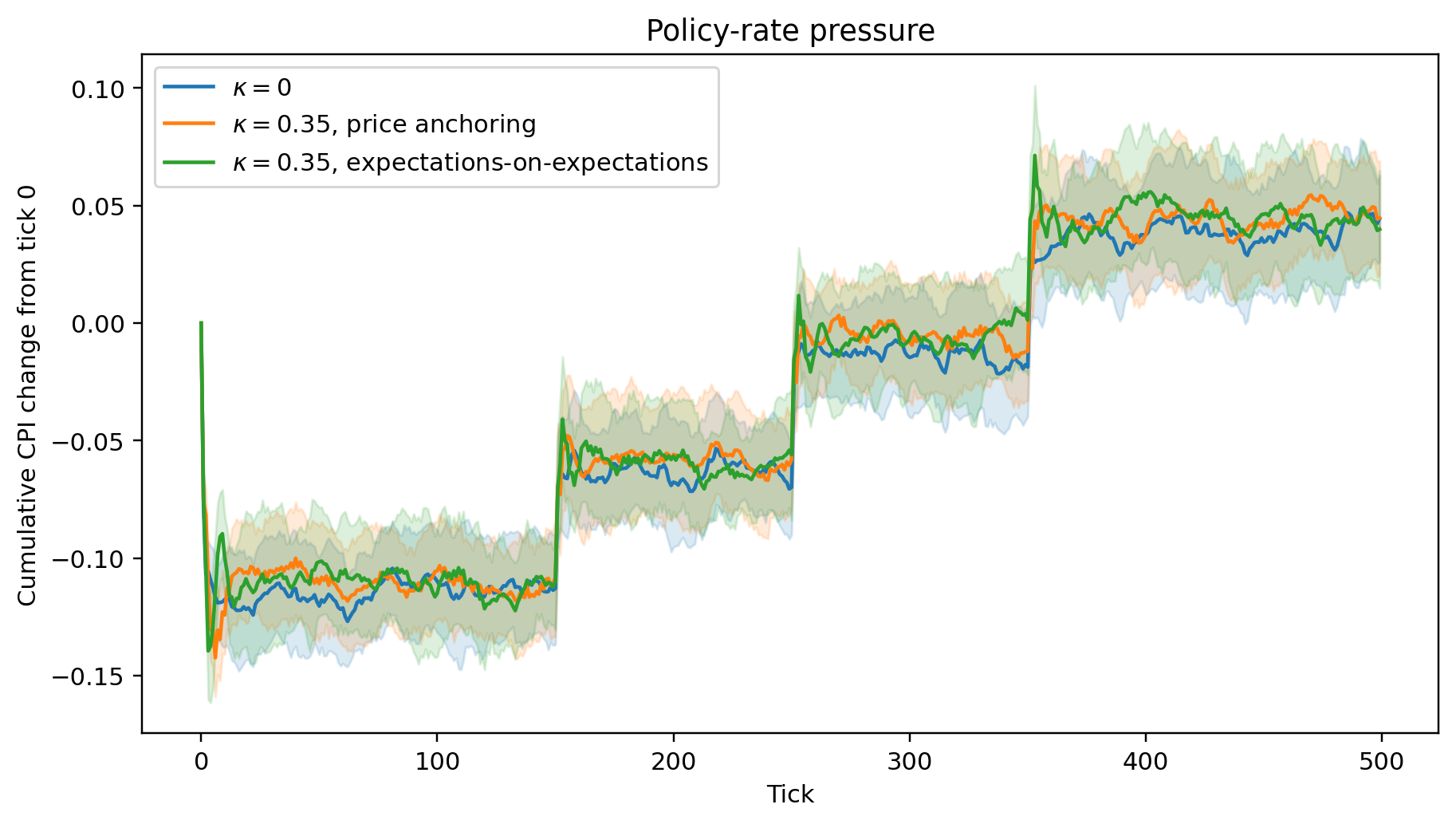}
	\caption{Cumulative CPI dynamics under policy-rate pressure.}
	\label{fig:kappa_cpi_policy}
\end{figure}

\begin{figure}[htbp]
	\centering
	\includegraphics[width=0.78\textwidth]
	{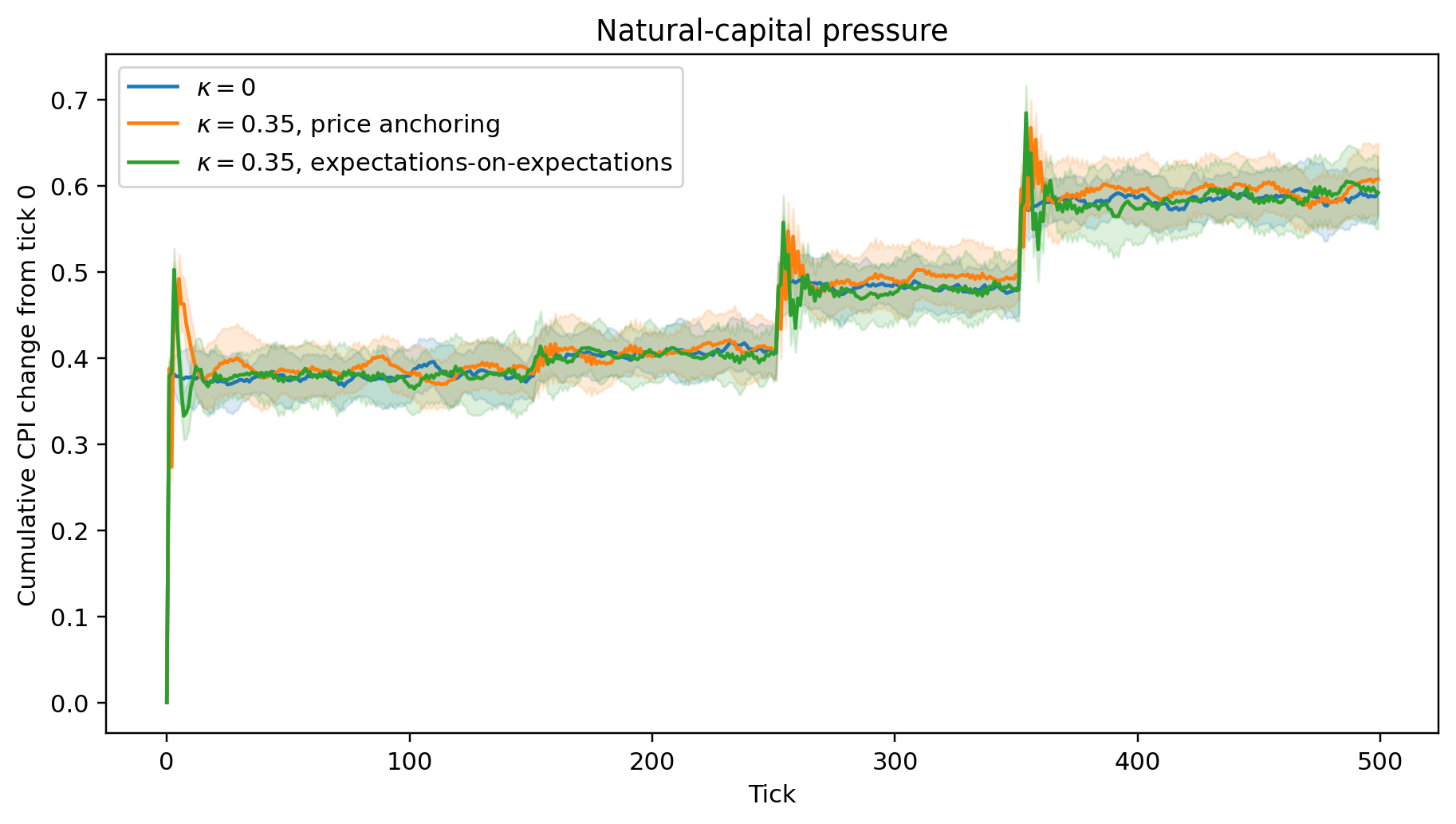}
	\caption{Cumulative CPI dynamics under natural-capital pressure.}
	\label{fig:kappa_cpi_natural}
\end{figure}

The context-specific trajectories reported in Figures~\ref{fig:kappa_cpi_alone}--\ref{fig:kappa_cpi_natural} make the timing of the expectations effect explicit. In each panel, the $\kappa=0$ trajectory provides the counterfactual path generated solely by realised costs and mark-ups, whereas the two $\kappa=0.35$ trajectories show how the alternative anchoring mechanisms modify the propagation of the same underlying pressure.

Taken together, these experiments provide a direct robustness test of the interpretation adopted in the main text. The $\kappa=0$ specification identifies the inflation generated by realised costs and mark-ups alone. The difference relative to $\kappa=0.35$ identifies the additional amplification attributable to expected inflation. Expectations are therefore evaluated not as an exogenous source of price growth, but as a mechanism that may alter the magnitude, timing, and persistence of inflation once an underlying pressure is present.

\newpage
\subsection*{A.7 Parameters Setting}
The full baseline parameterization and the scenario-specific deviations are reported in Tables~\ref{tab:baseline_parameters_1}--\ref{tab:scenario_deviations}.

\begingroup
\scriptsize
\setlength{\tabcolsep}{3pt}
\renewcommand{\arraystretch}{0.92}
\begin{longtable}{p{3.2cm}p{2.2cm}p{2.6cm}p{6.8cm}}
	\caption{Baseline parameterization: population, markets, technology, and production.}
	\label{tab:baseline_parameters_1}\\
	\toprule
	\textbf{Block} & \textbf{Symbol} & \textbf{Value} & \textbf{Description} \\
	\midrule
	\endfirsthead
	
	\toprule
	\textbf{Block} & \textbf{Symbol} & \textbf{Value} & \textbf{Description} \\
	\midrule
	\endhead
	
	\bottomrule
	\endfoot
	
	Population and markets 
	& $|\mathcal{B}|$ 
	& $10$ 
	& Number of banks. \\
	
	Population and markets 
	& $|\mathcal{H}|$ 
	& $1000$ 
	& Number of households. \\
	
	Population and markets 
	& $G_C$ 
	& $20$ 
	& Number of consumption-good markets. \\
	
	Population and markets 
	& $G_K$ 
	& $10$ 
	& Number of capital/intermediate-good markets. \\
	
	Population and markets 
	& $n_g^C$ 
	& $5$ 
	& Number of consumption-good firms per market. \\
	
	Population and markets 
	& $n_g^K$ 
	& $15$ 
	& Number of capital/intermediate-good firms per market. \\
	
	Population and markets 
	& $|\mathcal{I}|$ 
	& $250$ 
	& Total number of firms. \\
	
	Population and markets 
	& $\eta_g$ 
	& $1/20$ 
	& Equal CPI and consumption-budget weights over consumption-good markets. \\
	
	Production network 
	& $d_C\equiv|\mathcal{S}_{g_C}^{C}|$ 
	& $2$ 
	& Number of upstream $K$-good markets required by each consumption-good market in the baseline. \\
	
	Production network 
	& $d_K\equiv|\mathcal{S}_{g_K}^{K}|$ 
	& $0$ 
	& Number of upstream $K$-good markets required by each intermediate-good market in the baseline. \\
	
	Technology 
	& $a_N$ 
	& mean $0.60$, sd $0.05$ 
	& Labour coefficient. \\
	
	Technology 
	& $a_X$ 
	& mean $0.08$, sd $0.01$ 
	& Intermediate-input coefficient. \\
	
	Technology 
	& $a_{NK}$ 
	& mean $0.10$, sd $0.02$ 
	& Natural-capital coefficient for $K$-firms. \\
	
	Initial conditions 
	& $Q^{\mathrm{exp}}_{i,0}$ 
	& $2$ 
	& Initial expected production. \\
	
	Initial conditions 
	& $Q^i_{i,0}$ for $s(i)=C$ 
	& $0$ 
	& Initial inventories of consumption-good firms. \\
	
	Initial conditions 
	& $Q^i_{i,0}$ for $s(i)=K$ 
	& $2$ 
	& Initial inventories of capital/intermediate-good firms. \\
	
	Production plans 
	& $\alpha_Q$ 
	& $0.20$ 
	& Speed of adjustment of expected production. \\
	
\end{longtable}
\endgroup

\newpage
\begingroup
\scriptsize
\setlength{\tabcolsep}{3pt}
\renewcommand{\arraystretch}{0.92}
\begin{longtable}{p{3.2cm}p{2cm}p{2.8cm}p{6.8cm}}
	\caption{Baseline parameterization: pricing, expectations, credit, and households.}
	\label{tab:baseline_parameters_2}\\
	\toprule
	\textbf{Block} & \textbf{Symbol} & \textbf{Value} & \textbf{Description} \\
	\midrule
	\endfirsthead
	
	\toprule
	\textbf{Block} & \textbf{Symbol} & \textbf{Value} & \textbf{Description} \\
	\midrule
	\endhead
	
	\bottomrule
	\endfoot
	
	Prices and mark-ups 
	& $p_{i,0}$ 
	& mean $1.00$, sd $0.05$ 
	& Initial posted price. \\
	
	Prices and mark-ups 
	& $\mu_{i,0}$ 
	& mean $0.15$, sd $0.03$ 
	& Initial mark-up. \\
	
	Prices and mark-ups 
	& $\underline{p}$ 
	& $0.001$ 
	& Minimum admissible posted price. \\
	
	Prices and mark-ups 
	& $\underline{\mu}$ 
	& $0.001$ 
	& Minimum admissible mark-up. \\
	
	Mark-up dynamics 
	& $\zeta_{\mu}$ 
	& $0.03$ 
	& Response of mark-ups to changes in sectoral unit-sales shares. \\
	
	Mark-up dynamics 
	& $\zeta_g$ 
	& $0$ 
	& Success-contingent mark-up pressure in the baseline. \\
	
	Mark-up dynamics 
	& $\bar{\vartheta}$ 
	& $0.80$ 
	& Sell-through threshold activating success-contingent mark-up pressure. \\
	
	Mark-up dynamics 
	& $\zeta_u$ 
	& $0$ 
	& Mark-up response to unmet-demand pressure in the baseline. \\
	
	Mark-up dynamics 
	& $\zeta_I$ 
	& $0$ 
	& Inventory discipline in mark-up adjustment in the baseline. \\
	
	Choice rules 
	& $\psi$ 
	& $1$ 
	& Weight of mark-up/brand attractiveness in supplier and consumer choice. \\
	
	Choice rules 
	& $\varphi$ 
	& $1$ 
	& Price sensitivity in supplier choice. \\
	
	Expectations 
	& $\lambda_i$ 
	& mean $0.45$, sd $0.10$ 
	& Adaptive gain in price expectations. \\
	
	Expectations 
	& $\varsigma_i^{\max}$ 
	& $5$ 
	& Maximum memory length in belief correction. \\
	
	Expectations 
	& $\theta$ 
	& $0.65$ 
	& Geometric decay parameter in expectation weights. \\
	
	Expectations 
	& $\gamma$ 
	& $1$ 
	& Magnitude-sensitivity parameter in expectation weights. \\
	
	Expectations 
	& $\{\omega_{i,h,t}\}$ 
	& combined 
	& Baseline expectation-weighting rule. \\
	
	Expectations 
	& $\pi^e_{i,t}$ anchor 
	& price anchoring 
	& Baseline expected-inflation anchoring rule. \\
	
	Expectations 
	& $\chi_{\pi}$ 
	& $0.03$ 
	& Strength of the aggregate inflation signal in expectations. \\
	
	Pricing 
	& $\kappa$ 
	& $0.15$ 
	& Strength of expected price changes in the pricing rule. \\
	
	Pricing 
	& -- 
	& $0.25$ 
	& Bound on expected inflation entering the pricing rule. \\
	
	Wage process 
	& $w_0$ 
	& $1$ 
	& Initial wage level. \\
	
	Wage process 
	& $\omega_w$ 
	& $0.90$ 
	& Autoregressive coefficient of the wage process. \\
	
	Wage process 
	& $d$ 
	& $0.10$ 
	& Drift of the wage process. \\
	
	Wage process 
	& $\sigma_w$ 
	& $0.01$ 
	& Standard deviation of the wage shock. \\
	
	Wage process 
	& $\underline{w}$ 
	& $0.001$ 
	& Minimum wage. \\
	
	Natural capital 
	& $p_{NK,0}$ 
	& $1$ 
	& Baseline natural-capital price. \\
	
	Banking 
	& $\rho$ 
	& $0.02$ 
	& Policy rate. \\
	
	Banking 
	& $m_b$ 
	& mean $0.04$, sd $0.015$ 
	& Bank lending mark-up. \\
	
	Banking 
	& $\varepsilon_{i,t}$ 
	& sd $0.005$ 
	& Idiosyncratic loan-rate shock. \\
	
	Credit 
	& $\chi$ 
	& $0.60$ 
	& Fraction of variable costs financed through credit. \\
	
	Credit 
	& $\delta$ 
	& $0$ 
	& Baseline loan rejection/default-risk parameter. \\
	
	Credit 
	& -- 
	& probabilistic 
	& Baseline credit approval regime. \\
	
	Credit 
	& -- 
	& random 
	& Baseline bank-choice rule. \\
	
	Credit scoring 
	& -- 
	& $0.40$ 
	& Strength of endogenous credit scoring. \\
	
	Credit scoring 
	& -- 
	& $0.20$ 
	& Weight of financial need in credit scoring. \\
	
	Credit scoring 
	& -- 
	& $0.04$ 
	& Spread response to borrower score. \\
	
	Banking 
	& $N_b$ 
	& $10$ 
	& Bank labour requirement. \\
	
	Households 
	& -- 
	& $0.80$ 
	& Share of worker households. \\
	
	Households 
	& $c_W$ 
	& $0.85$ 
	& Mean propensity to consume of worker households. \\
	
	Households 
	& $c_X$ 
	& $0.55$ 
	& Mean propensity to consume of profit-recipient households. \\
	
	Households 
	& $\mathrm{sd}(c_z)$ 
	& $0.05$ 
	& Standard deviation of individual propensities to consume. \\
	
	Simulation controls 
	& $T$ 
	& $500$ 
	& Simulation horizon. \\
	
	Simulation controls 
	& $\varepsilon$ 
	& $10^{-9}$ 
	& Numerical tolerance. \\
	
\end{longtable}
\endgroup 
\newpage
\begingroup
\scriptsize
\setlength{\tabcolsep}{3pt}
\renewcommand{\arraystretch}{0.92}
\begin{longtable}{p{3.3cm}p{4.2cm}p{7.7cm}}
	\caption{Scenario-specific deviations from the baseline.}
	\label{tab:scenario_deviations}\\
	\toprule
	\textbf{Scenario} & \textbf{Deviating parameters} & \textbf{Description} \\
	\midrule
	\endfirsthead
	
	\toprule
	\textbf{Scenario} & \textbf{Deviating parameters} & \textbf{Description} \\
	\midrule
	\endhead
	
	\bottomrule
	\endfoot
	
	Mark-up pressure 
	& $\zeta_{\mu}=0.08$, $\zeta_g=0.01$, $\bar{\vartheta}=0.60$, $\psi=2.0$, $\varphi=0.7$ 
	& Firms' mark-ups become more responsive to competitive performance and success-contingent pricing power; brand/mark-up attractiveness is strengthened and supplier price sensitivity is reduced. \\
	
	Bank-cost steps 
	& $\chi=1.00$; $m_b \leftarrow m_b+0.05$ at ticks $150$, $250$, and $350$ 
	& Financial-cost scenario in which banks raise lending mark-ups stepwise. The shock directly increases firm-specific loan rates and financing costs. \\
	
	Policy-rate steps 
	& $\rho=\rho_0+0.05$, $\rho_0+0.10$, $\rho_0+0.15$ at ticks $150$, $250$, and $350$ 
	& Financial-cost scenario in which the cost increase originates from the policy rate. \\
	
	Policy-rate credit regimes 
	& probabilistic; endogenous scoring 
	& Policy-rate steps are simulated under both probabilistic credit approval and endogenous-scoring credit approval. In the endogenous-scoring case, $\delta=0.01$, scoring strength is $0.60$, financial-need weight is $0.25$, and score-spread strength is $0.08$. \\
	
	Natural-capital cost, low dependence 
	& $d_C=4$, $d_K=0$, $a_X$ mean $=0.14$, $a_{NK}$ mean $=0.10$; $p_{NK}=1.10$, $1.50$, $2.00$ at ticks $150$, $250$, and $350$ 
	& Moderate natural-capital dependence. The shock is upstream and pass-through depends on downstream exposure to intermediate inputs. \\
	
	Natural-capital cost, high dependence 
	& $d_C=4$, $d_K=0$, $a_X$ mean $=0.28$, $a_{NK}$ mean $=0.25$; $p_{NK}=1.10$, $1.50$, $2.00$ at ticks $150$, $250$, and $350$ 
	& Strong natural-capital dependence. Both intermediate-input intensity and natural-capital intensity are increased to make upstream resource-cost pass-through visible. \\
	
	Expectation-rule experiments 
	& price anchoring; expectations-on-expectations; combined weights; $\kappa=0.35$, $\chi_{\pi}=0.10$ 
	& Tests whether expectations generate inflation on their own or amplify existing mark-up, financial-cost, policy-rate, or natural-capital pressures. \\
	
Network grid 
&
\begin{tabular}[t]{@{}l@{}}
	$(d_C,d_K)\in\{(1,0),(2,0),(3,0),$\\
	$\qquad \,\,\, \quad \qquad (4,0),(5,0),(2,1),$\\
	$\qquad \,\,\, \quad \qquad (2,2),(4,1),(5,1),$\\
	$\qquad \,\,\, \quad \qquad (4,2),(5,2)\}$ \\
\end{tabular}
&
Comparative-statics exercise over alternative production-network architectures.
\\
	
	Network contexts 
	& alone; mark-up; bank-cost; policy-rate; natural-capital 
	& Network architectures are crossed with selected pressure contexts to distinguish propagation from autonomous sources of inflation. \\
	
\end{longtable}
\endgroup

\newpage
\subsection*{A. 8 Monte Carlo convergence}
\label{app:mc_convergence}

\begin{figure}[htbp]
	\centering
	\includegraphics[width=0.65\textwidth]{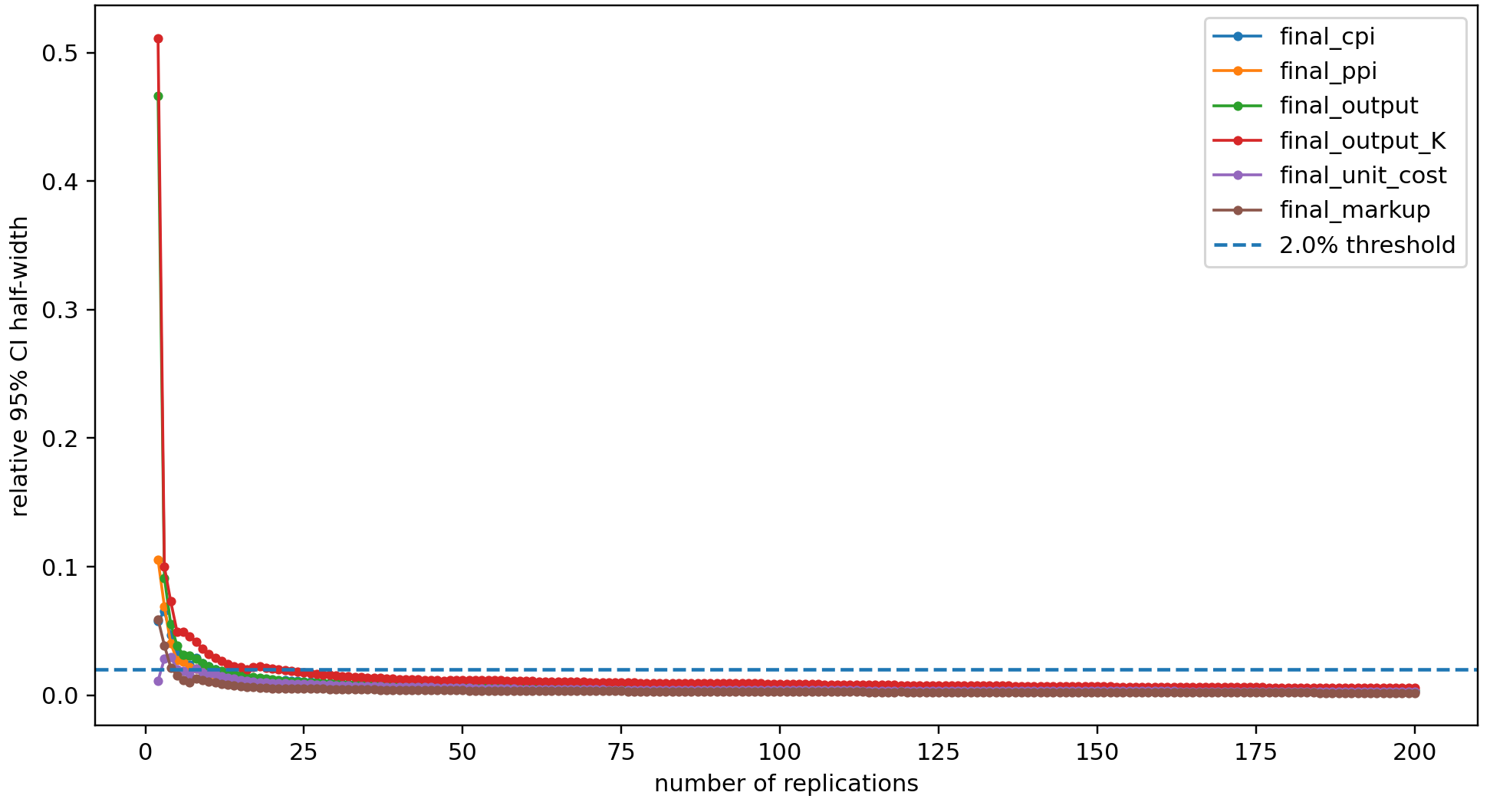}
	\caption{Monte Carlo convergence diagnostic. The figure reports the relative half-width of the $95\%$ confidence interval of the Monte Carlo mean as the number of replications increases. The dashed line marks the $2\%$ precision threshold.}
	\label{fig:mc_precision}
\end{figure}

Figure~\ref{fig:mc_precision} reports a Monte Carlo convergence diagnostic for the baseline economy. For each number of replications $N$, the figure plots the relative half-width of the $95\%$ confidence interval of the Monte Carlo mean, \(\frac{1.96\,s_N/\sqrt{N}}{|\bar{x}_N|}\), where $\bar{x}_N$ and $s_N$ are, respectively, the sample mean and standard deviation computed over the first $N$ replications. The dashed horizontal line marks the $2\%$ threshold. The diagnostic shows that, for the main variables used in the analysis, the estimated Monte Carlo mean stabilises rapidly and reaches acceptable precision after $N=24$ replications. We used $25$, for each simulation scenario.

\end{document}